\documentclass{llncs}

\usepackage{mathptmx}
\usepackage{helvet}
\usepackage{courier}
\usepackage{multicol}
\usepackage[bottom]{footmisc}
\usepackage{amssymb}
\usepackage{rotating}

\usepackage{epsfig}
\usepackage{latexsym}
\usepackage{graphicx}
\usepackage{color}
\usepackage{url}
\usepackage{amsfonts}
\usepackage{pgf,pgfarrows,pgfnodes,pgfautomata,pgfheaps,pgfshade}
\usepackage{tikz}

\usetikzlibrary{arrows,decorations.pathmorphing,backgrounds,positioning,fit,petri}
\usepackage{etex}

\usepackage{stmaryrd}
\usepackage{amsmath}

\usepackage{bussproofs}

\usepackage{lineno}

\newcommand{\post}[1]{\mbox{$#1^{\bullet}$}}
\newcommand{\pre}[1]{\mbox{$^{\bullet}#1$}}

\newcommand{\cons}{\mathcal{C}}

\newcommand{\fine}{{\mbox{ }\nolinebreak\hfill{$\Box$}}}

\newcommand{\dec}{\mbox{$dec$}}
\newcommand{\sost}[2]{\mbox{$\{#1/#2\}$}}

\newcommand{\deriv}[1]{{\mbox{${\:\stackrel{#1}{\longrightarrow}\:}$}}}

\newcommand{\Deriv}[1]{{\mbox{${\:\stackrel{#1}{\Longrightarrow}\:}$}}}

\newcommand{\Derivstar}[1]{{\mbox{${\:\stackrel{#1}{\Longrightarrow}\!\!\!\phantom{}^*\:}$}}}

\newcommand{\nderiv}[1]{\nrightarrow}

\newcommand{\eqdef}{ \doteq }

\newcommand{\bigfrac}[2]{
\renewcommand{\arraystretch}{1.5}
\begin{array}{c}#1\\
\hline
#2
\end{array}}
\newcommand{\restr}[1]{\mbox{$({\bf\nu} #1)$}}
\newcommand{\para}{\mbox{$\,|\,$}}

\newcommand{\nil}{\mbox{\bf 0}}

\newcommand{\const}[1]{\mbox{{\it Const}$(#1)$}}
\newcommand{\wf}[1]{\mbox{{\it wf}$(#1)$}}

\renewcommand{\mid}{\;\;\big|\;\;}

\newcommand{\nat}{{\mathbb N}}

\newcommand{\dom}{\mathit{dom}}

\newcommand{\MSync}{\mathit{MSync}}

\newcommand{\Sync}{\mathit{Sync}}

\newcommand{\term}[2]{\ensuremath{\mathcal{T}_{#2}(#1)}}

\begin{document}

 \pagestyle{headings}

\title{Compositional Semantics of Finite Petri Nets}
\author{Roberto Gorrieri\\
\institute{Dipartimento di Informatica --- Scienza e Ingegneria\\
Universit\`a di Bologna, \\Mura A. Zamboni, 7,
40127 Bologna, Italy}
\email{{\small roberto.gorrieri@unibo.it}}
}

\maketitle

\begin{abstract}
Structure-preserving bisimilarity \cite{G15} is a truly concurrent behavioral equivalence for finite Petri nets, 
which relates markings (of the same size only) 
generating the same causal nets, hence also the same partial orders of events. The process algebra FNM \cite{Gor17} truly represents
all (and only) the finite Petri nets, up to isomorphism. We prove that structure-preserving bisimilarity is a congruence  w.r.t. the FMN operators, 
In this way, we have defined 
a compositional semantics, fully respecting causality and the branching
structure of systems, for the class of all the finite Petri nets. Moreover, we study some algebraic properties of structure-preserving bisimilarity, that are at the base of a sound (but incomplete) axiomatization over FNM process terms.
\end{abstract}

%
\section{Introduction}
%

Structure-preserving bisimilarity \cite{G15} (sp-bisimilarity, for short) is a truly concurrent, bisimulation-based, behavioral equivalence for finite P/T Petri nets \cite{Pet81,DesRei98}, 
such that if two 
markings are related, then they generate the same causal nets \cite{GR83,BD87,Old,G15}, hence ensuring that they have the same size and 
generate the same partial orders of events. As discussed in \cite{G15}, sp-bisimilarity $\sim_{sp}$ is the coarsest behavioral equivalence that respects the following three aspects:
$(i)$ the branching 
structure (being based on the concept of bisimulation), $(ii)$ causality (being slightly finer than {\em fully-concurrent bisimilarity} \cite{BDKP91})
and $(iii)$ {\em inevitability} \cite{MOP89}, meaning that if two
systems are equivalent, and in one the occurrence of a certain action is inevitable, then so is it in the other one. Moreover,
$\sim_{sp}$ is decidable on bounded finite Petri nets \cite{CG21a}, while its decidability for unbounded nets is an open problem.

Finite P/T Petri nets can be represented, up to isomorphism, by means of the process algebra FNM \cite{Gor17}, which is a CCS-like
sub-calculus extended with an additional operator, called {\em strong prefixing}. This operator allows for 
atomic execution of action sequences, so that
a multi-party interaction can be modeled as an atomic execution of a sequence of binary interactions.

The main aim of this paper is to show that sp-bisimilarity is a congruence for the operators of FNM, thus yielding 
a compositional semantics, up to sp-bisimilarity, for the class of all the finite P/T Petri nets. Moreover, we study some algebraic properties of
sp-bisimilarity and, based on these, we introduce a sound (but incomplete) axiomatization over FNM process terms.

The paper is organized as follows. Section \ref{def-sec} introduces the basic definitions about Petri nets, some novel definitions (notably, the 
distinction between dynamically vs statically reachable subnet) and the definition of structure-preserving bisimilarity. 
Section \ref{fnm-sec} recalls from \cite{Gor17} the process algebra FNM, its (operational) semantics in terms of 
finite P/T Petri nets, and also
the representability theorem stating that each finite (statically reachable) marked Petri net can be represented by a suitable FNM term, up to isomorphism.
Section \ref{comp-sp-bis-sec} shows that sp-bisimilarity is a congruence for the FNM operators. 
Section \ref{comp-sp-alg-sec} presents a set of algebraic properties of sp-bisimilarity, some of which are conditional, and hints that 
these do not cover all the possible equalities that can be singled out.
Based on this set of algebraic properties, Section \ref{comp-sp-ax-sec} shows a sound, but incomplete, 
axiomatization of sp-bisimilarity over FNM process terms.
Finally, Section \ref{conc-comp-sec} is devoted to some conclusions.

%
\section{Basic Definitions} \label{def-sec}
%

\begin{definition}\label{multiset}{\bf (Multiset)}\index{Multiset}
Let $\nat$ be the set of natural numbers. 
Given a countable set $S$, a {\em finite multiset} over $S$ is a function $m: S \rightarrow\nat$ such 
that the {\em support} set $dom(m) = \{ s \in S \mid m(s) \neq 0\}$ is finite. 
The set of all finite multisets 
over $S$,  denoted by ${\mathcal M}_{fin}(S)$, is ranged over by $m$, possibly indexed. 
(The set of all finite subsets of $S$ is denoted by
$\mathcal{P}_{fin}(S)$.)
We write $s \in m$ if $m(s)>0$.
The {\em multiplicity} of $s$ in $m$ is given by the number $m(s)$. The {\em size} of $m$, denoted by $|m|$,
is the number $\sum_{s\in S} m(s)$, i.e., the total number of its elements.
A multiset $m$ such 
that $dom(m) = \emptyset$ is called {\em empty} and is denoted by $\theta$.
We write $m \subseteq m'$ if $m(s) \leq m'(s)$ for all $s \in S$. 
We also write $m \subset m'$ if 
$m \subseteq m'$ and $m(s) < m'(s)$ for some $s \in S$. 

{\em Multiset union} $\_ \oplus \_$ is defined as: $(m \oplus m')(s)$ $ = m(s) + m'(s)$.
This operation is commutative, associative and has $\theta$ as neutral element. 
{\em Multiset difference} $\_ \ominus \_$ is defined as: 
$(m_1 \ominus m_2)(s) = max\{m_1(s) - m_2(s), 0\}$.
The {\em scalar product} of a natural $j$ with $m$ is the multiset $j \cdot m$ defined as
$(j \cdot m)(s) = j \cdot (m(s))$. By $s_i$ we denote the multiset with $s_i$ as its only element.
Hence, a multiset $m$ over $S = \{s_1, \ldots, s_n\}$
can be represented as $k_1\cdot s_{1} \oplus k_2 \cdot s_{2} \oplus \ldots \oplus k_n \cdot s_{n}$,
where $k_j = m(s_{j}) \geq 0$ for $j= 1, \ldots, n$.
\fine
\end{definition}

\begin{definition}\label{pt-net-def}{\bf (Place/Transition Petri net)}
A labeled {\em Place/Transition} Petri net (P/T net, or Petri net, for short) is a tuple $N = (S, A, T)$, where
\begin{itemize}
\item 
$S$ is the countable set of {\em places}, ranged over by $s$ (possibly indexed),
\item 
$A \subseteq Lab$ is the finite set of {\em labels}, ranged over by $a$ (possibly indexed), and
\item 
$T \subseteq ({\mathcal M}_{fin}(S) \setminus \{\theta\}) \times A \times {\mathcal M}_{fin}(S)$ 
is the countable set of {\em transitions}, 
ranged over by $t$ (possibly indexed).
\end{itemize}
A Petri net $N = (S, A, T)$ is {\em finite} when $S$ and $T$ are finite sets.
Given a transition $t = (m, a, m')$,
we use the notation  $\pre t$ to denote its {\em pre-set} $m$ (which cannot be an empty multiset) of tokens to be consumed;
$\ell(t)$ for its {\em label} $a$, and
 $\post t$ to denote its {\em post-set} $m'$ of tokens to be produced.
Hence, transition $t$ can be also represented as $\pre t \deriv{\ell(t)} \post t$.
 \fine
\end{definition}

In the graphical description of P/T nets, 
places (represented as circles) and
transitions (represented as boxes) 
are connected by directed arcs.
The arcs may be labeled with the number representing how many tokens
are to be removed from (or produced into) that place; no label on
the arc is interpreted as the number one, i.e., one token flowing on the arc. 
This numerical label of the arc is called its {\em weight}.

\begin{definition}\label{net-system}{\bf (Marking, P/T net system)}
Given a P/T net  $N = (S, A, T)$, a finite multiset over $S$  is called a {\em marking}. Given a marking $m$ and a place $s$, 
we say that the place $s$ contains $m(s)$ {\em tokens}, graphically represented by $m(s)$ bullets
inside place $s$.
A {\em P/T net system} $N(m_0)$ is a tuple $(S, A, T, m_{0})$, where $(S,A, T)$ is a P/T net and $m_{0}$ is  
a marking over $S$, called
the {\em initial marking}. We also say that $N(m_0)$ is a {\em marked} net.
\fine
\end{definition}

\begin{definition}\label{firing-system}{\bf (Enabling, firing sequence, reachable marking)}
Given a P/T net $N = (S, A, T)$, a transition $t $ is {\em enabled} at $m$, 
denoted by $m[t\rangle$, if $\pre t \subseteq m$. 
The execution (or {\em firing}) of  $t$ enabled at $m$ produces the marking $m' = (m \ominus  \pre t) \oplus \post t$. 
This is written $m[t\rangle m'$. 
A {\em firing sequence} starting at $m$ is defined inductively as follows:
\begin{itemize}
\item $m[\epsilon\rangle m$ is a firing sequence (where $\epsilon$ denotes an empty sequence of transitions) and
\item if $m[\sigma\rangle m'$ is a firing sequence and $m' [t\rangle m''$, then
$m [\sigma t\rangle m''$ is a firing sequence. 
\end{itemize}

\noindent
The set of {\em reachable markings} from $m$, denoted by
$[m\rangle$, is $[m\rangle = \{m' \mid \exists \sigma.
m[\sigma\rangle m'\}$. 
\fine
\end{definition}

The reachable markings of a finite net can be countably infinitely many when the net is not bounded, 
i.e., when the number of tokens in some places 
can grow unboundedly.

\begin{definition}\label{net-classes}{\bf (Classes of finite P/T Nets)}
A {\em finite marked}  P/T net $N = (S, A, T, m_0)$ is: 
\begin{itemize}
    \item {\em safe} if each place contains at most one token in each reachable marking, i.e.,  
    $\forall m \in [m_0 \rangle,\forall s \in S,$ $ m(s) \leq 1 $.
   
    \item {\em bounded} if the number of tokens in each place is bounded by some $k$ for each reachable 
    marking, i.e., $\exists k \in \nat$ such that 
    $ \forall m \in [m_0 \rangle , \forall s \in S$ we have that $m(s) \leq k$.
    If this is the case, we say that the net is $k$-bounded (hence, a safe net is just a 1-bounded net). 
    \end{itemize} 
A {\em finite}  P/T net $N = (S, A, T)$ is a {\em BPP net} if $\forall t \in T$ we have $|\pre{t}| = 1$.
\fine
\end{definition}

We now recall a basic behavioral equivalence on P/T nets, derived from standard bisimilarity on labeled transition systems \cite{Mil89,GV15}.

\begin{definition}\label{def-int-bis}{\bf (Interleaving Bisimulation)}
Let $N = (S, A, T)$ be a P/T net. 
An {\em interleaving bisimulation} is a relation
$R\subseteq {\mathcal M}(S) \times {\mathcal M}(S)$ such that if $(m_1, m_2) \in R$
then
\begin{itemize}
\item $\forall t_1$ such that  $m_1[t_1\rangle m'_1$, $\exists t_2$ such that $m_2[t_2\rangle m'_2$ 
with $l(t_1) = l(t_2)$ and $(m'_1, m'_2) \in R$,
\item $\forall t_2$ such that  $m_2[t_2\rangle m'_2$, $\exists t_1$ such that $m_1[t_1\rangle m'_1$ 
with $l(t_1) = l(t_2)$ and $(m'_1, m'_2) \in R$.
\end{itemize}

Two markings $m_1$ and $m_2$ are {\em interleaving bisimilar}, 
denoted by $m_1 \sim_{int} m_2$, if there exists an interleaving bisimulation $R$ such that $(m_1, m_2) \in R$.
\fine
\end{definition}

Interleaving bisimilarity was proved undecidable in \cite{Jan95} for finite P/T nets having at least two unbounded places, 
with a proof based on the comparison of two {\em sequential} P/T nets, 
where a P/T net is sequential if it does not offer any concurrent behavior. Hence, interleaving bisimulation equivalence is 
undecidable even for the subclass of sequential finite P/T nets. Esparza observed in \cite{Esp98} that all the non-interleaving 
bisimulation-based equivalences (in the spectrum ranging from interleaving bisimilarity to fully-concurrent bisimilarity \cite{BDKP91})
collapse to interleaving bisimilarity over sequential P/T nets. Hence, the proof in \cite{Jan95} applies to all these
non-interleaving bisimulation equivalences as well.\\

Given a P/T net $N = (S, A, T)$ and a marking $m$, we say that two transitions $t_1, t_2$
are {\em concurrently enabled} at $m$ if $\pre t_1 \oplus \pre t_2 \subseteq m$. The {\em concurrent firing} of these
two transitions produces the marking $m' = (m \ominus (\pre t_1 \oplus \pre t_2)) \oplus (\post{t_1} \oplus \post{t_2})$.
We denote this fact by $m [\{t_1, t_2\}\rangle m'$.
It is also possible that the same transition is {\em self-concurrent} at some marking $m$, meaning that two or more 
occurrences of it are concurrently enabled at $m$. 

We can generalize the definition of concurrently enabled transitions to 
a finite, nonempty multiset $G$ over the set $T$, called a {\em step}.
A step $G: T \rightarrow \nat$ is enabled at marking $m$ if $\pre{G} \subseteq m$, where 
$\pre{G} =\bigoplus_{t \in T}G(t)\cdot \pre{t}$ and $G(t)$
denotes the number of occurrences of transition $t$ in the step $G$.
The execution of a step $G$ enabled at $m$ produces the marking $m' = (m \ominus \pre{G})\oplus \post{G}$, 
where $\post{G} = \bigoplus_{t \in T}G(t)\cdot\post{t}$. This is written $m[G\rangle m'$. 
We sometimes refer to
this as the {\em concurrent token game}, in opposition to the {\em sequential} token game of Definition \ref{firing-system}.
The label $l(G)$ of a step $G$ is the multiset $l(G):A \rightarrow \nat$ defined as follows:
$l(G)(a) = \sum_{t_i \in \dom(G). l(t_i) = a} G(t_i)$.

Now we define a notion of bisimulation based on the firing of steps, rather than of single transitions (as for interleaving bisimulation),
originally proposed in \cite{NT84}.

\begin{definition}\label{def-step-bis}{\bf (Step Bisimulation)}
Let $N = (S, A, T)$ be a P/T net. 
A {\em step bisimulation} is a relation
$R\subseteq {\mathcal M}(S) \times {\mathcal M}(S)$ such that if $(m_1, m_2) \in R$
then
\begin{itemize}
\item $\forall G_1$. $m_1[G_1\rangle m'_1$, $\exists G_2$ such that $m_2[G_2\rangle m'_2$ 
with $l(G_1) = l(G_2)$ and $(m'_1, m'_2) \in R$,
\item $\forall G_2$. $m_2[G_2\rangle m'_2$, $\exists G_1$ such that $m_1[G_1\rangle m'_1$ 
with $l(G_1) = l(G_2)$ and $(m'_1, m'_2) \in R$.
\end{itemize}

Two markings $m_1$ and $m_2$ are {\em step bisimilar} (or {\em step bisimulation equivalent}), 
denoted by $m_1 \sim_{s} m_2$, if there exists a step bisimulation $R$ such that $(m_1, m_2) \in R$.
\fine
\end{definition}

Of course, $\sim_{s}$ is finer than $\sim_{int}$; moreover,
also step bisimilarity is undecidable for P/T nets having at least two unbounded places \cite{Esp98}.

%
\subsection{Dynamically Reachable and Statically Reachable Subnets}\label{dyn-stat-reach-sec}
%


Now we introduce two different notions of {\em reachable subnet}, following \cite{Gor17}:
\begin{itemize}
\item {\em dynamically} reachable subnet, 
which refers to the set of places and transitions
dynamically reachable from the initial marking $m_0$ by the token game (as described in Definition \ref{firing-system}), and 
\item {\em statically} reachable subnet, which refers to the places and transitions which 
are reachable by a weaker token game, stating that
 a transition $t$ is statically enabled at $S'$ (i.e., the current set of the statically reached places 
 (initiallly $dom(m_0)$)), when $dom(\pre{t})$ is a subset of $S'$.
 \end{itemize}
 
 Based on the latter, we introduce the notion of {\em statically reduced} net, i.e., a net where all of its places and transitions 
are statically reachable from $dom(m_0)$. This notion is important in Section \ref{fnm-sec}, because we will 
show that the net semantics $Net(p)$ of an FNM term $p$ is
a finite, statically reduced P/T net. 

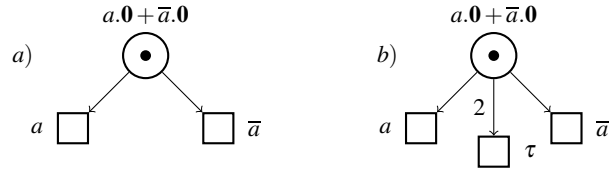
\begin{figure}[t]
\centering
\begin{tikzpicture}
[
every place/.style={draw,thick,inner sep=0pt,minimum size=6mm},
every transition/.style={draw,thick,inner sep=0pt,minimum size=4mm},
bend angle=45,
pre/.style={<-,shorten <=1pt,>=stealth,semithick},
post/.style={->,shorten >=1pt,>=stealth,semithick}
]
\def\eofigdist{4cm}
\def\eodist{0.75}

\node (a) [label=left:$a)\qquad \qquad $]{};

\node (p1) [place,tokens=1]  [label=above:$a.\nil + \overline{a}.\nil$] {};
\node (t1) [transition] [below left=\eodist of p1,label=left:$a\;$] {};
\node (t2)  [transition] [below right=\eodist of p1,label=right:$\;\overline{a}$] {};

\draw  [->] (p1) to (t1);
\draw  [->] (p1) to (t2);

  
 \node (b) [right={3.5cm} of a,label=left:$b)\;\;$] {};

\node (p2) [place,tokens=1]  [right=\eofigdist of p1,label=above:$a.\nil + \overline{a}.\nil$] {};
\node (t3) [transition] [below left=\eodist of p2,label=left:$a\;$] {};
\node (t4)  [transition] [below right=\eodist of p2,label=right:$\;\overline{a}$] {};
\node (t5)  [transition] [below=\eodist of p2,label=right:$\;\tau$] {};

\draw  [->] (p2) to (t3);
\draw  [->] (p2) to (t4);
\draw  [->] (p2) to node[auto,swap] {2} (t5);

\end{tikzpicture}
\caption{The dynamically reachable subnet (a) and the statically reachable subnet (b) for $a.\nil + \overline{a}.\nil$}
\label{static-fig}
\end{figure}

In fact, the
semantics of a term $p$ is not simply the net that can be dynamically reached from the initial marking $\dec(p)$ corresponding to $p$, 
rather it is the net describing all the potential behaviors of $p$, as if the number of tokens 
in $\dec(p)$ can be increased at will.
For instance, 
consider the FNM sequential process $p = a.\nil + \overline{a}.\nil$; its dynamically reachable subnet,
outlined in Figure \ref{static-fig}(a), shows the expected behavior that $p$ can perform either $a$ or its complementary action 
$\overline{a}$. On the contrary, its statically reachable subnet in (b)
describes additionally the potential
self-synchronization of $p$ with another copy of itself; in this way, the net semantics for $p$ and for the parallel term
$p \para p$
differ only for the form of the initial marking (one token for $p$ and two tokens for $p \para p$), but the 
underlying net is the same. 
The statically reachable subnet of a finite P/T net is always algorithmically computable, also for
Nonpermissive nets \cite{Gor17}, even though they are 
a Turing-complete model of computation, 
and so for them, as the reachability problem is undecidable,
it is not always possible to compute the dynamically reachable subnet. 

Now we provide the formal definitions for these two different forms of reachable subnet.

\begin{definition}\label{dyn-reach-def}\index{P/T Petri system!dynamically reachable subnet}{\bf (Dynamically reachable subnet, 
dynamically reduced net)}
Given a P/T Petri net system $N(m_0) =$  $ (S, A, T, m_{0})$, the {\em dynamically reachable subnet}
$Net_d(N(m_0))$ is $(S', A', T', m_0)$, where  

$
\begin{array}{rcl}
S' & = & \{s\in S \mid \exists m\in[m_0\rangle \mbox{ such that } m(s) \geq 1\},\\
T' &=& \{ t \in T \mid
\exists m\in[m_0\rangle \mbox{ such that } m[t\rangle \},\\
A' &=& \{a \mid \exists t \in T' \mbox{ such that } \ell(t) = a\}.
\end{array}
$

\noindent
A P/T net system $N(m_0) = (S, A, T, m_{0})$ is {\em dynamically reduced} if $N(m_0) = Net_d(N(m_0))$, i.e., the net 
system is equal to its dynamically reachable subnet.
\fine
\end{definition}

\begin{figure}[t]
\centering
\begin{tikzpicture}[
every place/.style={draw,thick,inner sep=0pt,minimum size=6mm},
every transition/.style={draw,thick,inner sep=0pt,minimum size=4mm},
bend angle=30,
pre/.style={<-,shorten <=1pt,>=stealth,semithick},
post/.style={->,shorten >=1pt,>=stealth,semithick}
]
\def\eofigdist{5cm}
\def\eodist{0.5cm}
\def\eodisty{0.9cm}

\node (p1) [place,tokens=1]  [label=above:$s_1$] {};
\node (p2) [place,tokens=2]  [right=\eodisty of p1,label=above:$s_2$] {};
\node (p3) [place]  [right=\eodist of p2,label=above:$s_3$] {};

\node (t1) [transition] [below left={1cm} of p1,label=left:$a\;$] {};
\node (t2) [transition] [below=\eodist of p1,label=right:$\;b$] {};
\node (t3) [transition] [below=\eodist of p2,label=right:$\;c$] {};
\node (t4) [transition] [below=\eodist of p3,label=right:$\;d$] {};

\node (p4) [place] [below=\eodist of t2,label=below:$s_4$]{};
\node (p5) [place] [below=\eodist of t3,label=below:$s_5$]{};

\node (a) [label=left:$a)\quad \qquad $]{};

\draw  [->] (p1) to (t2);
\draw  [->] (t2) to (p4);
\draw  [->, bend left] (p4) to (t1);
\draw  [->, bend left] (t1) to (p1);
\draw  [->] (p1) to node[auto] {2} (t3);
\draw  [->] (p2) to (t3);
\draw  [->] (t3) to (p5);
\draw  [->] (p3) to (t4);

  
\node (p'1) [place,tokens=1]  [right=\eofigdist of p1, label=above:$s_1$] {};
\node (p'2) [place,tokens=2]  [right=\eodisty of p'1,label=above:$s_2$] {};

\node (t'1) [transition] [below left={1cm} of p'1,label=left:$a\;$] {};
\node (t'2) [transition] [below=\eodist of p'1,label=right:$\;b$] {};

\node (p'4) [place] [below=\eodist of t'2,label=below:$s_4$]{};

\draw  [->] (p'1) to (t'2);
\draw  [->] (t'2) to (p'4);
\draw  [->, bend left] (p'4) to (t'1);
\draw  [->, bend left] (t'1) to (p'1);

\node (b) [right={4.5cm} of a,label=left:$b)\;\;$] {};

\end{tikzpicture}
\caption{A net system in (a) and its dynamically reachable subnet in (b)}
\label{subnet}
\end{figure}
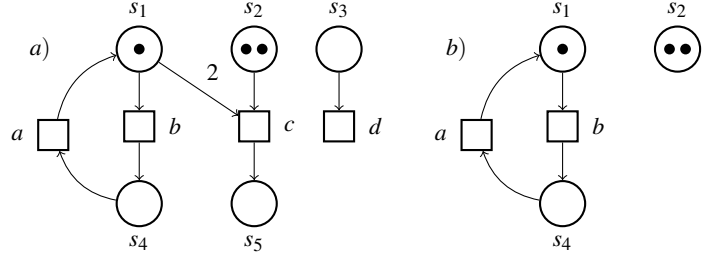

\noindent
The dynamically reachable subnet of the net in Figure \ref{subnet}(a) is outlined in 
Figure \ref{subnet}(b). Given a finite net $N(m_0)$, it is algorithmically
derivable $Net_d(N(m_0))$ by means of its coverability tree \cite{KM69}, but its complexity is 
exponential in the size of the net.

\begin{definition}\label{def-static-reach}\index{P/T Petri system!statically reachable subnet}
{\bf (Statically reachable subnet and statically reduced net)}
Given a P/T net  $N = (S, A, T)$, we say that a transition $t$ is {\em statically enabled} by a set of places
$S' \subseteq S$, denoted by $S' \llbracket t \rangle$, if $dom(\pre t) \subseteq S'$.

Given two sets of places $S_1, S_2 \subseteq S$,
we say that $S_2$ is {\em statically reachable in one step} from $S_1$
if there exists a transition $t \in T$, such that $S_1\llbracket t \rangle$, $dom(\post t) \not \subseteq S_1$ and 
$S_2 = S_1 \cup dom(\post t)$; this is denoted by $S_1 \Deriv{t} S_2$.
The {\em static reachability relation} $\Derivstar{} \subseteq \mathcal{P}_{fin}(S) \times \mathcal{P}_{fin}(S)$ is the 
least relation such that
\begin{itemize}
\item
$S_1\Derivstar{} S_1$ and
\item 
if $S_1 \Derivstar{} S_2$ and $S_2 \Deriv{t} S_3$, then $S_1 \Derivstar{} S_3$.
\end{itemize}
A set of places $S_k \subseteq S$ is the {\em largest} set statically reachable from $S_1$
if $S_1 \Derivstar{} S_k$ and for all $t \in T$ such that $S_k \llbracket t \rangle$, we have that
$dom(\post t) \subseteq S_k$.
Given a P/T net system $N(m_0) = (S, A, T, m_{0})$, we denote by $\llbracket dom(m_0) \rangle$
the largest set of places statically reachable from $dom(m_0)$, i.e., the largest $S_k$ such that
$dom(m_0) \Derivstar{} S_k$.

The {\em statically reachable subnet} $Net_s(N(m_0))$ is the net $(S', A', T', m_0)$, where

$
\begin{array}{rcl}
S' & = & \llbracket dom(m_0)\rangle,\\
T' &=& \{t \in T \mid S' \llbracket t \rangle \},\\
A' &=& \{a \mid \exists t \in T' \mbox{ such that } \ell(t) = a\}.
\end{array}
$

\noindent
A P/T  net system $N(m_0) = (S, A, T, m_{0})$ is {\em statically reduced} if $Net_s(N(m_0))$ $ = N(m_0)$,
i.e., the net system is equal to its statically reachable subnet.
\fine
\end{definition}

The statically reachable subnet 
$Net_s(N(m_0))$ of a finite Petri net $N(m_0)$ can be computed with an easy polynomial algorithm \cite{Gor17}.
We will show in Section \ref{fnm-sec} that, given an FNM process $p$, its net semantics is the statically reachable subnet from 
its initial marking $\dec(p)$, extracted from the infinite net describing the semantics of the whole FNM process algebra.



Finally, we outline some useful properties relating statically reduced nets and dynamically reduced ones.

\begin{proposition}\label{dyn-impl-stat}
Given a P/T net system $N(m_0) = (S, A, T, m_{0})$, if $N(m_0)$ is dynamically reduced, then it is also statically reduced.
%
\fine
\end{proposition}

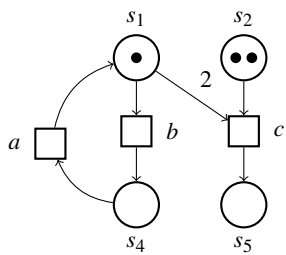
\begin{figure}[t]
\centering
\begin{tikzpicture}[
every place/.style={draw,thick,inner sep=0pt,minimum size=6mm},
every transition/.style={draw,thick,inner sep=0pt,minimum size=4mm},
bend angle=30,
pre/.style={<-,shorten <=1pt,>=stealth,semithick},
post/.style={->,shorten >=1pt,>=stealth,semithick}
]
\def\eofigdist{5cm}
\def\eodist{0.45cm}
\def\eodisty{0.8cm}

\node (p1) [place,tokens=1]  [label=above:$s_1$] {};
\node (p2) [place,tokens=2]  [right=\eodisty of p1,label=above:$s_2$] {};

\node (t1) [transition] [below left={1cm} of p1,label=left:$a\;$] {};
\node (t2) [transition] [below=\eodist of p1,label=right:$\;b$] {};
\node (t3) [transition] [below=\eodist of p2,label=right:$\;c$] {};

\node (p4) [place] [below=\eodist of t2,label=below:$s_4$]{};
\node (p5) [place] [below=\eodist of t3,label=below:$s_5$]{};

\draw  [->] (p1) to (t2);
\draw  [->] (t2) to (p4);
\draw  [->, bend left] (p4) to (t1);
\draw  [->, bend left] (t1) to (p1);
\draw  [->] (p1) to node[auto] {2} (t3);
\draw  [->] (p2) to (t3);
\draw  [->] (t3) to (p5);

\end{tikzpicture}
\caption{The statically reachable subnet of Figure \ref{subnet}(a)}
\label{subnet2}
\end{figure}

However, the converse implication is not true:
there are statically reduced P/T systems that are not dynamically reduced. E.g.,
the statically reduced P/T system $N(s_1) =$ $ (\{s_1, s_2, s_3\}, $ $\{a, b\}, \{(s_1, a, s_2), (2 \cdot s_1, b, s_3)\}, s_1)$
cannot dynamically reach  place $s_3$. 
Consider the net system in Figure \ref{subnet}(a); its statically reachable subnet is 
outlined in Figure \ref{subnet2}. If we compare it with its dynamically reachable subnet in 
Figure \ref{subnet}(b), we note
that the statically reachable subnet contains the dynamically reachable subnet. This holds in general.

\begin{proposition}
Given a P/T system $N(m_0) = (S, A, T, m_{0})$, if its dynamically reachable subnet 
$Net_d(N(m_0))$ is $(S', A', T', m_0)$ and its statically reachable subnet
$Net_s(N(m_0))$ is $(S'', A'', T'', m_0)$, then $S' \subseteq S''$, $T' \subseteq T''$
and $A' \subseteq A''$.
\fine
\end{proposition}

For some classes of nets, however, the two notions coincide, in particular for BPP nets, i.e., those nets whose transitions have singleton preset.

%
\subsection{Structure Preserving Bisimilarity}\label{sp-bis-sec}

We recall from \cite{G15} the definition of this behavioral equivalence, adapting some auxiliary notations to our aims.

\begin{definition}\label{link-def}{\bf (Linking)}
Let $N = (S, A, T)$ be a P/T net. 
A {\em link} is a pair $(s_1, s_2) \in S \times S$. 
A {\em linking} $l$ is a finite multiset of links, i.e.,
$l: S \times S \rightarrow \nat$ where $dom(l)$ is finite.
With abuse of notation, we denote by $\theta$ the empty linking, i.e., the empty multiset of links.
Given a set $L$ of links, we denote by $L^+$ the set of all the linkings over $L$, i.e., of all the multisets over $L$.

Given a linking $l$, the two projected markings $m_1$ and $m_2$ can be defined as
$m_1(s_1) = \pi_1(l)(s_1) =  \sum_{s_2 \in S} l(s_1, s_2)$ and 
$m_2(s_2) = \pi_2(l)(s_2) =  \sum_{s_1 \in S} l(s_1, s_2)$. 
For instance, if $l = \{(s_1, s_2), (s_1', s_2')\}$, then $\pi_1(l) = s_1 \oplus s_1'$
and $\pi_2(l) = s_2 \oplus s_2'$; instead, if $l = \theta$, then $\pi_1(l) = \theta$
and $\pi_2(l) = \theta$, i.e., $l$ is projected on the empty marking.

Given a linking $l$, its inverse $l^{-1}$ is defined as:  $l^{-1}(s_2,s_1) = l(s_1, s_2)$.
Given two linkings $l_1$ and $l_2$, their relational composition $l_1 \circ l_2$ 
defines a set of linkings, that can be empty (so the operation is not defined) in case $\pi_2(l_1) \neq \pi_1(l_2)$.
A linking $h \in l_1 \circ l_2$ if there exists a multiset $k: S \times S \times S \rightarrow \nat$
of triples of places, such that $l_1(s_1, s_2) = \sum_{s_3\in S} k(s_1, s_2, s_3)$,
$l_2(s_2, s_3) = \sum_{s_1\in S} k(s_1, s_2, s_3)$ and $h(s_1, s_3) = \sum_{s_2 \in S} k(s_1, s_2, s_3)$.
For instance, if $l_1 = \{(s_1, s_2), (s_1', s_2)\}$ and $l_2 = \{(s_2, s_3), (s_2, s_3')\}$, then
$l_1 \circ l_2 = \{h_1, h_2\}$, where $h_1 = \{(s_1, s_3), (s_1', s_3')\}$ and $h_2 = \{(s_1, s_3'), (s_1', s_3)\}$.

We will use $l$, $\overline{l}$, $h$, $\overline{h}$, $c$ and $\overline{c}$, possibly indexed, to range over linkings.
As linkings are multisets, we can use the operations defined over multisets, such as union $\oplus$ and difference $\ominus$.
For instance, $(l \oplus c)(s_1, s_2) = l(s_1, s_2) + c(s_1, s_2)$.
\fine
\end{definition}

 \begin{definition}\label{def-sp1-bis}{\bf (Structure Preserving Bisimulation)}
Let $N = (S, A, T)$ be a P/T net. 
A {\em structure-preserving bisimulation} (sp-bisimulation, for short)
is a set $R$ of linkings such that if $l \in R$, then $\forall c \subseteq l$, 
\begin{enumerate}
\item $\forall t_1$ such that $\pre{t_1} = \pi_1(c)$, there exist a transition $t_2$ 
such that $\ell(t_1) = \ell(t_2)$, $\pre{t_2} = \pi_2(c)$, and a linking $\overline{c}$ such that 
$\post{t_1} = \pi_1(\overline{c})$, $\post{t_2} = \pi_2(\overline{c})$ and $\overline{l} = (l \ominus c) \oplus \overline{c} \in R$; 
\item $\forall t_2$ such that $\pre{t_2} = \pi_2(c)$, there exist a transition $t_1$ 
such that $\ell(t_1) = \ell(t_2)$, $\pre{t_1} = \pi_1(c)$, and a linking $\overline{c}$ such that 
$\post{t_1} = \pi_1(\overline{c})$, $\post{t_2} = \pi_2(\overline{c})$ and $\overline{l} = (l \ominus c) \oplus \overline{c} \in R$.
\end{enumerate}
Two markings $m_1$ and $m_2$ are  {\em structure-preserving bisimilar}, denoted by
$m_1 \sim_{sp} m_2$, if there exists a linking $l$ in a structure preserving bisimulation $R$ such that $m_1 = \pi_1(l)$
and $m_2 = \pi_2(l)$.
\fine
\end{definition}
 
Note that the two matching transitions $t_1$ and $t_2$ in the structure-preserving bisimulation game must have the same shape 
because $(i)$ $\pre{t_1} = \pi_1(c)$ and $\pre{t_2} = \pi_2(c)$ implies that
$|\pre{t_1}| = |\pre{t_2}|$, then $(ii)$ $\ell(t_1) = \ell(t_2)$ and, finally, $(iii)$  from $\post{t_1} = \pi_1(\overline{c})$ and $\post{t_2} = \pi_2(\overline{c})$
we derive that $|\post{t_1}| = |\post{t_2}|$.

\begin{remark}{\bf (Structure-preserving bisimilarity implies interleaving bisimilarity)}\label{rem-int-sp}
It is easy to observe that if $R_1$ is an sp-bisimulation, then 

$R_2 = \{(\pi_1(l), \pi_2(l)) \mid l \in R_1\}$ 

\noindent is an interleaving bisimulation such that
the matching transitions in the bisimulation game have the same shape. Therefore, if $m_1 \sim_{sp} m_2$, then
\begin{enumerate}
\item $\forall t_1$ such that $m_1[t_1\rangle m_1'$, there exists $t_2$ such that $m_2[t_2\rangle m_2'$,
$|\pre{t_1}| = |\pre{t_2}|$, $\ell(t_1) = \ell(t_2)$, $|\post{t_1}| = |\post{t_2}|$ and $m_1' \sim_{sp} m_2'$;
\item $\forall t_2$ such that $m_2[t_2\rangle m_2'$, there exists $t_1$ such that $m_1[t_1\rangle m_1'$,
$|\pre{t_1}| = |\pre{t_2}|$, $\ell(t_1) = \ell(t_2)$, $|\post{t_1}| = |\post{t_2}|$ and $m_1' \sim_{sp} m_2'$.\\[-1cm]
\end{enumerate}
\fine
\end{remark}
 
\begin{remark}{\bf (Structure-preserving bisimilarity implies step bisimilarity)}\label{rem-step-sp}
It is easy to observe that if $R$ is an sp-bisimulation, then it is also 
able to match steps rather than single transitions. In fact, we want to prove that if $l \in R$, then $\forall c \subseteq l$
 \begin{enumerate}
\item $\forall G_1$ such that $\pre{G_1} = \pi_1(c)$, there exist a step $G_2$ 
such that $\ell(G_1) = \ell(G_2)$, $\pre{G_2} = \pi_2(c)$, and a linking $\overline{c}$ such that 
$\post{G_1} = \pi_1(\overline{c})$, $\post{G_2} = \pi_2(\overline{c})$ and $\overline{l} = (l \ominus c) \oplus \overline{c} \in R$; 
\item $\forall G_2$ such that $\pre{G_2} = \pi_2(c)$, there exist a step $G_1$ 
such that $\ell(G_1) = \ell(G_2)$, $\pre{G_1} = \pi_1(c)$, and a linking $\overline{c}$ such that 
$\post{G_1} = \pi_1(\overline{c})$, $\post{G_2} = \pi_2(\overline{c})$ and $\overline{l} = (l \ominus c) \oplus \overline{c} \in R$.
\end{enumerate}
This can be proved as follows. For simplicity' sake, we consider a step $G_1 = \{t_1, t_2\}$ composed of two transitions only; 
the general proof is only notationally more complex. By symmetry we consider only the first item.

If $\pre{G_1} = \pi_1(c)$, this means that $c = c_1 \oplus c_2$ such that $\pi_1(c_i) = \pre{t_i}$ for $i = 1, 2$.
Since $l \in R$ and $R$ is an sp-bisimulation, for transition $t_1$, there exist a transition $t_1'$ such that
$\ell(t_1) = \ell(t_1')$, $\pi_2(c_1) = \pre{t_1'}$, and $\overline{c_1}$ such that $\pi_1(\overline{c_1}) = \post{t_1}$,
$\pi_2(\overline{c_1}) = \post{t_1'}$ and, moreover, $(l \ominus c_1) \oplus \overline{c_1} \in R$.
Hence, for transition $t_2$, there exist a transition $t_2'$ such that $\ell(t_2) = \ell(t_2')$,
$\pi_2(c_2) = \pre{t_2'}$, and $\overline{c_2}$ such that $\pi_1(\overline{c_2}) = \post{t_2}$,
$\pi_2(\overline{c_2}) = \post{t_2'}$ and $(((l \ominus c_1) \oplus \overline{c_1}) \ominus c_2) \oplus \overline{c_2}$
$= (l \ominus (c_1 \oplus c_2)) \oplus (\overline{c_1} \oplus \overline{c_2}) $ $= (l \ominus c) \oplus \overline{c} \in R$,
by taking $\overline{c_1} \oplus \overline{c_2} = \overline{c}$.
Therefore, If $\pre{G_1} = \pi_1(c)$, then a step $G_2 = \{t_1', t_2'\}$ exists such that 
$\ell(G_1) = \ell(G_2)$, $\pre{G_2} = \pi_2(c)$, and a linking $\overline{c}$ such that 
$\post{G_1} = \pi_1(\overline{c})$, $\post{G_2} = \pi_2(\overline{c})$ and $\overline{l} = (l \ominus c) \oplus \overline{c} \in R$,
as required.

As a consequence of this observation, it follows that if $R_1$ is an sp-bisimulation, then 
$R_2 = \{(\pi_1(l), \pi_2(l)) \mid l \in R_1\}$  is a step bisimulation.
\fine
\end{remark}

Actually, it can be proved \cite{G15} that if two markings are sp-bisimilar, then they generate the same {\em causal nets} 
(also called {\em occurrence nets})
\cite{GR83,BD87,Old}, so that this behavioral semantics is very concrete and slightly finer than {\em fully-concurrent bisimilarity} \cite{BDKP91},
an adaptation to Petri nets of {\em history-preserving bisimilarity} \cite{RT88,vGG89,DDM89}.

Interestingly enough, sp-bisimilarity is {\em resource-aware}:
as a token is an instance of a 
sequential process to be executed over one processor, if two markings 
have different size, then a different number of processors is necessary. 
Hence, a behavioral semantics, such as sp-bisimilarity, equates distributed systems only if they require the same amount of execution resources.
Van Glabbeek \cite{G15} argued that structure-preserving bisimilarity (hence, also its alternative process-oriented
characterization, called {\em causal-net bisimilarity} \cite{G15,Gor22}) 
is the most appropriate behavioral equivalence for Petri nets, as it is the only one
respecting a list of 9 desirable requirements he proposed, among which there is the observation that it is the coarsest
equivalence respecting {\em inevitability} \cite{MOP89}, meaning that if two
systems are equivalent, and in one the occurrence of a certain action is inevitable, then so is it in the other one. 

Finally, note that structure-preserving bisimilarity $\sim_{sp}$ is such that it relates (dynamically) reachable markings only, i.e., 
it is a relation actually defined over $Net_d(N(m_0))$.

\begin{figure}[t]
\centering

\begin{tikzpicture}[
every place/.style={draw,thick,inner sep=0pt,minimum size=6mm},
every transition/.style={draw,thick,inner sep=0pt,minimum size=4mm},
bend angle=42,
pre/.style={<-,shorten <=1pt,>=stealth,semithick},
post/.style={->,shorten >=1pt,>=stealth,semithick}
]
\def\eofigdist{5cm}
\def\eodist{0.5cm}
\def\eodisty{1.7cm}

\node (p1) [place,tokens=1]  [label=above:$P_1$] {};
\node (p2) [place,tokens=1]  [right=\eodist of p1,label=above:$C_1$] {};
\node (t1) [transition] [below=\eodist of p1,label=left:{\em prod}] {};
\node (p3) [place] [below=\eodist of t1,label=left:$D_1$] {};
\node (t2) [transition] [below=\eodist of p3,label=left:{\em del}] {};
\node (p4) [place] [below right=\eodist of t2,label=left:$C_1'$] {};
\node (t3) [transition] [below=\eodist of p4,label=left:{\em cons}] {};

\draw  [->, bend right] (p1) to (t1);
\draw  [->] (t1) to (p3);
\draw  [->, bend right] (t1) to (p1);
\draw  [->] (p3) to (t2);
\draw  [->, bend left] (p2) to (t2);
\draw  [->] (t2) to (p4);
\draw  [->] (p4) to (t3);
\draw  [->, bend right] (t3) to (p2);

\node (q1) [place,tokens=1]  [right=\eofigdist of p1,label=above:$P_2$] {};
\node (q2) [place,tokens=1]  [right=\eodisty of q1,label=above:$C_2$] {};
\node (s1) [transition] [below left=\eodist of q1,label=left:{\em prod}] {};
\node (s'1) [transition] [below right=\eodist of q1,label=right:{\em prod}] {};

\node (q3) [place] [below=\eodist of s1,label=left:$D_2'$] {};
\node (q'3) [place] [below=\eodist of s'1,label=left:$D_2''$] {};

\node (s2) [transition] [below=\eodist of q'3,label=left:{\em del}] {};
\node (s'2) [transition] [below right={1cm} of q'3,label=left:{\em del}] {};

\node (q4) [place] [below right={0.8cm} of s2,label=left:$C_2'$] {};
\node (s3) [transition] [below=\eodist of q4,label=left:{\em cons}] {};

\draw  [->, bend left] (q1) to (s1);
\draw  [->, bend right] (q1) to (s'1);
\draw  [->, bend left] (s1) to (q1);
\draw  [->, bend right] (s'1) to (q1);
\draw  [->] (s1) to (q3);
\draw  [->] (s'1) to (q'3);
\draw  [->] (q3) to (s2);
\draw  [->] (q2) to (s2);
\draw  [->] (q'3) to (s'2);
\draw  [->] (q2) to (s'2);
\draw  [->] (s2) to (q4);
\draw  [->] (s'2) to (q4);
\draw  [->] (q4) to (s3);
\draw  [->, bend right] (s3) to (q2);

\end{tikzpicture}
\caption{Two net bisimilar unbounded producer-consumer systems}
\label{upcs-fig}
\end{figure}
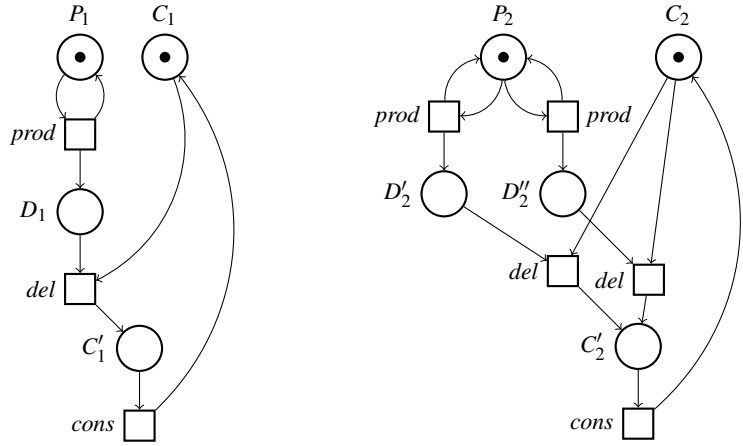

\begin{example}\label{examp-1}
Consider the two nets in Figure \ref{upcs-fig}, representing two unbounded producer-consumer systems, where {\em prod} is 
the action of producing an item,
{\em del} of delivering an item, {\em cons} of consuming an item.
We want to prove that $P_1 \oplus C_1 \sim_{sp} P_2 \oplus C_2$. This can be easily achieved by defining
the following set of links $L = \{(P_1, P_2), (C_1, C_2), (C_1', C_2'),$ $ (D_1, D_2'), (D_1, D_2'')\}$,
and then by considering the set $L^+$ of all the multisets over $L$: it is easy to realize that 
$L^+$ is a structure-preserving bisimulation, that can be used to prove not only 
that $P_1 \oplus C_1 \sim_{sp} P_2 \oplus C_2$
but also that, e.g., $2 \cdot P_1 \oplus 3 \cdot C_1 \oplus D_1 \sim_{sp} 2 \cdot P_2 \oplus 3 \cdot C_2 \oplus D_2''$
as well as $D_1 \oplus C_1' \sim_{sp} D_2' \oplus C_2'$.
As a matter of fact, $L$ is a {\em place bisimulation} \cite{ABS91,Gor21}, a behavioral relation slightly finer than
structure-preserving bisimilarity (see \cite{Gor21} for examples distinguishing place bisimilarity from sp-bisimilarity).
\fine
\end{example}

In order to show that $\sim_{sp}$ is an equivalence relation, we now list some useful properties of sp-bisimulation relations. 

\begin{proposition}\label{sp-bis-prop-bis}
For each P/T net $N = (S, A, T)$, the following hold:
\begin{enumerate}
\item Given the set $Id = \{(s,s) \mid s \in S\}$, the set $Id^+$ is an sp-bisimulation;
\item the inverse $R^{-1} = \{ l^{-1} \mid l \in R\}$ of an sp bisimulation $R$ is an sp-bisimulation;
\item the relational composition $R_1 \circ R_2 = \bigcup_{l_1 \in R_1, l_2 \in R_2} l_1 \circ l_2$ of
two sp-bisimulations $R_1$ and $R_2$ is an sp-bisimulation;
\item given a family $\{R_i\}_{i\in I}$ of  structure-preserving bisimulations, their union $\cup_{i\in I} R_i$ is an sp-bisimulation.
\end{enumerate}
\proof 
The proof is obvious, except for case 3. Assume $h \in R_1 \circ R_2$, so that there exist $l_i \in R_i$, for $i = 1, 2$ such that
$h \in l_1 \circ l_2$, as by Definition \ref{link-def}.
We have to prove that 
\begin{enumerate}
\item $\forall t_1$ such that there exists $c \subseteq h$ with $\pre{t_1} = \pi_1(c)$,
there exist a transition $t_3$ 
such that $\ell(t_1) = \ell(t_3)$, $\pre{t_3} = \pi_2(c)$, and a linking $\overline{c}$ such that  
$\post{t_1} = \pi_1(\overline{c})$, $\post{t_3} = \pi_2(\overline{c})$ and $\overline{h} = (h \ominus c) \oplus \overline{c} \in R_1 \circ R_2$; 
\item and the symmetric condition for all $t_3$.
\end{enumerate}

However, this is really possible because $l_1$ is a linking in $R_1$, $l_2$ is a linking of $R_2$ and there exist $c_1 \subseteq l_1$,
$c_2 \subseteq l_2$ such that $c \in c_1 \circ c_2$, and $\overline{c_1}, \overline{c_2}$ such that 
$\overline{c} \in \overline{c_1} \circ \overline{c_2}$. In fact, by $l_1 \in R_1$ we have that

\begin{enumerate}
\item $\forall t_1$ such that $\pre{t_1} = \pi_1(c_1)  = \pi_1(c)$, there exist a transition $t_2$ 
such that $\ell(t_1) = \ell(t_2)$, $\pre{t_2} = \pi_2(c_1)$, and a linking $\overline{c_1}$ such that 
$\post{t_1} = \pi_1(\overline{c_1}) = \pi_1(\overline{c})$, $\post{t_2} = \pi_2(\overline{c_1})$ and 
$\overline{l_1} = (l_1 \ominus c_1) \oplus \overline{c_1} \in R_1$; 
\item and the symmetric condition for all $t_2$.
\end{enumerate}
Then, since $l_2$ is a linking in $R_2$ and $\pi_2(l_1) = \pi_1(l_2)$,
\begin{enumerate}
\item $\forall t_2$ such that $\pre{t_2} = \pi_2(c_1) = \pi_1(c_2)$, there exist a transition $t_3$ 
such that $\ell(t_2) = \ell(t_3)$, $\pre{t_3} = \pi_2(c_2) = \pi_2(c)$, and a linking $\overline{c_2}$ such that 
$\post{t_2} = \pi_2(\overline{c_1}) = \pi_1(\overline{c_2})$, $\post{t_3} = \pi_2(\overline{c_2}) = \pi_2(\overline{c})$ and 
$\overline{l_2} = (l_2 \ominus c_2) \oplus \overline{c_2} \in R_2$; 
\item and the symmetric condition for all $t_3$.
\end{enumerate}
Finally, $\overline{h} \in \overline{l_1} \circ \overline{l_2}$, so that $\overline{h} \in R_1 \circ R_2$.
And this completes the argument.
\fine
\end{proposition}

\begin{proposition}\label{dis-bis-bis-eq}
For each P/T net $N = (S, A, T)$, $\sim_{sp} \; = \; 
\bigcup \{R \mid $ $R \mbox{ is an sp-bisimulation}\}$ is an equivalence relation
and the largest sp-bisimulation.
\proof
The proof, based on the observations in Proposition \ref{sp-bis-prop-bis}, is  standard.
\fine
\end{proposition}

%
\section{FNM: Syntax and Net Semantics} \label{fnm-sec}
%

In this section we briefly recall some definitions and theorems from \cite{Gor17}, where the reader can find more details.

Let ${\mathcal L}$ be a finite set of {\em names}, ranged over by 
$a, b, c, \ldots$, also called the {\em input actions}. Let  $\overline{\mathcal L}$ be the set of {\em co-names}, ranged over by 
$\overline{a}, \overline{b},\overline{c},\ldots$, also called the {\em output actions}. 
The set  ${\mathcal L}\cup\overline{\mathcal L}$, ranged over by $\alpha, \beta, \ldots$,
is the set of {\em observable actions}. 
Let $Act = {\mathcal L} \cup \overline{\mathcal L} \cup\{\tau\}$,
such that $\tau\not\in{\mathcal L}\cup\overline{\mathcal L}$, be the finite set of {\em actions} (or {\em labels}), ranged over by 
$\mu$. Action $\tau$ denotes an invisible, internal activity. 
Let $\cons$ be a finite set of process {\em  constants}, disjoint from 
$Act$, ranged over by $A, B, C,\ldots$, possibly indexed. 

{\em Finite-Net Multi-CCS} (FNM for short) is the calculus whose terms are generated from actions and constants
as described by the following abstract syntax:\\

$\begin{array}{lclccccccccllcl}
s &  ::= &  \nil  & | &  \mu.t  & | &\;  \underline{a}.s \; & | &\; s+s & \hspace{1.5 cm} \mbox{{\em guarded terms}}\ \\
q & ::= & s  & | &  C &&&&& \hspace{1.5 cm} \mbox{{\em sequential terms}}\\
t & ::= & q  & | &  t \para t  &&&&&\hspace{1.3 cm} \mbox{{\em restriction-free terms}}\\
p & ::= & t  & | & \restr{a}p  &&&&&\hspace{1.3 cm} \mbox{{\em general terms}}\\\\
\end{array}$

\noindent 
where we assume that a constant $C$ is defined by a {\em guarded} process, i.e., a process in syntactic category $s$ (i.e., $C \eqdef s$).
An FNM term $p$ is an FNM {\em process} if the set  \const{p} of constants used by $p$ is finite and each constant in \const{p} 
is equipped with a defining equation. 
The set of FNM processes is denoted by 
$\mathcal{P}_{FNM}$, the set of its sequential processes, i.e.,
of the processes in syntactic category $q$, by $\mathcal{P}_{FNM}^{seq}$, and the set of its guarded processes, i.e.,
of the processes in syntactic category $s$, by $\mathcal{P}_{FNM}^{grd}$.

The FNM operators are those of CCS \cite{Mil89,GV15}, used in constrained manner, with the addition of the  {\em strong prefixing} 
operator:
$ \underline{a}.s$ is a strongly prefixed process, where the strong prefix $a$ is the first input action of a transaction 
that continues with the {\em sequential} process $s$
(provided that $s$ can complete the transaction). 
%
%
 %
Hence, the strong prefixing operator allows for the creation of transitions labeled by an {\em atomic} sequence $\sigma$, 
which is a sequence of {\em visible} 
actions, composed of inputs only, but possibly ending with an output, that are executed atomically; more precisely, $\sigma$ ranges over 
the set of {\em labels} 
 ${\mathcal A} = \{\tau\} \cup \mathcal{L}^*\cdot (\mathcal{L} \cup \overline{\mathcal L})$.
E.g., $\underline{a}.(b.\nil + \overline{c}.\nil)$ can perform two transitions reaching $\nil$:
one labeled by the atomic sequence $ab$, the other one by the atomic sequence $a\overline{c}$.

A consequence of the fact that transitions may be labeled by atomic sequences is the need for a new, more general,
discipline of synchronization, which extends the basic CCS rule \cite{Mil89,GV15} for  synchronizing two complementary actions.
%
Two sequences $\sigma_1$ and $\sigma_2$ can be synchronized, and the result is $\sigma$, if 
relation $\Sync(\sigma_1,  \sigma_2, \sigma)$ holds. This relation, formally defined in Table \ref{fnm-sync},
holds if at least one of the two sequences is a single output action,
say $\sigma_1 = \overline{a}$, and the other one is either the complementary input action $a$ or an atomic
sequence starting with $a$.
Hence, it is not possible to synchronize two atomic sequences. 

A {\em well-formed} FNM process \cite{Gor17} $p$ is an FNM process satisfying a simple syntactic condition, denoted by {\em wf(p)}, ensuring 
that its executable atomic sequences
are composed of input actions only. This ensures that
a multi-party synchronization can take place only among one {\em leader}, i.e., the process performing
the atomic sequence of inputs, and as many other components (the {\em servants}), as the length of the atomic sequence, 
where each servant executes one output action. 
We will show that the Petri net associated with a well-formed FNM process $p$ is finite.
In the following we restrict our attention to well-formed processes only.

\begin{table}[t]
\hrulefill\\[-.7cm]
\begin{center}
$\begin{array}{lclclclclcl}
\bigfrac{}{\Sync(\alpha, \overline{\alpha}, \tau)} & \quad  
\bigfrac{\sigma \neq \epsilon}{\Sync(a\sigma, \overline{a}, \sigma)} & \quad
 \bigfrac{\sigma \neq \epsilon}{\Sync(\overline{a}, a\sigma, \sigma)}\\[-.2cm]
\end{array}$
\end{center}
\hrulefill
\caption{Synchronization relation $\Sync$
}\label{fnm-sync}
\end{table}\index{Synchronization relation}

%
\subsection{Extended Terms} \label{fnm-extend-sec}
%

The FNM processes are built upon the set  
${\mathcal L}\cup\overline{\mathcal L}$, ranged over by $\alpha$, of visible actions. 
We assume we also have  sets ${\mathcal L}' = \{a' \, | \,  a\in \mathcal{L}\}$
and $\overline{{\mathcal L}'} =$ $ \{\overline{a'} \, | \,  $ $\overline{a}\in \overline{\mathcal{L}}\}$,
where ${\mathcal L}' \cup\overline{\mathcal L'}$, 
ranged over by $\alpha'$, is the set of auxiliary {\em restricted} actions, i.e., actions that are only allowed to synchronize.
By definition, 
each restricted action $\alpha'$ corresponds to exactly  one
visible action $\alpha$. 
Let $\mathcal{G} =$ $ {\mathcal L}\cup{\mathcal L}'$, ranged over by $\gamma$, be
 the set of input actions and their restricted counterparts. 
$\overline{\mathcal{G}} = \overline{\mathcal L} \cup \overline{{\mathcal L}'}$
is the set of output actions and their restricted counterparts. 
The set $Act_{\gamma} = {\mathcal G} \cup \overline{\mathcal{G}} \cup \{\tau\}$, ranged over by $\mu$ (with abuse of notation), 
is used to build the set of {\em extended terms}, whose syntax is defined as for FNM,
where the prefixes are taken from the set $Act_{\gamma}$, the strong prefixes from the set $\mathcal{G}$ and the bound action $a$ is in $\mathcal{L}$. 
An extended general FNM term $p = \restr{L}t$ is an {\em extended process} if \const{p} is finite, each constant in \const{p} is defined and
$t$ is {\em admissible}, i.e.,

$\hspace{4cm} \forall a \in \mathcal{L},\, \{a, a'\} \not\subseteq fn(t)$, 

\noindent
where the function $fn(-)$, computing the free names, is defined on extended terms in the obvious way.
The admissibility condition expresses a sort of sanity check on any restriction-free, extended term $t$: 
it is not possible that, for each action $a \in \mathcal{L}$, 
there are occurrences in $t$ of both $a$ and its associated restricted action $a'$; this because each action type can 
occur in $t$ only in one of the two modalities: either restricted or {\em unrestricted} (i.e., normally visible). For instance,
$a.a'.\nil$ is not admissible, while $a.\nil \para b'.\nil$ is admissible.
By the notation 
$ad(t)$ we mean that $t$ is admissible.

By $ {\mathcal P}_{FNM}^{\gamma}$ we denote the set of all extended FNM processes.
By $ {\mathcal P}_{FNM}^{\gamma, par}$ we denote the set of all restriction-free, extended 
FNM processes, i.e., those extended processes in syntactic category $t$.
By $ {\mathcal P}_{FNM}^{\gamma, seq}$ we denote the set of all sequential, extended 
FNM processes, i.e., those extended processes
in syntactic category $q$.
By $ {\mathcal P}_{FNM}^{\gamma, grd}$ we denote the set of all guarded, extended 
FNM processes, i.e., those extended processes
in syntactic category $s$.

%
\subsection{Net Semantics} \label{fnm-sem-sec}
%

In this section, we summarize a technique (operational in style), proposed in \cite{Gor17}, for building an infinite 
P/T net for the whole of FNM,
starting from a description of its places and its net transitions. 
The resulting net $N_{FNM} = (S_{FNM}, \mathcal{A}, T_{FNM})$ is such that,
for each $p \in {\mathcal P}_{FNM}$, the net system $N_{FNM}(\dec(p))$ statically reachable from the initial
marking $\dec(p)$ is a statically reduced P/T net, which is finite if $p$ is well-formed; such a net system is denoted by $Net(p)$.

\begin{table}[!t]

{\renewcommand{\arraystretch}{1.5}
\hrulefill\\[-.65cm]
\begin{center}
$\begin{array}{rclrclllll}
\dec(\nil) & = &  \theta & \qquad \dec(p + q) & = &  \{p + q\} & \dec(\mu.p) & = & \{ \mu.p\}\\
\dec(\underline\gamma.p, I) & = & \{\underline\gamma.p\} & \qquad \dec(p \para q) & = & \dec(p) \oplus dec(q) \\
\dec(C)  & = &  \{C\}   & \qquad
\dec(\restr{a}p) & = & \dec (p)\sost{a'}{a} \quad   a' \in {\mathcal L}' \\[-.3cm]
\end{array}$
\end{center}}
\hrulefill

\caption{Decomposition function} \protect\label{dec-fnm}
\end{table}

The set of FNM places, ranged over by $s$, is 
$S_{FNM} = {\mathcal P}_{FNM}^{\gamma, seq} \setminus \{\nil\}$, i.e., the set of all sequential, {\em extended}
FNM processes, except $\nil$.

Function $\dec: {\mathcal P}_{FNM}^{\gamma} \rightarrow {\mathcal M}_{fin}(S_{FNM})$, which
maps extended processes into markings, is outlined in Table \ref{dec-fnm}.
Process $\nil$ is mapped to the empty marking $\theta$.
A sequential process $p$ is mapped to one place with name $p$. This is the case
of $\mu.p$ (where $\mu$ can be any action in $Act_{\gamma}$), a constant $C$,  $p + q$ and 
$\underline\gamma.p$ (where $\gamma \in \mathcal{G} = \mathcal{L} \cup \mathcal{L}'$).
Note that when $C \eqdef \nil$, we have that $\dec(\nil) = \theta \neq \{C\} = \dec(C)$.
Note also that $\dec(\nil) = \theta \neq \{ \nil + \nil\} = \dec(\nil+\nil)$.

Parallel composition is interpreted as multiset union;  e.g., the decomposition of $a.\nil \para a.\nil$ 
produces the marking $a.\nil \oplus a.\nil = 2 \cdot a.\nil$.
The decomposition of
a general process $\restr{a}p$ -- where $a \in \mathcal{L}$ -- generates the multiset obtained from the 
decomposition of $p$, to which the substitution $\sost{a'}{a}$ is applied;
the application of the substitution $\sost{a'}{a}$ to a multiset is performed element-wise.
We assume that, in decomposing $\restr{a}p$, the choice of the restricted name 
is fixed by the rule that associates with a visible action
$a$ its {\em unique} corresponding restricted action $a'$.

\begin{proposition}\label{fin-dec-m0-fnm}
For each $p \in {\mathcal P}_{FNM}^{\gamma}$, $\dec(p)$ is a finite multiset of places. 
\fine
\end{proposition}

A marking $m \in  {\mathcal M}_{fin}(S_{FNM})$ is {\em admissible}, denoted by $ad(m)$, if for all $a \in \mathcal{L}$, $\{a, a'\} \not \subseteq fn(m)$,
where $fn(m) = \bigcup_{s\in dom(m)} fn(s)$, with $fn(\theta) = \emptyset$.

A marking $m \in  {\mathcal M}_{fin}(S_{FNM})$ is {\em complete} if an
FNM process $p \in \mathcal{P}_{FNM}$ exists such that $\dec(p) = m$.

\begin{theorem}\label{adm=complete-th-fnm}\index{Marking!admissible}\index{Marking!complete}
A marking $m \in  {\mathcal M}_{fin}(S_{FNM})$ is admissible iff it is complete.
\fine
\end{theorem}

Hence, this theorem states not only that function $\dec$ maps FNM processes to admissible markings over $S_{FNM}$, but also 
that $\dec$ is surjective over this set.

\begin{definition}\label{wf-marking}{\bf (Well-behaved)}\index{Marking!well-behaved}
A set of places $S \subseteq S_{FNM}$ is {\em well behaved} if for all $s \in S$ we have that
$\wf{s}$ holds, i.e., the sequential FNM extended term $s$ is well-formed.
\fine
\end{definition}

In order to define the set $T_{FNM}$ of all the FNM net transitions, we need some auxiliary definitions.
Let ${\mathcal A}^{\gamma}  = \{\tau\} \cup \mathcal{G}^* \cdot (\mathcal{G} \cup \overline{\mathcal{G}})$, 
ranged over by $\sigma$ with abuse of notation, be the set of labels; hence, a label $\sigma$ can be the invisible action $\tau$,
or a (possibly empty) sequence of inputs (or restricted inputs) followed by an input or an output (or its restricted counterpart).
Let $\rightarrow \; \subseteq  {\mathcal M}_{fin}(S_{FNM}) \times {\mathcal A}^{\gamma} \times {\mathcal M}_{fin}(S_{FNM})$
be the least set of transitions generated by the axiom and
rules in Table~\ref{netrules-fnm}.

\begin{table}[t]
{\renewcommand{\arraystretch}{2.5}
\hrulefill\\[-.7cm]

\begin{center}
$\begin{array}{lcllcl}
\mbox{(pref)}  &\bigfrac{}{\{\mu.p\}\deriv{\mu}\dec(p)} & \quad & \; \; \;
\mbox{(cons)} & \bigfrac{\dec(p)\deriv{\sigma}m}{\{C\}\deriv{\sigma}m}& C \eqdef p \\
\mbox{(sum$_1$)}  & \bigfrac{\dec(p)\deriv{\sigma}m}{\{p+q\}\deriv{\sigma}m} & \quad & \; \; \;
\mbox{(s-pref)} & \bigfrac{\dec(p) \deriv{\sigma}m }{ \{\underline\gamma.p\} \deriv{\gamma \diamond\sigma}m}\\
\end{array}$

$\begin{array}{lcllcl}
\mbox{(s-com)} & \bigfrac{m_1 \deriv{\sigma_1}m_1' \; \; m_2  \deriv{\sigma_2}m_2' }{ 
m_1 \oplus m_2  \deriv{\sigma} m_1' \oplus m_2'} &  ad(m_1 \oplus m_2) \wedge Sync(\sigma_1, \sigma_2, \sigma)\\
\end{array}$

\hrulefill
\end{center}}
\caption{Rules for net transitions (symmetric rule (sum$_2$) omitted)}\label{netrules-fnm}
\end{table}

Let us comment on the rules of Table \ref{netrules-fnm}. 
Axiom (pref) states that if one token is present in the place $\mu.p$, then a $\mu$-labeled transition is derivable
from marking $\{\mu.p\}$,
producing the marking $\dec(p)$. This holds for each $\mu$, i.e., for the invisible action $\tau$,
for each visible action $\alpha$ as well as for each restricted action $\alpha'$.
By rule (sum$_1$), the transitions from the place $p + q$ are those from the marking $\dec(p)$; as $p$ is sequential,
$\dec(p)$ is $\{p\}$ if $p \neq \nil$, while, if $p = \nil$, $\dec(p) = \theta$, but no transition is derivable 
from the empty marking, so that the rule is really applicable only when $\dec(p) = \{p\}$.
Similarly, rule (cons) states that the transitions derivable from $\{C\}$ are those derivable from
the place $\{p\}$, if $C \eqdef p$ with $p \neq \nil$.
In rule (s-pref), $\gamma$ may be any input action $a$ or any {\em restricted} action $a'$, and the auxiliary function 
$\gamma \diamond \sigma$ returns $\gamma$ if $\sigma = \tau$,
or $\gamma \sigma$ otherwise. This rule requires that the premise transition 
$\dec(p) \deriv{\sigma} m$ be derivable 
by the rules, which is really possible only when $\dec(p) = \{p\}$. 
Rule (s-com) requires that the transition pre-set be admissible in order to avoid producing
transitions that have no counterpart in the LTS semantics of FNM, described in \cite{Gor17}.
Rule (s-com) 
explains how a synchronization takes place: it is required that $m_1$ and $m_2$ perform
synchronizable sequences $\sigma_1$ and $\sigma_2$, producing $\sigma$; here we assume that relation
$\Sync$ has been extended also to restricted actions in the obvious way, i.e., a 
restricted output action $\overline{a'}$
can be synchronized only with its complementary restricted input action $a'$ or 
with an atomic sequence beginning with $a'$.
As an example, the net transition
$\{\underline{a}.b'.p, \overline{a}.q, \overline{b'}.r\} \deriv{\tau}  \dec(p) \oplus \dec(q) \oplus \dec(r)$
is derivable by the rules, as shown in Table \ref{net-trans-proof}. 

\begin{table}[t]
\begin{center}
\begin{prooftree}
                    \AxiomC{}
                   \LeftLabel{\scriptsize{\mbox{(pref)}}}
                  \UnaryInfC{$\{b'.p\} \deriv{b'}  \dec(p)$}
                    \LeftLabel{\scriptsize{\mbox{(s-pref)}}}
                 \UnaryInfC{$\{\underline{a}.b'.p\} \deriv{ab'}  \dec(p)$}
                  

                    \AxiomC{}
                   \LeftLabel{\scriptsize{\mbox{(pref)}}}
                  \UnaryInfC{$\{\overline{a}.q\} \deriv{\overline{a}}  \dec(q)$}


%
                \LeftLabel{\scriptsize{\mbox{(s-com)}}}
                \BinaryInfC{$\{\underline{a}.b'.p, \overline{a}.q\} \deriv{b'}  \dec(p) \oplus \dec(q)$}
                                    \AxiomC{}
                   \LeftLabel{\scriptsize{\mbox{(pref)}}}
                  \UnaryInfC{$\{\overline{b'}.r\} \deriv{\overline{b'}}  \dec(r)$}

%
                   \LeftLabel{\scriptsize{\mbox{(s-com)}}}
                \BinaryInfC{$\{\underline{a}.b'.p, \overline{a}.q, \overline{b'}.r\} \deriv{\tau}  \dec(p) \oplus \dec(q) \oplus \dec(r)$}

\end{prooftree}
\end{center}
\caption{The proof of a net transition}\label{net-trans-proof}
\end{table}

Transitions with labels containing restricted actions must not be taken in the resulting net, as we 
accept only transitions labeled over ${\mathcal A} = \{\tau\} \cup \mathcal{L}^* \cdot (\mathcal{L} \cup \overline{\mathcal{L}})$. 
However, they are useful in 
producing acceptable transitions, as two complementary restricted actions can synchronize, 
producing a $\tau$-labeled transition or shortening the synchronized sequence. 
For instance, in the example above,
the derivable transition $\{b'.p\} \deriv{b'}  \dec(p)$
is not an acceptable transition because its label is not in ${\mathcal A}$, while $\{\underline{a}.b'.p, \overline{a}.q, \overline{b'}.r\} \deriv{\tau}  \dec(p) \oplus \dec(q) \oplus \dec(r)$ is so.
Hence, the P/T net for FNM is the 
triple $N_{FNM} = (S_{FNM}, {\mathcal A},$ $ T_{FNM})$, where the set 

$T_{FNM} = \{ (m_1, \sigma, m_2) \mid m_1 \deriv{\sigma}m_2$ $ \mbox{ is derivable by the rules and } \sigma \in {\mathcal A} \}$ 

\noindent is obtained by filtering out 
those transitions derivable by the rules whose label  $\sigma$ contains some
restricted name $a'$ or $\overline{a'}$. 

We now want to show that for each finite, well-behaved set of places $S \subseteq S_{FNM}$, the set of transitions 
statically enabled at $S$ is finite.
Given a place $s \in S$, by $s \vdash t$ we mean that transition $t = (\{s\}, \sigma, m)$ is derivable
by the rules in Table \ref{netrules-fnm}, hence with $\sigma \in \mathcal{A}^{\gamma}$.

\begin{lemma}\label{ts-finite-fnm}
The set $T_s = \{ t \mid s \vdash t\}$ is finite, for each $s \in S_{FNM}$.

\proof By induction on the axiom and rules in Table \ref{netrules-fnm}. 
\fine
\end{lemma}

Given a finite, well-behaved set of places $S \subseteq S_{FNM}$, let $T^S_1$ be $\bigcup_{s \in S} T_s$, i.e., the 
set of all transitions, with a singleton pre-set in $S$, derivable by the rules with labeling in $\mathcal{A}^{\gamma}$. 
The set $T^S_1$ is
finite, being the finite union (as $S$ is finite) of finite sets (as $T_s$ is finite for each $s \in S$, by Lemma \ref{ts-finite-fnm}).

Let $k \in \nat$ be the length of the longest label of the transitions in $T^S_1$. 
It is possible to argue (details in \cite{Gor17}) that if a multi-party transition $t$ is derivable
by the rules from the well-behaved set $S$, then its proof contains $k$ synchronizations at most, each 
one between a transition (labeled with a sequence of inputs) and a {\em singleton-pre-set} transition 
(labeled with a {\em single} output action); hence, at most $k + 1$ participants can take part in a multi-party
synchronization.
Therefore, the set of all the transitions statically enabled at a finite, well-behaved set $S$ can be defined by means
of a sequence of sets  $T^S_i$ of transitions, for $2 \leq i \leq k+1$,
where each transition $t \in T^S_i$ has a pre-set $\pre t$ of size $i$, as follows:\\

$
\begin{array}{rclll}
T^S_i   & = &   \{(m_1 \oplus m_2, \sigma, m_1' \oplus m_2') \mid ad(m_1 \oplus m_2), \\
& & \qquad \qquad \exists \sigma_1 \exists \gamma. (m_1, \sigma_1, m_1') \in T^S_{i-1}, (m_2, \overline{\gamma}, m_2') \in T^S_1, \Sync(\sigma_1, \overline{\gamma}, \sigma)\}.
\end{array}$\\

\noindent
Note that $T^S_2$ is finite, because $T^S_1$ is finite; inductively, 
$T^S_{i+1}$, for $2 \leq i \leq k$ is finite, because $T^S_{i}$ and $T_1$ are finite.
So, the set $T_S$ of all the transitions statically enabled at $S$ is
\[
T_S = \{t \mid t \in \bigcup_{i = 1}^{k+1} T^S_i \wedge  \ell(t) \in \mathcal{A}\},\]
\noindent where only transitions labeled over $\mathcal{A}$
are considered. $T_S$ is finite, being a finite union of finite sets; therefore, we have the following result.

\begin{theorem}\label{fin-trans-fnm}
If $S \subseteq S_{FNM}$ is a finite, well-behaved set of places, 
then set $T_S \subseteq T_{FNM}$ of all the transitions statically enabled at $S$ is finite.
\fine
\end{theorem}

\begin{table}[t]
\begin{center}
\begin{prooftree}
                    \AxiomC{}
                   \LeftLabel{\scriptsize{\mbox{(pref)}}}
                  \UnaryInfC{$\{dec.\nil\} \deriv{dec}  \theta$}
                    \LeftLabel{\scriptsize{\mbox{(s-pref)}}}
                 \UnaryInfC{$\{\underline{c'}.dec.\nil\} \deriv{c'dec}  \theta$}
 
                     \LeftLabel{\scriptsize{\mbox{(s-pref)}}}
                 \UnaryInfC{$\{\underline{c'}.\underline{c'}.dec.\nil\} \deriv{c'c'dec}  \theta$}
   
                       \LeftLabel{\scriptsize{\mbox{(sum$_1$)}}}
                 \UnaryInfC{$\{s_2\} \deriv{c'c'dec}  \theta$}
              

                    \AxiomC{}
                   \LeftLabel{\scriptsize{\mbox{(pref)}}}
                  \UnaryInfC{$\{\overline{c'}.\nil\} \deriv{\overline{c'}} \theta$}

                   \LeftLabel{\scriptsize{\mbox{(sum$_2$)}}}
                  \UnaryInfC{$\{s_2\} \deriv{\overline{c'}} \theta$}

                \LeftLabel{\scriptsize{\mbox{(s-com)}}}
                \BinaryInfC{$\{s_2, s_2\} \deriv{c'dec}  \theta$}
                                    \AxiomC{}
                \LeftLabel{\scriptsize{\mbox{(pref)}}}
                  \UnaryInfC{$\{\overline{c'}.\nil\} \deriv{\overline{c'}} \theta$}

                   \LeftLabel{\scriptsize{\mbox{(sum$_2$)}}}
                  \UnaryInfC{$\{s_2\} \deriv{\overline{c'}} \theta$}

                   \LeftLabel{\scriptsize{\mbox{(s-com)}}}
                \BinaryInfC{$\{s_2, s_2, s_2\} \deriv{dec}  \theta$}

\end{prooftree}
\end{center}
\caption{The proof of a net transition, where $s_2 =  \underline{c'}.\underline{c'}.dec.\nil + \overline{c}'.\nil$}\label{net-trans2-proof}
\end{table}

\begin{figure}[t]
\centering
\begin{tikzpicture}[
every place/.style={draw,thick,inner sep=0pt,minimum size=6mm},
every transition/.style={draw,thick,inner sep=0pt,minimum size=4mm},
bend angle=45,
pre/.style={<-,shorten <=1pt,>=stealth,semithick},
post/.style={->,shorten >=1pt,>=stealth,semithick}
]
\def\eofigdist{2.5cm}
\def\eodist{0.5}

\node (q1) [place,tokens=1]  [label=right:$\;s_1$] {};
\node (s1) [transition] [below left=\eodist of q1,label=left:$inc\;$] {};
\node (q2) [place]  [below right=\eodist of s1,label=right:$\;s_2$] {};
\node (s3) [transition] [below left=\eodist of q2,label=left:$dec\;$]{};

\draw  [->,bend right] (q1) to (s1);
\draw  [->,bend right] (s1) to (q1);
\draw  [->] (s1) to (q2);
\draw  [->] (q2) to node[auto,swap]{3} (s3);
  
\end{tikzpicture}
\caption{The P/T net for the $1/3$ semi-counter}
\label{sec3}
\end{figure}
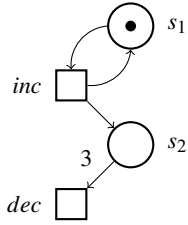

\begin{example}\label{one-tird-sc-ex} {\bf (1/3 Semi-counter)}
Let us consider a semi-counter such that three occurrences of $inc$ are needed to enable one $dec$,
whence the name 1/3 semi-counter. 
The well-formed process $p = \restr{c}A$, where 

\[A \eqdef inc.(A \para (\underline{c}.\underline{c}.dec.\nil + \overline{c}.\nil))\] 

\noindent is a 1/3 semi-counter.
The initial marking 
$m_0$ is $\dec(p) = \dec(\restr{c}A) = $ $\dec(A)\sost{c'}{c} = \{A\}\sost{c'}{c} = \{s_1\}$; place $s_1$ is the extended, sequential process
$A_{\sost{c'}{c}}$,
where the constant $A_{\sost{c'}{c}}$ is obtained by applying the substitution $\sost{c'}{c}$ to the body of $A$: 

$A_{\sost{c'}{c}} \eqdef inc.(A_{\sost{c'}{c}} \para (\underline{c'}.\underline{c'}.dec.\nil + \overline{c'}.\nil))$.

\noindent
Then, let $S_0 = dom(m_0) = \{s_1\}$. 
The set of transitions statically enabled at $S_0$ is
$T^{S_0}_1 = T_{s_1} = \{t_1\}$, where the only transition is $t_1 = \{s_1\} \deriv{inc} \{s_1, s_2\}$,
with 
$s_2 =  \underline{c'}.\underline{c'}.dec.\nil + \overline{c}'.\nil$.
Therefore, the new set of statically reachable places is $S_1 = \{s_1, s_2\}$.
Note that $s_2$ can produce two transitions in $T_{s_2}$, namely $t' = \{s_2\} \deriv{c'c'dec} \theta$ and $t'' =  \{s_2\} \deriv{\overline{c'}} \theta$, but neither is labeled over $\mathcal{A}$. 
Since the longest label has length $3$, we have to compute
the sets $T^{S_1}_i$ for $i = 1, \ldots, 4$:

$T^{S_1}_1 = \{t_1, t', t''\}$,

$T^{S_1}_2 = \{t'''\}$, where $t''' = (\{s_2, s_2\}, c' dec, \theta)$,

$T^{S_1}_3 = \{t_2\}$, where $t_2 = (\{s_2, s_2, s_2\}, dec, \theta)$, whose proof is shown in Table \ref{net-trans2-proof}, 

$T^{S_1}_4 = \emptyset$.

Hence, $T_{S_1} = \{t_1, t_2\}$, as these two are the only transitions labeled over $\mathcal{A}$.
As $t_2$ does not add any new reachable place, we have that $S_1$ is the set of places
statically reachable from the initial marking, 
and  $T_{S_1}$ is the set of transitions statically enabled at $S_1$. This net is depicted in Figure \ref{sec3}.
\fine
\end{example}

The P/T net system associated with a process $p \in \mathcal{P}_{FNM}$ is the subnet of 
$N_{FNM}$ statically reachable from the initial marking $\dec(p)$, denoted by $Net(p)$.

\begin{definition}\label{net-p-fnm}
Let $p$ be a process in $\mathcal{P}_{FNM}$. 
The P/T net system statically associated with $p$ is $Net(p) = (S_p, A_p, T_p, m_0)$, where $m_0 = \dec(p)$ and \\

$
\begin{array}{rcl}
S_p & = & \llbracket dom(m_0) \rangle \quad \mbox{computed in $N_{FNM}$,}\\
T_p & = & \{ t \in T_{FNM}  \mid S_p\llbracket t\rangle\},\\
A_p & = & \{\sigma \in {\mathcal A} \mid \exists t \in T_p \mbox{ such that }  \ell(t) = \sigma\}.\\[-.3cm]
\end{array}
$

\fine
\end{definition}

\noindent
The following propositions present three facts that are obviously true by construction of the net $Net(p)$
associated with an FNM process $p$.

\begin{proposition}\label{stat-red-net5}
For each $p \in {\mathcal P}_{FNM}$, $Net(p)$ is a statically reduced P/T net.
\fine
\end{proposition}

\begin{proposition}\label{netp=netq-fnm}
If $\dec(p) = \dec(q)$, then $Net(p) = Net(q)$.
\fine
\end{proposition}

\begin{proposition}\label{par-net-self-fnm}
For each restriction-free $t \in \mathcal{P}_{FNM}$ and for each $L \subseteq \mathcal{L}$:

\begin{itemize}
\item[$i)$] If $Net(t) = (S, A, T, m_0)$,
then, for each $n \geq 1$, $Net(t^n) = (S, A, T, n \cdot m_0)$, where $t^1 = t$ and $t^{n+1} = t \para t^{n}$.

\item[$ii)$] If $Net(\restr{L}t) = (S, A, T, m_0)$,
then $Net(\restr{L}(t^n)) = (S, A, T, n \cdot m_0)$, for $n \geq 1$.
\fine
\end{itemize}

\end{proposition}

Definition \ref{net-p-fnm} suggests a way of
generating $Net(p)$ with an algorithm based on the inductive definition of the static reachability relation 
(see Definition \ref{def-static-reach}):
Start with the initial set of places $S_0 = dom(\dec(p))$, 
and then apply the rules in Table \ref{netrules-fnm} in order to produce the set $T_{S_0}$ of transitions
(labeled over $\mathcal{A}$) statically enabled at $S_0$, as well as the additional places statically reachable 
by means of such transitions. 
Then repeat this procedure from the set of places statically reached so far. An instance of this procedure was  given in 
Example \ref{one-tird-sc-ex}.
There are two problems with this algorithm:
\begin{itemize}
\item the obvious {\em halting condition} is ``until no new places are statically reachable''; of course, the algorithm terminates if we know that
the set $S_p$ of places statically reachable from $dom(\dec(p))$ is finite; additionally,
\item at each step of the algorithm, we have to be sure that the set of transitions derivable from the current set of statically
reachable places is finite.
\end{itemize}

We are going to prove only the first requirement --- $S_p$ is finite for each $p \in \mathcal{P}_{FNM}$ --- because it 
implies also the second one
for well-formed processes.
As a matter of fact, if $p$ is well formed, then $dom(\dec(p))$ is well behaved, and so is each 
set $S$ of places statically reachable 
from $dom(\dec(p))$ (as proved in \cite{Gor17}); since $S \subseteq S_p$, $S$ is also finite, and so, by Theorem \ref{fin-trans-fnm},  
the set $T_S$ of transitions statically enabled at the finite, well-behaved set $S$ is finite, too.

\begin{theorem}\label{fin-s-places-fnm}
For each $p \in \mathcal{P}_{FNM}$, let $Net(p) = (S_p, A_p, T_p, m_0)$ be defined  as in Definition \ref{net-p-fnm}.
Then, the set $S_p$ is finite.

\proof By induction on the static reachability relation $\Derivstar{}$ (details in \cite{Gor17}).
\fine
\end{theorem}

\begin{theorem}\label{net(p)-finite-fnm}\index{P/T Petri net!finite}\cite{Gor17}
For each well-formed FNM process $p$, $Net(p) = (S_p, A_p, T_p, \dec(p))$ is a finite P/T net.
\fine
\end{theorem}


%
\subsection{Representing All Finite P/T Net} \label{fnm-sem-sec}
%

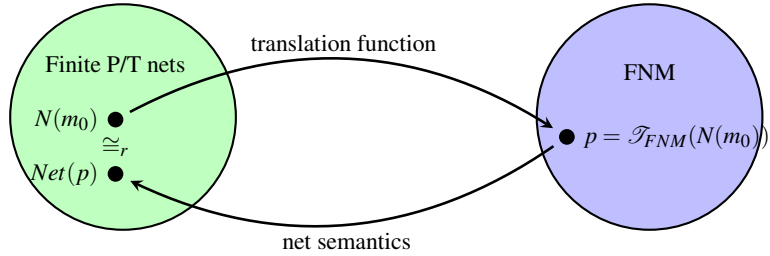
\begin{figure}[t]
\centering

\begin{tikzpicture}
  \tikzset{%
    mythick/.style={%
        line width=.35mm,>=stealth
    }
  }
  \tikzset{%
    mynode/.style={
      circle,
      fill,
      inner sep=2.1pt
    },
    shorten >= 3pt,
    shorten <= 3pt
  }
\def\eodiaglabeldist{0mm}
\def\eolabeldist{0mm}
\def\eofigdist{4.5cm}
\def\rrrel{$\cong_r$}
\def\eodist{0.5cm}
\def\eodisty{0.4cm}
\def\eodistw{0.8cm}

\draw [thick,fill=green!25] (-1,-1) circle [radius=1.5cm];
\draw (-1.1,-0.3) node (p0) {Finite P/T nets};
\node (p1) [mynode,below =\eodisty of p0, label={[label distance=\eodiaglabeldist]left:$N(m_0)$}] {};
\node (p2) [mynode,below =\eodist of p1, label={[label distance=\eodiaglabeldist]left:$Net(p)$}] {};
 \path (p1) -- node (R3) [inner sep=1pt] {\rrrel} (p2);

\draw [thick,fill=blue!25] (6,-1) circle [radius=1.5cm];
\draw (6,-0.4) node (q0) {FNM};
\node (q1) [mynode,below left =\eodistw of q0, label={[label distance=\eodiaglabeldist]right:$p = \term{N(m_0)}{FNM}$}] {};

\draw (p1) edge[mythick,->, bend left] node[above] {translation function} (q1);
\draw (q1) edge[mythick,->, bend left] node[below] {net semantics} (p2);

\end{tikzpicture}
\caption{Graphical description of the representability theorem, up to isomorphism}
\label{rep-fnm-pt}
\end{figure}

Theorem \ref{net(p)-finite-fnm} ensures that {\em only} finite P/T nets can be represented by FNM processes.
It is not completely obvious that {\em all} finite P/T nets can be represented by FNM processes.
However, we now hints (details in \cite{Gor17}), that this is the case, indeed. As illustrated in Figure \ref{rep-fnm-pt},
given a P/T net $N(m_0)$, we define a translation to an FNM term $\term{N(m_0)}{FNM} = p$,
such that its associated net $Net(p)$, according to the net semantics described in the previous section, is a P/T net {\em (rooted)
isomorphic} \cite{Gor17} to $N(m_0)$, i.e., $N(m_0) \cong_r Net(p)$.

The translation from nets to processes defines a constant $C_i$ in correspondence with 
each place $s_i$; the constant $C_i$ has a summand $c^j_i$ for each transition $t_j$, which is $\nil$ when $s_i$ is not in the pre-set of $t_j$. 
The FNM process $\term{N(m_0)}{FNM}$ associated with the finite net system $N(m_0)$,
labeled over $\mathcal{L} \cup \{\tau\}$,
has a bound name $x^j_i$ for each pair $(s_i, t_j)$, where $s_i$ is a place
and $t_j$ is a transition; such bound names are used to force synchronization among the components
participating in transition $t_j$ with pre-set of cardinality two or more. Among the many places in the pre-set of $t_j$, 
the one with least index $i$ (as we assume that places are indexed) plays the role of {\em leader} of the  
synchronization; the corresponding leader constant $C_i$ has a summand $c_i^j$ 
containing the atomic input sequence needed for the 
multi-party synchronization,
such that each strong input prefix $x_h^j$ (for $h > i$) is synchronized with the corresponding output  $\overline{x}_h^j$ 
performed by the {\em servant} participant of index $h$; in case $\pre{t_j}(s_i) \geq 2$, then the summand $c_i^j$
is actually a sum of $\overline{x}^j_{i}.\nil$ with the atomic input sequence, 
so that one instance of $C_i$ acts as the leader, while the others are servants.

\begin{definition}\label{translate-fnm}{\bf (Translating finite P/T nets into well-formed FNM processes)}
Given $A \subseteq \mathcal{L} \cup \{\tau\}$, let $N(m_0) = (S, A, T, m_0)$ --- with $S = \{s_1, \ldots, s_n\}$, 
$T = \{t_1, \ldots, t_k\}$, and $\ell(t_j) = \mu_j$  --- be a finite P/T net.
Function $\term{-}{FNM}$, from finite P/T nets to well-formed FNM processes, is defined as

\[
\term{N(m_0)}{FNM} = \restr{L} (\underbrace{C_1 | \cdots | C_1}_{m_0(s_1)} | \cdots | 
                    \underbrace{C_n | \cdots | C_n}_{m_0(s_n)} )
\]

\noindent    
where $L = \{x^1_1,\ldots, x^1_n, x^2_1, \ldots, x^2_n, \ldots ,x^k_1, \ldots ,x^k_n\}$ is such that $L \cap A = \emptyset$,
each $C_i$ is equipped with a defining equation $C_i \; \eqdef{} \; c_i^1 + \cdots + c_i^{k}$ (with 
$C_i \eqdef \nil$ if $k = 0$), and each summand $c_i^j$, for $j = 1, \ldots, k$, is equal to 

\begin{itemize}
\item $\nil$, if $s_i \not\in \pre t_j$;
\item $\mu_{j}.\Pi_{j}$, 
              if  $\pre t _{j} = \{s_i\}$;
\item $\overline{x}^j_{i}.\nil$, 
             if $\pre t_{j} (s_i) > 0$ and $\pre t_{j} (s_{i'}) > 0$ for some  $i' < i$
             (i.e., $s_i$ is not the leader for the synchronization on $t_{j}$);
\item $\underbrace{ \underline{x}^j_{i+1}. \cdots . \underline{x}^j_{i+1}}_{\pre t_j(s_{i+1})}. \ldots .\underbrace{ \underline{x}^j_{n}. \cdots .\underline{x}^j_{n}}_{\pre t_j(s_{n})}.\mu_{j}.\Pi_{j}$,
              if $\pre t_{j} (s_i) = 1$ and $s_i$ is the 
              leader of the synchronization (i.e., $\pre t_{j} (s_{i'}) > 0$ for no  $i' < i$, while  $\pre t_{j} (s_{i'}) > 0$ for some $i' > i$); 
\item $\overline{x}^j_{i}.\nil + \underbrace{ \underline{x}^j_{i}. \cdots . \underline{x}^j_{i}}_{\pre t_j(s_{i}) -1}.
\underbrace{ \underline{x}^j_{i+1}. \cdots . \underline{x}^j_{i+1}}_{\pre t_j(s_{i+1})}. \ldots .
\underbrace{ \underline{x}^j_{n}. \cdots .\underline{x}^j_{n}}_{\pre t_j(s_{n})}.\mu_{j}.\Pi_{j}$,
              otherwise (i.e., $s_i$ is the leader and $\pre t_{j} (s_i) \geq 2$).
\end{itemize}
Finally, process $\Pi_{j}$ is 
$\underbrace{ C_1 | \cdots | C_1}_{ \post{t_{j}} (s_1) } | \cdots | 
                    \underbrace{ C_n | \cdots | C_n}_{ \post{t_{j}} (s_n) }$, meaning that $\Pi_{j} = \nil$ if $\post{t_j} = \theta$.
\fine
\end{definition}

Note that $\term{N(m_0)}{FNM}$ is an FNM process: in fact, the restriction operator occurs only at the top level, 
applied to the parallel composition of a finite number of constants; each constant has a body that is sequential and restriction-free.
Note also that $\term{N(m_0)}{FNM}$ is a {\em well-formed} process: in fact, each {\em strong prefix} is a bound
 input $x^j_i$, and each
sequence ends with an action $\mu_j \in A$, which is either an input or $\tau$; hence, no 
atomic sequence ends with an output.
Therefore, the following proposition holds by Theorem \ref{net(p)-finite-fnm} and Proposition \ref{stat-red-net5}.

\begin{proposition}\label{inet-finite-fnm}
For each finite P/T Petri net $N(m_0)$, the net $Net(\term{N(m_0)}{FNM})$ is
a finite, statically reduced P/T net.
\fine
\end{proposition}

\begin{theorem}\label{represent-fnm}{\bf (Representability Theorem)}\cite{Gor17}
Let $N(m_0) = (S, A, T, m_0)$ be a finite, statically reduced P/T net system 
such that $A \subseteq \mathcal{L} \cup \{\tau\}$, 
and let $p = \term{N(m_0)}{FNM}$. Then, $Net(p)$ is isomorphic to $N(m_0)$.
\fine
\end{theorem}

\begin{example}
Consider Figure \ref{upcs-fig}(a). After clean-up (i.e., removing all the unnecessary $\nil$ summands and all the unused restricted
actions), its FNM representation is  
$p = \restr{a}(P_1 \para C_1) $, where $P_1 \eqdef prod.(P_1 \para D_1)$, $D_1 \eqdef \overline{a}.\nil$, 
$C_1 \eqdef \underline{a}.del.C_1'$ and $C_1' \eqdef cons.C_1$. Indeed, $Net(p)$ is isomorphic to the net in Figure \ref{upcs-fig}(a),
where place $P_1$ corresponds to the term $P_1\sost{a'}{a}$ (i.e., the constant $P_{1\sost{a'}{a}} \eqdef prod.(P_{1\sost{a'}{a}} \para D_{1\sost{a'}{a}})$), 
place $D_1$ to the term $D_1\sost{a'}{a}$ (i.e., the constant $D_{1\sost{a'}{a}} \eqdef \overline{a}'.\nil$), and so on.
\fine
\end{example}

%
\section{Structure Preserving Bisimilarity is a Congruence} \label{comp-sp-bis-sec}
%

Now we prove that structure-preserving bisimilarity $\sim_{sp}$ is a congruence for the FNM operators, or, in other words, that the 
FNM operators are compositional up to $\sim_{sp}$. 
We extend the definition of sp-bisimilarity to FNM terms, 
i.e., we write $p \sim_{sp} q$, meaning that there exists an sp-bisimulation $R$  in the
net obtained by the union of $Net(p)$ and $Net(q)$, that contains a linking $l$ such that
$\dec(p) = \pi_1(l)$ and $\dec(q) = \pi_2(l)$.
In the following, given an FNM term $r$, by $Id_r$ we denote the set of links

$Id_r = \{(s, s) \mid \exists m \in [\dec(r)\rangle \text{ and } s\in m\}$

\noindent
so that $Id_r^+$ is clearly a structure-preserving bisimulation justifying $\dec(r) \sim_{sp} \dec(r)$.

\begin{proposition}\label{prop1-cong-sp}{\bf (Congruence for strong prefixing and choice)}
Let $p$ and $q$ be FNM guarded processes, i.e., processes in syntactic category $s$. If $p \sim_{sp} q$, then 

$
\begin{array}{lrcll}
(i) &  \underline{a}.p & \sim_{sp} & \underline{a}.q, & \mbox{ for all $a \in \mathcal{L}$},\\
(ii) &  p + r & \sim_{sp} & q + r, & \mbox{ for each process $r$ in syntactic category $s$.}\\
\end{array}$

\proof
Assume $R$ is an sp-bisimulation containing the linking $(\{p\}, \{q\})$. (If $p = \nil = q$, then $R = \{\theta\}$, i.e., $R$ contains 
only one linking, that is the empty multiset of links, whose projections are the empty marking.)

Case $(i)$ can be proven by considering relation  $R_1 = R \cup \{(\{\underline{a}.p\}, \{\underline{a}.q\})\}$,
which is an sp-bisimulation. In fact, consider the new linking (actually a link) $(\{\underline{a}.p\}, \{\underline{a}.q\})$.
Note that $l = c$, so that $(l \ominus c) \oplus \overline{c} = \overline{c}$.
Transition $\{\underline{a}.p\} \deriv{a\diamond\sigma} m_1$ is derivable, by rule (s-pref),
only if $\{p\} \deriv{\sigma} m_1$.
As $R$ is an sp-bisimulation containing the linking $(\{p\}, \{q\})$, there exist a 
linking $\overline{c} \in R$ such that $\{q\} \deriv{\sigma} m_2$ with $\pi_1(\overline{c}) = m_1$ and
$\pi_2(\overline{c}) = m_2$. 
Hence, also $\{\underline{a}.q\} \deriv{a\diamond\sigma} m_2$
with $\overline{c} \in R_1$ with $\pi_1(\overline{c}) = m_1$ and
$\pi_2(\overline{c}) = m_2$, as required. 
The symmetric case when $\underline{a}.q$ moves first
is analogous and so omitted.

Case $(ii)$ can be proven, for each $r$  in syntactic category $s$, by showing that $R_2 = \{(\{p + r\}, \{q + r\})\} \cup 
R \cup Id_r^+$ is an sp-bisimulation. Since $R$ and $Id_r^+$ are already sp-bisimulations and the union 
of sp-bisimulations is an sp-bisimulation, we have only to focus on the
linking $(\{p + r\}, \{q + r\})$. Note that also in this case $l = c$, so that $(l \ominus c) \oplus \overline{c} = \overline{c}$.
If $\{p + r\} \deriv{\sigma} m_1$, then this is due to $\{p\} \deriv{\sigma} m_1$ or to $\{r\} \deriv{\sigma} m_1$.
In the former case, since $R$ is an sp-bisimulation containing the linking $(\{p\}, \{q\})$, there exist a 
linking $\overline{c} \in R$ such that $\{q\} \deriv{\sigma} m_2$ with $\pi_1(\overline{c}) = m_1$ and
$\pi_2(\overline{c}) = m_2$. Hence, also
$\{q + r\} \deriv{\sigma} m_2$ with $\overline{c} \in R_2$ with $\pi_1(\overline{c}) = m_1$ and
$\pi_2(\overline{c}) = m_2$, as required. 
In the latter case, as $Id_r^+$ is an sp-bisimulation containing the linking $(\{r\}, \{r\})$,
we have that there exists a linking $\overline{c}' \in Id_r^+$ such that $\pi_1(\overline{c}') = m_1$ and
$\pi_2(\overline{c}') = m_1$. Hence,  also $\{q + r\} \deriv{\sigma} m_1$ 
with with $\overline{c}' \in R_2$ with $\pi_1(\overline{c}') = m_1$ and
$\pi_2(\overline{c}') = m_1$, as required.  
The symmetric case when $q + r$ moves first
is analogous and so omitted.
\fine
\end{proposition}

\begin{proposition}\label{prop2-cong-sp}{\bf (Congruence for prefixing)}
Let $p$ and $q$ be restriction-free FNM processes. If $p \sim_{sp} q$, then 
$ \mu.p  \sim_{sp}  \mu.q$, for all $\mu \in Act$.

\proof
Assume $R$ is a structure-preserving bisimulation containing a link $l$ such that
$\dec(p) = \pi_1(l)$ and $\dec(q) = \pi_2(l)$. 
Relation $R_3 = R \cup \{(\{\mu.p\}, \{\mu.q\})\}$ is an sp-bisimulation.
In fact, if $\{\mu.p\}\deriv{\mu} \dec(p)$, then $\{\mu.q\}\deriv{\mu} \dec(q)$, where $l \in R$ (and so, $l \in R_3$) is
such that $\dec(p) = \pi_1(l)$ and $\dec(q) = \pi_2(l)$, as required. The symmetric case when $\mu.q$ moves first
is analogous and so omitted.
\fine
\end{proposition}

In order to prove the congruence property for parallel composition on restriction-free FNM process terms, we have to
recall that a transition $t$ may be the result of a synchronization of many parallel transitions, say a step $G = \{t_1, \ldots, t_k\}$
(where each transition has preset of size 1),
such that one transition is the leader of the synchronization, performing a sequence of inputs, and all the others are servant
transitions, performing  one output action only. Note that if $t$, derivable by repeated application of rule (s-com), is enabled, then
also all the transitions in the step $G$ are derivable by the rules and enabled, as the process terms are restriction-free, so that no
restricted action may occur in the label of such transitions.
We now introduce an auxiliary relation, called $\MSync$ and outlined in Table \ref{fnm-m-sync}, 
that explains how the synchronization of such a step $G$ takes place.
 
\begin{table}[t]
\hrulefill\\[-.7cm]
\begin{center}
$\begin{array}{lcr}
 \bigfrac{}{\MSync(\{\sigma\}, \{\sigma\})} & \qquad \quad & \bigfrac{\Sync(\sigma_1, \sigma_2, \sigma) \; \;  \; \MSync(M\oplus\{\sigma\}, M')}
{\MSync(M \oplus \{\sigma_1, \sigma_2\}, M' )}  
\end{array}$
\end{center}
\hrulefill
\caption{Multiset synchronization relation $\MSync$}\label{fnm-m-sync}
\end{table}

\begin{proposition}\label{prop3-cong-sp}{\bf (Congruence for parallel composition)}
Let $p$ and $q$ be restriction-free FNM processes. If $p \sim_{sp} q$, then 
$p \para r  \sim_{sp}  q \para r$, for each restriction-free $r \in \mathcal{P}_{FNM}$.

\proof
Assume $R$ is a structure-preserving bisimulation containing a link $l$ such that
$\dec(p) = \pi_1(l)$ and $\dec(q) = \pi_2(l)$. 
Relation 
$R_4  =  \{ l_1 \oplus l_2 \mid l_1 \in R \wedge l_2 \in Id_r^+\}$
 is an sp-bisimulation containing the linking $l \oplus h \in R_4$ (where 
 $l \in R$ is such that $\dec(p) = \pi_1(l)$ and $\dec(q) = \pi_2(l)$ and
 $h \in Id_r^+$ is such that $\pi_1(h) = \dec(r) = \pi_2(h)$)
 is such that $\pi_1(l \oplus h) = \dec(p) \oplus \dec(r)$ and $\pi_2(l \oplus h) = \dec(q) \oplus \dec(r))$.   
 In fact, consider a linking $l_1 \oplus l_2 \in R_4$, where $l_1 \in R$ (with $\pi_1(l_1) = m_1$ and $\pi_2(l_1) = m_2$)
 and $l_2 \in Id_r^+$ (with $\pi_1(l_2) = m = \pi_2(l_2)$),
 and assume $c_i \subseteq l_i$ for $i = 1, 2$, and transition $t_1$ such that $\pi_1(c_1 \oplus c_2) = \pre{t_1}$
 and $m_1 \oplus m [t_1\rangle \overline{m}_1$. We have to  consider the following three cases:

 \begin{itemize}
 \item 
 $m_1[t_1\rangle m_1'$, so that $\overline{m}_1 = m_1' \oplus m$. In such a case, $c_2 = \theta$ and, since $l_1 \in R$,
 we have that there exist a transition $t_2$ such that $\ell(t_1) = \ell(t_2)$, $\pi_2(c_1) = \pre{t_2}$,
 and a linking $\overline{c_1}$ such that $\pi_1(\overline{c_1}) = \post{t_1}$, $\pi_2(\overline{c_1}) = \post{t_2}$
 and $(l_1 \ominus c_1) \oplus \overline{c_1} \in R$.
 Therefore, by considering $c_2 = \theta$ and $\overline{c_2} = \theta$, 
 we have that $c_1 \oplus c_2$ is such that  $\pi_2(c_1 \oplus c_2) = \pre{t_2}$
 and $\overline{c_1} \oplus \overline{c_2}$ is such that $\pi_1(\overline{c_1} \oplus \overline{c_2}) = \post{t_1}$ and 
 $\pi_2(\overline{c_1} \oplus \overline{c_2}) = \post{t_2}$
 and $((l_1 \oplus l_2) \ominus (c_1 \oplus c_2)) \oplus (\overline{c_1} \oplus \overline{c_2}) $ 
 $= ((l_1 \ominus c_1) \oplus \overline{c_1})  \oplus l_2\in R_4$.

 \item 
 $m[t_1\rangle m'$, so that $\overline{m}_1 = m_1 \oplus m'$. In such a case, $c_1 = \theta$
 and, since $l_2 \in Id_r^+$,
 we have that the same transition $t_1$ is such that $\pi_2(c_2) = \pre{t_1}$,
 and there exists a linking $\overline{c_2}$ such that $\pi_1(\overline{c_2}) = \post{t_1}$ and $\pi_2(\overline{c_2}) = \post{t_1}$
 and $(l_2 \ominus c_2) \oplus \overline{c_2} \in Id_r^+$.
 Therefore, by considering $c_1 = \theta$ and $\overline{c_1} = \theta$, 
 we have that $c_1 \oplus c_2$ is such that  $\pi_2(c_1 \oplus c_2) = \pre{t_1}$
 and $\overline{c_1} \oplus \overline{c_2}$ is such that $\pi_1(\overline{c_1} \oplus \overline{c_2}) = \post{t_1}$ and 
 $\pi_2(\overline{c_1} \oplus \overline{c_2}) = \post{t_1}$
 and $((l_1 \oplus l_2) \ominus (c_1 \oplus c_2)) \oplus (\overline{c_1} \oplus \overline{c_2})$
 $= l_1 \oplus ((l_2 \ominus c_2) \oplus \overline{c_2})\in R_4$.

 \item 
 there exist a parallel step $G = \{\underline{t}_1, \ldots, \underline{t}_k\}$, such that  
 $\pre{t_1} = \pre{G} = \pre{\underline{t}_1} \oplus \ldots \oplus \pre{\underline{t}_k}$, $\post{t_1} = \post{G} = 
 \post{\underline{t}_1} \oplus \ldots \oplus \post{\underline{t}_k}$ and
 $\MSync(\ell(G), \ell(t_1))$ holds, with $\ell(G) = \{\ell(\underline{t}_1), \ldots, \ell(\underline{t}_k)\}$.
 Assume, w.l.o.g., that $G = G_1 \oplus G_2$, where $G_1 = \{\underline{t}^1_1, \ldots, \underline{t}^1_{k_1}\}$
 and $G_2 = \{\underline{t}^2_1, \ldots, \underline{t}^2_{k_2}\}$, with $\pre{G_1} \subseteq m_1$ and
 $\pre{G_2} \subseteq m$.
 We have  $\pi_1(c_1) = \pre{G_1}$ and $\pi_1(c_2) = \pre{G_2}$.
  In such a case,  since $l_1 \in R$, by Remark \ref{rem-step-sp}
 we have that there exist a step $G_1'$ such that $\ell(G_1') = \ell(G_1)$, $\pi_2(c_1) = \pre{G'}$,
 and a linking $\overline{c_1}$ such that $\pi_1(\overline{c_1}) = \post{G_1}$ and $\pi_2(\overline{c_1}) = \post{G_1'}$
 and $(l_1 \ominus c_1) \oplus \overline{c_1} \in R$.
Similarly, as $l_2 \in Id_r^+$, we have that the same step $G_2$ is such that $\pi_2(c_2) = \pre{G_2}$,
 and there exists a linking $\overline{c_2}$ such that $\pi_1(\overline{c_2}) = \post{G_2} = \pi_2(\overline{c_2})$
 and $(l_2 \ominus c_2) \oplus \overline{c_2} \in Id_r^+$. 
 Hence,  there exists a step $G' = G_1' \oplus G_2$, giving origin to a transition $t_2$, such that $\pre{t_2} = \pre{G'} = \pre{G_1'} \oplus \pre{G_2}$,
 $\post{t_2} = \post{G'} = \post{G_1'} \oplus \post{G_2}$, $\MSync(\ell(G'), \ell(t_1))$ holds (because $\ell(G) = \ell(G')$), so that $\ell(t_2) = \ell(t_1)$ 
  and
 $m_2 \oplus m [t_2\rangle m_2' \oplus m'$; moreover, $\pi_1(c_1 \oplus c_2) = \pre{t_1}$,
 $\pi_2(c_1 \oplus c_2) = \pre{t_2}$, $\pi_1(\overline{c_1} \oplus \overline{c_2}) = \post{t_1}$,
 $\pi_2(\overline{c_1} \oplus \overline{c_2}) = \post{t_2}$ and, finally, 
 $((l_1 \oplus l_2) \ominus (c_1 \oplus c_2)) \oplus (\overline{c_1} \oplus \overline{c_2}) = 
 ((l_1 \ominus c_1) \oplus \overline{c_1}) \oplus ((l_2 \ominus c_2) \oplus \overline{c_2}) \in R_4$, as required.
  \end{itemize}
  
 The symmetric case when $m_2 \oplus m$ moves first is analogous, and so omitted. 
  \fine
\end{proposition}

Note that it is easy to generalize the congruence property for parallel composition as follows: If $p_i \sim_{sp} q_i$ for $i = 1, 2$, then 
$p_1 \para p_2 \sim_{sp} q_1 \para q_2$. In fact, if $R_i$ is an sp-bisimulation for $p_i \sim_{sp} q_i$, for $i = 1, 2$,
then it is easy to see that $R =  \{l_1 \oplus l_2 \mid
l_1 \in R_1, l_2 \in R_2\}$ is an sp-bisimulation.

\begin{proposition}\label{prop3-cong-nb}{\bf (Congruence for restriction)}

\noindent
Let $p$ and $q$ be general FNM processes. If $p \sim_{sp} q$, then 
$\restr{a}p  \sim_{sp}  \restr{a}q$\,  for all $a \in \mathcal{L}$.

\proof Let $R$ be a structure-preserving bisimulation containing a link $l$ such that $\dec(p) = \pi_1(l)$ and $\dec(q) = \pi_2(l)$.
Relation $R_5 =  \{l\sost{a'}{a} \mid l \in R\}$, where the substitution $\sost{a'}{a}$ is applied element-wise,\footnote{This means that
$\theta\sost{a'}{a} = \theta$ and $((p_1, p_2) \oplus l)\sost{a'}{a} = (p_1\sost{a'}{a}, p_2\sost{a'}{a}) \oplus l\sost{a'}{a}$.}
is the required sp-bisimulation. 
 
 In fact, assume $c\sost{a'}{a} \subseteq l\sost{a'}{a} \in R_5$ and a transition $t_1$ such that $\pi_1(c\sost{a'}{a}) = \pre{t_1}$.
 By the net semantics, this is 
 possible only if 
 
 $t_1 = t_1'\sost{a'}{a} = (\pre{t_1'}\sost{a'}{a}, \ell(t_1'), \post{t_1'}\sost{a'}{a})$ 
 
 \noindent
 and $\ell(t_1')$ does not contain
 any occurrence of action $a$. Therefore, $c \subseteq l \in R$ is such that  $\pi_1(c) = \pre{t_1'}$.
 Since $R$ is an sp-bisimulation, there exist a transition $t_2'$ such that $\ell(t_1') = \ell(t_2')$, $\pi_2(c) = \pre{t_2'}$,
 and a linking $\overline{c}$ such that $\pi_1(\overline{c}) = \post{t_1'}$, $\pi_2(\overline{c}) = \post{t_2'}$
 and $(l \ominus c) \oplus \overline{c} \in R$.
 Therefore, by the net semantics it is possible to derive $t_2 = t_2'\sost{a'}{a}$
 with the property that $\ell(t_1) = \ell(t_2)$, $\pi_2(c\sost{a'}{a}) = \pre{t_2}$,
 $\pi_1(\overline{c}\sost{a'}{a}) = \post{t_1}$, $\pi_2(\overline{c}\sost{a'}{a}) = \post{t_2}$
 and $(l\sost{a'}{a} \ominus c\sost{a'}{a}) \oplus \overline{c}\sost{a'}{a} \in R_5$, as required.
 The symmetric case when a transition $t_2$ is such that $\pi_2(c\sost{a'}{a}) = \pre{t_2}$ is analogous, and so omitted.
 \fine
\end{proposition}

Still there is one construct missing: recursion, defined over guarded terms only. 
Consider an extension of FNM where terms can be constructed using variables, such as $x, y, \ldots$:
this defines
an ``open'' FNM. 

\begin{definition}{\bf (Open FNM)}\label{open-fnm-def}
Let $Var = \{x, y, z, \ldots\}$ be a finite set of variables. The FNM {\em open terms} 
are generated from actions, constants and variables by the 
following abstract syntax (using three syntactic categories):

$\begin{array}{lclccccccccllcl}
s &  ::= &  \nil  & | &  \mu.t  & | &\;  \underline{a}.s \; & | &\; s+s  & \hspace{1.3 cm} \mbox{{\em guarded open terms}}\\
q & ::= & s  & | &  C & | & x &&& \hspace{1.4 cm} \mbox{{\em sequential open terms}}\\
t & ::= & q  & | &  t \para t  &&&&&\hspace{1.3 cm} \mbox{{\em restriction-free open terms}}\\
p & ::= & t  & | & \restr{a}p  &&&&&\hspace{1.3 cm} \mbox{{\em general open terms}}\\
\end{array}$

\noindent
where $x$ is any variable taken from $Var$. 
The {\em open} net semantics for open FNM extends the 
net semantics in Section \ref{fnm-sec} with $Net(x)   =   (\{x\}, \emptyset, \emptyset, x)$, 
so that, e.g., the semantics of $a.x$ is the net $(\{a.x, x\}, \{a\},$ $\{(a.x,a,x)\}, a.x)$. 
\fine
\end{definition}

Sometimes we use the notation $p(x_1, \ldots, x_n)$ to state explicitly that term $p$ is open on the tuple 
of variables $(x_1, \ldots, x_n)$.
For instance, $p_1(x) = a.(b.\nil + c.x) + d.x$ and $p_2(x) = d.x + a.(c.x + b.\nil)$ are open guarded FNM terms.

Structure-preserving bisimulation equivalence can be extended to open terms as follows.
An open term $p(x_1, \ldots, x_n)$ can be {\em closed} by means of a substitution

$p(x_1, \ldots, x_n)\{r_1 / x_1, \ldots, r_n / x_n\}$

\noindent
with the effect that each occurrence of the variable $x_i$ (within $p$ and the body of each
constant in $\const{p}$)
is replaced by the {\em closed} FNM sequential 
process $r_i$, for $i = 1, \ldots, n$. For instance,
$p_1(x)\{d.\nil / x\} = a.(b.\nil + c.d.\nil) + d.d.\nil$.

A natural extension of structure-preserving bisimilarity $\sim_{sp}$ over open {\em sequential} terms is as follows:
 $p(x_1, \ldots, x_n) \sim_{sp} q(x_1, \ldots, x_n)$
if for all tuples of (closed) FNM 
terms $(r_1, \ldots, r_n)$, we have that

$p(x_1, \ldots, x_n)\{r_1 / x_1, \ldots, r_n / x_n\}$ $\sim_{sp}$ $q(x_1, \ldots, x_n)\{r_1 / x_1, \ldots, r_n / x_n\}$. 

\noindent
E.g., it is easy to see that $p_1(x) \sim_{sp} p_2(x)$. As a matter of fact, for all $r$, 

$p_1(x)\{r / x\} = a.(b.\nil + c.r) +d.r$ $\sim_{sp}$ $d.r + a.(c.r + b.\nil)$ = $p_2(x)\{r / x\}$,

\noindent which can be easily 
proved. 

This definition can be extended to open markings (which are multisets of (open) sequential terms).
If $m(x_1, \ldots, x_n) = s_1(x_1, \ldots, x_n) \oplus \ldots \oplus s_k(x_1, \ldots, x_n)$, then 

$m(x_1, \ldots, x_n)\{r_1 / x_1, \ldots, r_n / x_n\} = $

$\hspace{1.5cm} s_1(x_1, \ldots, x_n)\{r_1 / x_1, \ldots, r_n / x_n\} \oplus \ldots \oplus s_k(x_1, \ldots, x_n)\{r_1 / x_1, \ldots, r_n / x_n\}$.

\noindent
Therefore, we state $m_1(x_1, \ldots, x_n) \sim_{sp} m_2(x_1, \ldots, x_n)$ if for all tuples of (closed) FNM 
terms $(r_1, \ldots, r_n)$, we have that

$m_1(x_1, \ldots, x_n)\{r_1 / x_1, \ldots, r_n / x_n\}$ $\sim_{sp}$ $m_2(x_1, \ldots, x_n)\{r_1 / x_1, \ldots, r_n / x_n\}$.

For simplicity's sake,
let us now restrict our attention to open guarded terms using a single undefined variable.
We can {\em recursively close} an open term $p(x)$ by means of a recursively defined constant. For instance, 
$A \eqdef p(x)\{A / x\}$. The resulting process constant $A$ is a closed FNM sequential process. 
By saying that net bisimilarity is a congruence for
recursion we mean what is stated in the following.  
For simplicity's sake, in the following a term $p$ open on a variable $x$ is no longer annotated as $p(x)$. 

\begin{theorem}\label{recurs-cong-sp-th}\index{Congruence}
Let $p$ and $q$ be two open guarded FNM terms, with one variable $x$ at most. 
Let $A \eqdef p\{A / x\}$, $B \eqdef q\{B / x\}$
and $p \sim_{sp} q$. Then $A \sim_{sp} B$.
\proof
Let $R = \{(s\{A / x\}, s\{B / x\}) \mid 
s \in S_{op}\}$, where 
$S_{op} = {\mathcal P}_{oFNM}^{seq} \setminus \{\nil\}$, i.e., the set of all sequential,
open FNM processes, except $\nil$.\footnote{Note that we are not considering {\em extended} terms, but only `normal' FNM sequential terms, 
because the restriction operator cannot occur within the body of a recursively defined constant. Hence, each marking $m$ over $S_{op}$ is admissible.}
Note that when $s$ is $x$, we get $(A, B) \in R$.

We want to prove that $R^+$ is an sp-bisimulation (up to $\sim_{sp}$), that can be used to prove that
$m\{A / x\} \sim_{sp} m\{B / x\}$ for all the markings $m \in \mathcal{M}_{fin}(S_{op})$.
By symmetry, it is enough to prove that if $l \in R^+$, then for all $c \subseteq l$
and for all $t_1$ such that $\pi_1(c) = \pre{t_1}$, there exist $t_2$ such that $\ell(t_1) = \ell(t_2)$, $\pi_2(c) = \pre{t_2}$, 
and $\overline{c}$ such that
$\pi_1(\overline{c}) = \post{t_1}$, $\pi_2(\overline{c}) = \post{t_2}$ and, moreover, $(l \ominus c) \oplus \overline{c}$ 
is such that
$\pi_1((l \ominus c) \oplus \overline{c}) \sim_{sp} m\{A/x\}$,  $\pi_2((l \ominus c) \oplus \overline{c}) \sim_{sp} m\{B/x\}$,
so that there exists $\overline{l} \in R^+$ such that $\pi_1(\overline{l}) = m\{A/x\}$ and $\pi_2(\overline{l}) = m\{B/x\}$.
The proof proceeds by induction on the size of $m$.
When $|m| = 1$, i.e., $m = s$ for some (open) sequential FNM process $s$, the proof is by induction on the definition of the net for $s\{A / x\}$. 
For this base case, we assume $l = \{(s\{A / x\},s\{B / x\})\}$, so that $c = l$ and $(l \ominus c) \oplus \overline{c} = \overline{c}$.

\begin{itemize}
\item 
$s = \mu.r$. In this case, $s\{A / x\} = \mu.r\{A / x\} \deriv{\mu} \dec(r)\{A / x\}$. Similarly,  $s\{B / x\} =$ 
$ \mu.r\{B / x\} $ $ \deriv{\mu} \dec(r)\{B / x\}$,
and clearly there exists $\overline{c} \in R^+$ such that
$\pi_1(\overline{c}) = \dec(r)\{A / x\}$ and $\pi_2(\overline{c}) = \dec(r)\{B / x\}$.

\item 
$s = \underline{a}.r$. In this case, transition $s\{A / x\} = \underline{a}.r\{A / x\} \deriv{a \diamond \sigma} m_1$ 
is derivable only if
$r\{A / x\} \deriv{\sigma}m_1$. Since $r$ is guarded, $r\{A / x\} \deriv{\sigma} m_1$ is possible only if $r \deriv{\sigma} m$
with $m_1 = m\{A / x\}$. So, $r\{B / x\} \deriv{\sigma} m\{B / x\}$ is derivable, and also 
$s\{B / x\}  \deriv{a \diamond \sigma} m\{B / x\}$, and clearly there exists $\overline{c} \in R^+$ such that
$\pi_1(\overline{c}) = m\{A / x\}$ and $\pi_2(\overline{c}) = m\{B / x\}$.

\item  
$s = D$, with $D \eqdef r$. So, $s\{A / x\}  \eqdef r\{A / x\}$ and 
$s\{B / x\}  \eqdef r\{B / x\}$.
If $ s\{A / x\} \deriv{\sigma} m_1$, then this is possible only if 
$r\{A / x\} \deriv{\sigma} m_1$. Since $r$ is guarded, $r\{A / x\} \deriv{\sigma} m_1$ is possible only if $r \deriv{\sigma} m$
with $m_1 = m\{A / x\}$. Therefore, also $r\{B / x\} \deriv{\sigma} m\{B / x\}$ is derivable, and also 
$s\{B / x\}  \deriv{\sigma} m\{B / x\}$, and clearly there exists $\overline{c} \in R^+$ such that
$\pi_1(\overline{c}) = m\{A / x\}$ and $\pi_2(\overline{c}) = m\{B / x\}$.

\item 
$s = s_1 + s_2$. In this case, $s\{A / x\} = s_1\{A / x\}  + s_2\{A / x\} $. A transition from $s\{A / x\}$, e.g.,
$s_1\{A / x\}  + s_2\{A / x\}   \deriv{\sigma} m_1$,
is derivable only if $s_i\{A / x\}  \deriv{\sigma} m_1$ for some $i = 1, 2$. Without loss of generality, assume
the transition is due to $s_1\{A / x\}  \deriv{\sigma} m_1$. Since $s_1$ is guarded, transition
$s_1\{A / x\}  \deriv{\sigma} m_1$ is derivable because $s_1  \deriv{\sigma} m$, with $m_1 = m\{A / x\}$.
Therefore, also $s_1\{B / x\}  \deriv{\sigma} m\{B / x\}$ is derivable, as well $s\{B / x\} = s_1\{B / x\}  + s_2\{B / x\} $
$\deriv{\sigma} m\{B / x\}$, and clearly there exists $\overline{c} \in R^+$ such that
$\pi_1(\overline{c}) = m\{A / x\}$ and $\pi_2(\overline{c}) = m\{B / x\}$.

\item 
$s = x$. We have $s\{A / x\} = A$ and $s\{B / x\} = B$. We prove that for each 
 $A  \deriv{\sigma} m_1$, there exists $m_2$ such that 
$B \deriv{\sigma} m_2$ with $m_1 \sim_{sp} m\{A / x\}$ and $m_2 \sim_{sp} m\{B / x\}$, so that
there exists $\overline{l} \in R^+$ such that $\pi_1(\overline{l}) = m\{A / x\}$ and $\pi_2(\overline{l}) = m\{B / x\}$.

By hypothesis, $A \eqdef p\{A / x\}$, hence also $p\{A / x\} \deriv{\sigma} m_1$ is a transition in the net for 
$p\{A / x\}$; since $p$ is guarded, $p\{A / x\} \deriv{\sigma} m_1$ is possible only if $p \deriv{\sigma} m$
with $m_1 = m\{A / x\}$. Therefore, also $p\{B / x\} \deriv{\sigma} m\{B / x\}$ is derivable.

But we also have $p \sim_{sp} q$, so $p \deriv{\sigma} m$ can be matched by $q \deriv{\sigma} m'$ with $m \sim_{sp} m'$ (by Remark \ref{rem-int-sp}).
Hence, $q\{B / x\} \deriv{\sigma} m'\{B / x\}$ is a derivable transition, with $m\{B / x\} \sim_{sp} m'\{B / x\}$. 
As $B \eqdef q\{B / x\}$, also $B \deriv{\sigma} m'\{B / x\}$ is a transition with 
$m_1 \sim_{sp} m\{A / x\}$, $m\{B / x\} \sim_{sp} m'\{B/ x\}$, so that 
there exists $\overline{l} \in R^+$ such that $\pi_1(\overline{l}) = m\{A / x\}$ and $\pi_2(\overline{l}) = m\{B / x\}$, as required.
\end{itemize}

When $|m| = n+1$, then a place $s$ and a marking $m'$ exist such that $m = s \oplus m'$. Hence, 
we have that $(s\{A / x\}, s\{B / x\}) \in R^+$ and there exists a linking $l_1 \in R^+$ such that $\pi_1(l_1) = m'\{A / x\}$
and $\pi_2(l_1) = m'\{B / x\}$. Note that $l_2= (s\{A / x\}, s\{B / x\}) \oplus l_1 \in R^+$ is a linking such that $\pi_1(l_2) = m\{A / x\}$
and $\pi_2(l_2) = m\{B / x\}$. Assume $c_2 \subseteq l_2$ and a transition $t_1$ such that $\pi_1(c_2) = \pre{t_1}$. $R^+$ is
an sp-bisimulation (up to $\sim_{sp}$) if there exist $t_2$ such that $\ell(t_1) = \ell(t_2)$ and $\pi_2(c_2) = \pre{t_2}$, and a linking $\overline{c_2}$
such that $\pi_1(\overline{c_2}) = \post{t_1}$, $\pi_2(\overline{c_2}) = \post{t_2}$,
$\pi_1((l_2 \ominus c_2) \oplus \overline{c_2}) \sim_{sp} \overline{m}\{A/x\}$,  $\pi_2((l \ominus c) \oplus \overline{c}) \sim_{sp} \overline{m}\{B/x\}$,
so that there exists $\overline{l} \in R^+$ such that $\pi_1(\overline{l}) = \overline{m}\{A/x\}$ and $\pi_2(\overline{l}) = \overline{m}\{B/x\}$.
We have to distinguish three cases.

\begin{itemize}
\item $c_2 = \{(s\{A / x\}, s\{B / x\})\}$. This is what we have considered for the base case. Hence, we know that 
if $t_1 = s\{A / x\}  \deriv{\sigma} m_1$, then there exists $m_2$ such that 
$t_2 = s\{B / x\} \deriv{\sigma} m_2$ with $m_1 \sim_{sp} \underline{m}\{A / x\}$ and $m_2 \sim_{sp} \underline{m}\{B / x\}$, so that
there exists $\overline{l}' \in R^+$ such that $\pi_1(\overline{l}') = \underline{m}\{A / x\}$ and $\pi_2(\overline{l}') = \underline{m}\{B / x\}$.
Therefore, for $t_1$ such that $\pi_1(c_2) = \pre{t_1}$,
there exist $t_2$ such that $\ell(t_1) = \ell(t_2)$, $\pi_2(c_2) = \pre{t_2}$, 
and $\overline{c_2}$ such that
$\pi_1(\overline{c_2}) = \post{t_1} = m_1$, $\pi_2(\overline{c_2}) = \post{t_2} = m_2$ and, moreover, $\overline{c_2} \oplus l_1$ 
is such that
$\pi_1(\overline{c_2} \oplus l_1) \sim_{sp} \overline{m}\{A/x\} = \underline{m}\{A / x\} \oplus m'\{A / x\}$,  
$\pi_2(\overline{c_2} \oplus l_1) \sim_{sp} \overline{m}\{B/x\} = \underline{m}\{B / x\} \oplus m'\{B / x\}$,
so that \footnote{Note 
that $\pi_1(\overline{c_2} \oplus l_1) = m_1 \oplus m'\{A / x\}$ is structure preserving bisimilar to  
$\overline{m}\{A/x\} = \underline{m}\{A / x\} \oplus m'\{A / x\}$, because $m_1 \sim_{sp} \underline{m}\{A / x\}$ and
$m'\{A / x\} \sim_{sp} m'\{A / x\}$. In fact, by Theorem \ref{adm=complete-th-fnm}, 
there exist (open) FNM {\em restriction-free} processes
$p_1, p_2,$ $q$ such that $m_1 = \dec(p_1)$,  $\underline{m}\{A / x\} = \dec(p_2)$, $m'\{A / x\} = \dec(q)$.
Therefore, the thesis follows by compositionality of $\sim_{sp}$ w.r.t. parallel composition, 
because $m_1 \oplus m'\{A / x\} = \dec(p_1 \para q)$ and $\underline{m}\{A / x\} \oplus m'\{A / x\} = \dec(p_2 \para q)$. 
This observation applies also in the following items.}
there exists $\overline{l}' \oplus l_1 \in R^+$ such that 
$\pi_1(\overline{l}'\oplus l_1) = \overline{m}\{A/x\}$ and $\pi_2(\overline{l}' \oplus l_1) = \overline{m}\{B/x\}$, as required.

\item $c_2 \subseteq l_1$. By induction (as $|m'| = n$), we know that for all $t_1$ such that $\pi_1(c_2) = \pre{t_1}$, there exist 
$t_2$ such that $\ell(t_1) = \ell(t_2)$, $\pi_2(c_2) = \pre{t_2}$, 
and $\overline{c_2}$ such that
$\pi_1(\overline{c_2}) = \post{t_1}$, $\pi_2(\overline{c_2}) = \post{t_2}$ and, moreover, $(l_1 \ominus c_2) \oplus \overline{c_2}$ 
is such that
$\pi_1((l_1 \ominus c_2) \oplus \overline{c_2}) \sim_{sp} \underline{m}'\{A/x\}$,  
$\pi_2((l_1 \ominus c_2) \oplus \overline{c_2}) \sim_{sp} \underline{m}'\{B/x\}$,
so that there exists $\overline{l}' \in R^+$ such that $\pi_1(\overline{l}') = \underline{m}'\{A/x\}$ and $\pi_2(\overline{l}') = \underline{m}'\{B/x\}$.
Therefore, for $t_1$ such that $\pi_1(c_2) = \pre{t_1}$,
there exist $t_2$ such that $\ell(t_1) = \ell(t_2)$, $\pi_2(c_2) = \pre{t_2}$, 
and $\overline{c_2}$ such that
$\pi_1(\overline{c_2}) = \post{t_1}$, $\pi_2(\overline{c_2}) = \post{t_2}$ and, moreover, setting $\underline{l} = 
(s\{A / x\}, s\{B / x\}) \oplus (l_1 \ominus c_2) \oplus \overline{c_2}$, we have that
$\pi_1(\underline{l}) \sim_{sp} \overline{m}\{A/x\} = s\{A / x\} \oplus \underline{m}'\{A / x\}$,  
$\pi_2(\underline{l}) \sim_{sp} \overline{m}\{B/x\} = s\{A / x\} \oplus \underline{m}'\{B / x\}$,
so that there exists $(s\{A / x\}, s\{B / x\}) \oplus \overline{l}'  \in R^+$ such that 
$\pi_1((s\{A / x\}, s\{B / x\}) \oplus \overline{l}')$ $ = \overline{m}\{A/x\} $ $= s\{A / x\} \oplus \underline{m}'\{A/x\}$ and 
$\pi_2((s\{A / x\}, s\{B / x\}) \oplus \overline{l}') = \overline{m}\{B/x\} = s\{B / x\} \oplus \underline{m}'\{B/x\}$, as required.

\item $c_2 = (s\{A / x\}, s\{B / x\}) \oplus c_2'$, with $c_2' \subseteq l_1$, $c_2' \neq \theta$.
Assume a transition $t_1$ is such that $\pi_1(c_2) = \pre{t_1}$. 
Then, there exist $t_1'$ and a step $G$ such that $\pre{t_1'} =  s\{A / x\}$, $\pre{G} = \pi_1(c_2')$,
 (hence, $\pre{t_1} = \pre{t_1'} \oplus \pre{G}$), $\post{t_1} = \post{t_1'} \oplus \post{G}$ and
 $\MSync(\ell(t_1') \oplus \ell(G), \ell(t_1))$ holds.
 
 In such a case,  since $(s\{A / x\}, s\{B / x\}) \in R$, by induction
 we have that there exist a transition $t_2'$ such that $\ell(t_1') = \ell(t_2')$, $\pre{t_2'} = s\{B / x\}$,
 and a linking $\overline{c_1}$ such that $\pi_1(\overline{c_1}) = \post{t_1'}$ and $\pi_2(\overline{c_1}) = \post{t_2'}$,
 $\pi_1(\overline{c_1}) \sim_{sp} \underline{m}\{A/x\}$,  $\pi_2(\overline{c_1}) \sim_{sp} \underline{m}\{B/x\}$,
so that there exists $\overline{l} \in R^+$ such that $\pi_1(\overline{l}) = \underline{m}\{A/x\}$ and $\pi_2(\overline{l}) = \underline{m}\{B/x\}$.
 
Since $c_2' \subseteq l_1$ and $l_1 \in R^+$, by induction (and by Remark \ref{rem-step-sp}), 
there exist a step $G'$ such that $\ell(G) = \ell(G')$, $\pi_2(c_2') =\pre{G'}$,
 and a linking $\overline{c_2}'$ such that $\pi_1(\overline{c_2}') = \post{G}$ and $\pi_2(\overline{c_2}') = \post{G'}$,
 $\pi_1((l_1 \ominus c_2') \oplus \overline{c_2}') \sim_{sp} \underline{m}'\{A/x\}$,  
 $\pi_2((l_1 \ominus c_2') \oplus\overline{c_2}') \sim_{sp} \underline{m}'\{B/x\}$,
so that there exists $\overline{l}' \in R^+$ such that $\pi_1(\overline{l}') = \underline{m}'\{A/x\}$ and $\pi_2(\overline{l}') = \underline{m}'\{B/x\}$.

 Hence, there exists a step $t_2' \oplus G'$, giving origin to transition $t_2$,
  such that $\pre{t_2} = \pre{t_2'} \oplus \pre{G'}$, $\post{t_2} = \post{t_2'} \oplus \post{G'}$, 
  $\MSync(\ell(t_2') \oplus \ell(G'), \ell(t_1))$ holds (because $\ell(t_1') = \ell(t_2')$ and $\ell(G) = \ell(G')$), so that $\ell(t_1) = \ell(t_2)$,
  $\pi_1(c_2) = \pre{t_1}$, $\pi_2(c_2) = \pre{t_2}$, and, letting $\overline{c_2} = \overline{c_1} \oplus \overline{c_2}'$, we have
 also that $\pi_1(\overline{c_2}) = \post{t_1}$, $\pi_2(\overline{c_2}) = \post{t_2}$ and, finally, 
 letting $h = \overline{c_1} \oplus (l_1 \ominus c_2') \oplus\overline{c_2}'$, 
 we have that
 $\pi_1(h) \sim_{sp} \underline{m}\{A/x\} \oplus \underline{m}'\{A/x\}$,
 $\pi_2(h) \sim_{sp} \underline{m}\{B/x\} \oplus \underline{m}'\{B/x\}$,
and $\overline{l} \oplus \overline{l}' \in R^+$ is such that 
$\pi_1(\overline{l} \oplus\overline{l}') = \underline{m}\{A/x\} \oplus \underline{m}'\{A/x\}$ and 
$\pi_2(\overline{l} \oplus \overline{l}') = \underline{m}\{B/x\} \oplus \underline{m}'\{B/x\}$, as required.\\[-1.2cm]
\end{itemize}
\fine
\end{theorem}

The extension to the case of open terms with multiple undefined constants, e.g., $p(x_1, \ldots, x_n)$ 
can be obtained in a standard way \cite{Mil89,GV15}.

%
\section{Algebraic Properties}\label{comp-sp-alg-sec}
%

Now we propose a set of algebraic properties that hold for structure-preserving bisimilarity. For some of these, 
we take inspiration from similar laws for
standard bisimulation equivalence on the labeled transition system semantics of CCS-like process algebras (see, e.g., \cite{Mil89,GV15,SW}).

\begin{proposition}\label{aci-+-fnm}{\bf (Laws of the choice operator)}
 For all $p, q, r \in \mathcal{P}_{FNM}^{grd}$, the following hold:
 
 $\begin{array}{lrcllllll}
\qquad & p + (q + r)  &\sim_{sp} & (p + q) + r & \quad &\; \mbox{(associativity)}\\
\qquad & p + q  &\sim_{sp} & q + p & \quad &\;  \mbox{(commutativity)}\\
\qquad & p + \nil  &\sim_{sp} & p & \mbox{if $p \neq \nil$\; } &\;  \mbox{(identity)}\\
\qquad & p + p  &\sim_{sp} & p & \mbox{if $p \neq \nil$\; } & \; \mbox{(idempotency)}\\
\end{array}$
\proof For each law, it is enough to exhibit a suitable structure-preserving bisimulation relation.
For instance, for idempotency, for each $p$ guarded ($p \neq \nil$), take 
relation $R_p \; = \; \{(p + p, p) \} \cup \mathcal{I}_p^+ $
where 
$\mathcal{I}_p = \{(s, s) \mid \exists m \in [dec(p)\rangle, s \in m\}$. It is an easy exercise to 
check that $R_p$ is a structure-preserving bisimulation.
In fact, if $p \deriv{\sigma} m$, then (according to rule (sum$_1$)) also $p + p \deriv{\sigma} m$ and there exists
$l \in \mathcal{I}_p^+$ such that $\pi_1(l) = m = \pi_2(l)$, and so $l \in R_p$. Symmetrically, if $p + p \deriv{\sigma} m$, 
then (according to the rule (sum$_1)$) 
this is possible only if  $p \deriv{\sigma} m$ is derivable and the condition $l \in R_p$ such that $\pi_1(l) = m = \pi_2(l)$ 
is trivially satisfied.
As a further example, for the associativity law, the candidate sp-bisimulation relation is
$R_{(p,q,r)} \; = \; \{(p + (q +r), (p+ q) +r) \} \cup \mathcal{I}_{(p,q,r)}^+$,
where 
$\mathcal{I}_{(p,q,r)} = \{(s, s) \mid \exists m \in [\dec(p + (q +r))\rangle, s \in m\}$. 
\fine
 \end{proposition}
 
 Note that the identity law and the idempotency law have the side condition $p \neq \nil$. This is strictly necessary because
 $\nil + \nil$ and $\nil$ have a completely different semantics: the former originates a stuck place, while the latter the empty marking. Of course,
 these are not sp-bisimilar as $|\dec(\nil+\nil)| = 1$ while $|\dec(\nil)| = 0$.
 
 As we have shown that the choice operator $+$ is associative (and commutative)
w.r.t. $\sim_{sp}$, 
we sometimes use the $n$-ary version of this operator.
Hence, we shorten the term $( \ldots (p_1 + p_2)  + \ldots + p_n)$ as $\sum_{i = 1}^{n} p_i$.

\begin{proposition}\label{aci-spref-fnm}{\bf (Laws of the strong prefixing operator)}
 For all $p, q \in \mathcal{P}_{FNM}^{grd}$ and for each $a \in \mathcal{L}$, the following hold:
 
   $\begin{array}{lrcllllll}
&  \underline{a}.(p + q)  &\sim_{sp} & \underline{a}.p + \underline{a}.q & \quad & \;  \mbox{(distribution)}\\
&  \underline{a}.\tau.p  &\sim_{sp} & a.p & \quad & \; \mbox{(absorption)}\\
\mbox{if $p \sim_{sp} \nil +\nil$ or $p = \nil$, then} & \quad \underline{a}.p   &\sim_{sp} & \nil +\nil & \quad &\; \mbox{(annihilation)}\\
\end{array}$
\proof For each law, it is enough to exhibit a suitable sp-bisimulation relation.
For distribution, take 
relation $R_1 \; = \; \{( \underline{a}.(p + q),  \underline{a}.p + \underline{a}.q) \} \cup \mathcal{I}_{(p,q)}^+$, which is clearly an sp-bisimulation.
For absorption, take 
relation $R_2 \; = \;  \{( \underline{a}.\tau.p,  a.p) \} \cup \mathcal{I}_{p}^+$, which is clearly an sp-bisimulation.
For annihilation, take 
relation $R_3 \; = \; \{( \underline{a}.p, \nil + \nil) \}$, which is clearly an sp-bisimulation.
\fine
 \end{proposition}
 
 Note that in the annihilation law we have equated $\underline{a}.p$ to $\nil + \nil$, as both are 
 stuck places, and not to $\nil$, whose semantics is the empty marking.

\begin{proposition}\label{rec-laws-fnm}{\bf (Laws of the constant)}
 For each $p \in \mathcal{P}_{FNM}^{grd}$, and each $C \in \cons$, 
 the following hold:
 
  $\begin{array}{lrcllllll}
\mbox{if $C \eqdef \nil$, then} & \quad C  &\sim_{sp} & \nil + \nil & \quad &\; \mbox{(stuck)}\\
\mbox{if $C \eqdef p$ and $p \neq \nil$, then} & \quad C  &\sim_{sp} & p & \quad &\; \mbox{(unfolding)}\\
 \mbox{if $C \eqdef p\{C/x\}$  
 and $q \sim_{sp} p\{q/x\} $ then} & \quad C & \sim_{sp} & q & \quad &\;  \mbox{(folding)}\\
\end{array}$\\

\noindent where, in the third law, $p$ is actually open on $x$ (while $q$ is closed).

\proof The stuck property is trivial: since the decomposition of a constant is a place, if the body is stuck,
it corresponds to a stuck place, such as $\nil + \nil$.

The required structure-preserving bisimulation proving the unfolding property is 
$R_{C,p} \; = \; \{(C, p)\} \cup \mathcal{I}_C^+$,
where $\mathcal{I}_C = \{(s, s) \mid \exists m \in [\dec(C)\rangle, s \in m\}$. 
In fact, if $C \deriv{\sigma} m$, then (according to the operational net semantics for $C \eqdef p$) this means that
also $p \deriv{\sigma} m$, so that there exists $l \in \mathcal{I}_C^+$ such that $\pi_1(l) = m = \pi_2(l)$, and so $l \in R_{C,p}$
as required. Symmetrically if $p$ moves first.

For the folding property, observe that 
the statement is implied by the following:
if $q_1 \sim_{sp} p\{q_1/x\}$ and $q_2 \sim_{sp} p\{q_2/x\}$ then $q_1 \sim_{sp} q_2$. In fact, if we choose $q_1 = C$,
then $C = q_1 \sim_{sp} p\{q_1/x\} = p\{C/x\}$ (which holds by hypothesis, due to the unfolding property) 
and $C = q_1 \sim_{sp} q_2$, which is the thesis. Note that $q_1$ and $q_2$ must be sequential processes.
This statement can be proven by showing that, given

$
R = \{(s\{q_1/x\}, s\{q_2/x\}) \mid \exists m \in [\dec(p)\rangle, s  \in  m\}, 
$

\noindent
we have that $R^+$ is a structure-preserving bisimulation up to $\sim_{sp}$.
Clearly, when $s = x$, we have that $(q_1, q_2) \in R$.
So, it remains to prove the sp-bisimulation (up to) conditions. 

If $s\{q_1/x\} \deriv{\sigma} t$, then this 
can be due to one of the following:
\begin{itemize}
\item 
$s \deriv{\sigma} m$ and so $t = m\{q_1/x\}$, where the substitution is applied element-wise to 
each place in $m$.
In this case, also $s\{q_2/x\} \deriv{\sigma} m\{q_2/x\}$ is 
derivable such that there exists $l \in R^+$ such that $\pi_1(l) = m\{q_1/x\}$ and $\pi_2(l) = m\{q_2/x\}$.
\item 
$s = x$ and $q_1 \deriv{\sigma} m_1$, and so $t = m_1$. 
Since $q_1 \sim_{sp} p\{q_1/x\}$ and $p$ is guarded, we have that there exists 
$m$ such that $p \deriv{\sigma} m$, 
$p\{q_1/x\} \deriv{\sigma} m\{q_1/x\}$ and (by Remark \ref{rem-int-sp}) $m_1 \sim_{sp} m\{q_1/x\}$. 
Therefore, $p\{q_2/x\} \deriv{\sigma} m\{q_2/x\}$ is derivable, too, and  $l \in R^+$ exists such that
$\pi_1(l) = m\{q_1/x\}$ and $\pi_2(l) = m\{q_2/x\}$.
Since $q_2 \sim_{sp} p\{q_2/x\}$, it follows that  there 
exists a marking $m_2$  such that
$q_2 \deriv{\sigma} m_2$ with $m_2 \sim_{sp} m\{q_2/x\}$ (by Remark \ref{rem-int-sp}). 

Summing up, if $x\{q_1/x\} = q_1 \deriv{\sigma} m_1$, 
then $x\{q_2/x\} = q_2 \deriv{\sigma} m_2$
such that $m_1 \sim_{sp} m\{q_1/x\}$, $l \in R^+$ exists such that
$\pi_1(l) = m\{q_1/x\}$ and $\pi_2(l) = m\{q_2/x\}$ and, moreover, 
$m\{q_2/x\} \sim_{sp} m_2$, as required 
by the structure-bisimulation bisimulation up to condition. 
\end{itemize}
Simmetrically, if $s\{q_2/x\}$ moves first. Hence, $R^+$ is a structure-preserving bisimulation up to $\sim_{sp}$.
\fine
 \end{proposition}

\begin{remark}\label{rem-unf}
Note that the stuck property and the unfolding property can be summarized by one single unfolding law as follows:
\[
\mbox{if $C \eqdef p$, then \quad $C  \sim_{sp}  p + \nil$}
\]
because in case $p = \nil$, we have that $C$ is sp-bisimilar to $\nil + \nil$, while in case $p \neq \nil$, the addition of the summand
$\nil$ is irrelevant, as it can be absorbed (by the identity law in Proposition \ref{aci-+-fnm}).
\fine
\end{remark}
 
 \begin{proposition}\label{aci-parallel-fnm}{\bf (Laws of the parallel operator)}
 For all restriction-free FNM processes $p, q, r$, the following hold:

 $\begin{array}{lrclllll}
\qquad & \quad p \para (q \para r)  & \; \sim_{sp} \; & (p \para q) \para r & \quad & \mbox{(associativity)}\\
\qquad & \quad p \para q  &\; \sim_{sp} \; & q \para p & \quad & \mbox{(commutativity)}\\
\qquad & \quad p \para \nil  & \; \sim_{sp} \; & p & \quad & \mbox{(identity)}\\
\end{array}$
\proof
To prove that each law is sound, it is enough to observe that the net for the process
in the left-hand-side is exactly the same as the net for the process in the right-hand-side.
For instance, $Net(p \para q) = Net(q \para p)$, because 
$\dec(p \para q) = \dec(p) \oplus \dec(q) = \dec(q) \oplus \dec(p) = \dec (q \para p)$ (cf. Proposition \ref{netp=netq-fnm}). 
Therefore, taken the identity relation $\mathcal{I}$ on places, the relation $\mathcal{I}^+$ is enough to prove that
$p \para q   \sim_{sp}  q \para p$.
 \fine
 \end{proposition}
 
As we have shown that parallel composition is associative (and commutative)
w.r.t. $\sim_{sp}$, 
we sometimes use the $n$-ary version of this operator.
Hence, we shorten the term $( \ldots (p_1 \para p_2)  \para \ldots \para p_n)$ as $\Pi_{i = 1}^{n} p_i$.

In order to define some algebraic properties for the restriction operator, we assume the reader familiar with the notion of {\em free names} of a process $p$ (denoted by ${\it fn}(p)$), 
{\em bound names} (denoted by ${\it bn}(p)$) and {\em syntactic substitution} of action $b$ in place of $a$ within $p$, denoted by 
$p\sost{b}{a}$  (see, e.g., \cite{GV15,SW} for introductory books on these topics).

\begin{proposition}\label{restr-law-fnm}{\bf (Laws of the restriction operator)}
For each process $p \in \mathcal{P}_{FNM}$ and for all $a, b \in \mathcal{L}$, the following hold:
 
  $\begin{array}{lrclllll}
\qquad & \restr{a}p   &\sim_{sp} & p  & \mbox{if $a \not\in \mbox{{\it fn}}(p)$} & \mbox{(remove)}\\
\qquad &  \restr{a}(\restr{b}p) & \sim_{sp} & \restr{b}(\restr{a}p) & \mbox{if $a \neq b$} & \mbox{(reorder)}\\
\qquad &  \restr{a}p & \sim_{sp} &  \restr{b}(p\sost{b}{a}) &  \mbox{if $b \not\in \mbox{{\it fn}}(p) \cup bn(p)$ } & \mbox{(alpha-conversion)}\\
 \end{array}$
\proof
The proof of the remove law is easy because, if $a \not\in \mbox{{\it fn}}(p)$, then $\dec(\restr{a}p) = \dec(p)$, so that 
by Proposition \ref{netp=netq-fnm} we have that $Net(\restr{a}p) = Net(p)$.
Hence, taken the identity relation $\mathcal{I}$ on places, the relation $\mathcal{I}^+$ is enough to prove this property.

Similarly, one can argue that the reorder law holds, because $\dec(\restr{a}(\restr{b}p)) = \dec(\restr{b}(\restr{a}p))$.

To prove the alpha-conversion law, consider that $\dec(\restr{a}p) = \dec(p)\sost{a'}{a}$ and that 
$\dec(\restr{b}(p\sost{b}{a})) = \dec(p\sost{b}{a})\sost{b'}{b}$.
Hence, consider the relation 

$R = \{(s\sost{a'}{a}, s\sost{b'}{a}) \mid \exists m \in \dec(p), s \in m\}$.

\noindent
It is easy to observe that, under the assumption that $b \not\in \mbox{{\it fn}}(p) \cup bn(p)$, relation 
$R^+$ is an sp-bisimulation proving the alpha-conversion law.
 \fine
 \end{proposition}

Note that the remove law above implies also the following law:

$\quad  \restr{a}(\restr{a}p)  \sim_{sp} \restr{a}p$

\noindent stating that duplicated application of the restriction operator are inessential.
This law, together with the reorder law, 
justifies the notational convention, that we often adopt, of using the restriction operator
over a {\em set} of names; e.g., for $\restr{a}(\restr{b}(\restr{a}p))$, we can use the notation $\restr{\{a,b\}}p$.

Now we introduce some laws about the interplay between some operators, related to manipulation 
of the preset or of the postset of net transitions.
However, we anticipate that these laws are rather specific, less general and less elegant than those above.
The following example illustrates the issue for the presets.
 
\begin{example}\label{ex-preset1} 
Consider the term $p = \restr{a}(\restr{b}((\underline{a}.\underline{b}.c.p_1 \para \overline{a}.p_2) \para \overline{b}.p_3))$. 
The only initial transition
performable by $\dec(p)$ is a three-way synchronization, labeled $c$, reaching the marking
$((\dec(p_1) \oplus \dec(p_2) \oplus \dec(p_3))\sost{b'}{b})\sost{a'}{a}$.
The same behavior is also possible by the term $q =  \restr{a}(\restr{b}((\underline{b}.\underline{a}.c.p_1 \para \overline{a}.p_2) \para \overline{b}.p_3))$,
where the leader process has permuted the two strong prefixes.
As a matter of fact, the restricted actions that occur in the input sequence of the leader process, that is able to perform a multi-party synchronization in full, are to be considered 
as a multiset of inputs rather than a sequence of inputs. Moreover,
also the term $r = \restr{a}(\restr{b}((\underline{a}.\underline{b}.c.p_1 \para \overline{b}.p_2) \para \overline{a}.p_3))$, where the output guards have been exchanged, 
is sp-bisimilar to both $p$ and $q$.
The following laws focus on these aspects.
\fine
\end{example}

\begin{proposition}\label{preset-law-fnm}{\bf (Laws of the preset)} 
For each $n \geq 2$, for each restriction-free process $p \in \mathcal{P}_{FNM}$,
for each $a_0, \ldots a_n \in \mathcal{L}$, for each $A \subseteq \mathcal{L}$ such that $\{a_1, \ldots a_n\} \subseteq A$ ($a_0 \not\in A$),
for each $r \in \mathcal{P}^{grd}_{FNM}$ such that $fn(r) \cap \{a_1, \ldots, a_n\} = \emptyset$,
for each restriction-free process $q \in \mathcal{P}_{FNM}$ 
such that $fn(q) \cap \{a_1, \ldots, a_n\} = \emptyset$, 
for each $\delta$ permutation on $\{1, \ldots, n\}$,
the following holds:\\
 
 $\begin{array}{lrcrllll}
(i) & \restr{A}(((\underline{a}_n.\ldots \underline{a}_1.a_0.p + r) \para  \Pi_{i=1}^n \overline{a}_i.p_i) \para q) &  \sim_{sp}  \\
& \restr{A}(((\underline{a}_{\delta(n)}.\ldots \underline{a}_{\delta(1)}.a_0.p + r) \para  \Pi_{i=1}^n \overline{a}_i.p_i) \para q)\\\\
(ii) & \restr{A}(((\underline{a}_n.\ldots \underline{a}_1.a_0.p + r) \para  \Pi_{i=1}^n \overline{a}_i.p_i) \para q) &  \sim_{sp}  \\
& \restr{A}(((\underline{a}_{n}.\ldots \underline{a}_{1}.a_0.p + r) \para  \Pi_{i=1}^n \overline{a}_{\delta(i)}.p_i) \para q)\\
\end{array}$\\

For each $n \geq 1$, for each restriction-free process $p \in \mathcal{P}_{FNM}$,
for each $a_0, \ldots a_n \in \mathcal{L}$, for each $A \subseteq \mathcal{L}$ such that $\{a_0, \ldots a_n\} \subseteq A$,
for each $r \in \mathcal{P}^{grd}_{FNM}$ such that $fn(r) \cap \{a_0, \ldots, a_n\} = \emptyset$,
for each restriction-free process $q$ 
such that $fn(q) \cap \{a_0, \ldots, a_n\} = \emptyset$, 
for each
$\delta$ permutation on $\{0, \ldots, n\}$
the following holds:\\

 $\begin{array}{lrcrllll}
(iii) &  \restr{A}(((\underline{a}_n.\ldots \underline{a}_1.a_0.p + r) \para  \Pi_{i=0}^n \overline{a}_i.p_i) \para q) &  \sim_{sp}  \\
&  \restr{A}(((\underline{a}_{\delta(n)}.\ldots \underline{a}_{\delta(1)}.a_{\delta(0)}.p + r) \para  \Pi_{i=0}^n \overline{a}_i.p_i)\para q)\\\\
(iv) &  \restr{A}(((\underline{a}_n.\ldots \underline{a}_1.a_0.p + r) \para  \Pi_{i=0}^n \overline{a}_i.p_i) \para q) &  \sim_{sp}  \\
&  \restr{A}(((\underline{a}_{n}.\ldots \underline{a}_{1}.a_{0}.p + r) \para  \Pi_{i=0}^n \overline{a}_{\delta(i)}.p_i)\para q)\\
\end{array}$

\proof 
Let $\rho$ be the substitution that replaces $a_i$ by $a_i'$ for each $a_i \in A$. Moreover, 
let $Id_{p} = \{(s', s') \mid $ $ \exists m \in [\dec(p)\rangle, s' \in m\}$ and
$Id_{r} = \{(s'', s'') \mid \exists m \in [\dec(r)\rangle, s'' \in m\}$.
For law (i), let $Q = \Pi_{i=1}^n \overline{a}_i.p_i \para q$ and $Id_{Q} = \{(s, s) \mid \exists m \in [\dec(Q)\rangle, s \in m\}$.
Let us consider $s_1 = (\underline{a}_n.\ldots \underline{a}_1.a_0.p + r)\rho$
and $s_2 = (\underline{a}_{\delta(n)}.\ldots \underline{a}_{\delta(1)}.a_0.p + r)\rho$.
Then, 

$\begin{array}{lcllll}
R & = & \{(s_1, s_2) \oplus l\rho \mid l \in Id_{Q}^+\}  \cup\\ 
& & \{l_1\rho \oplus l_2\rho \mid l_1 \in Id^+_{p}, l_2 \in Id^+_{Q}\} \cup\\
& &  \{l_1\rho \oplus l_2\rho \mid l_1 \in Id^+_{r}, l_2 \in Id^+_{Q}\}\\
\end{array}$ 

\noindent
is an sp-bisimulation proving law (i) because it contains the linking $(s_1, s_2) \oplus l\rho$, where $l \in Id_{Q}^+$ is such that
$ \pi_1(l\rho) = \dec(Q)\rho = \pi_2(l\rho)$.
In a similar manner, law (iii) can be proved.

For law (ii), consider the linking $\overline{l} = \{( \overline{a}_1.p_1,  \overline{a}_{\delta(1)}.p_1), \ldots, ( \overline{a}_n.p_n,  \overline{a}_{\delta(n)}.p_n)\}$,
 the identity link $(s, s)$, where $s = (\underline{a}_n.\ldots \underline{a}_1.a_0.p + r)\rho$, the family of relations 
 $Id_{p_i} =  \{(s', s') \mid $ $ \exists m \in [\dec(p_i)\rangle, s' \in m\}$ and the relation 
 $Id_{q} = \{(s'', s'') \mid $ $ \exists m \in [\dec(q)\rangle, s'' \in m\}$. Let us denote by 
 $R' = \{l_1 \oplus \ldots \oplus l_n \mid l_i \in Id_{p_i}^+, \mbox{ for } i = 1, \ldots, n\}$.
 Then, relation

$\begin{array}{lcllll}
R'' & = & \{(s, s) \oplus \overline{l}\rho \oplus l\rho \mid l \in Id_{q}^+\}  \cup\\ 
& & \{l_1\rho \oplus l_2\rho \oplus l_3\rho \mid l_1 \in Id^+_{p}, l_2 \in R', l_3 \in Id_{q}^+\} \cup\\
& & \{l_1\rho \oplus \overline{l}\rho \oplus l_3\rho \mid l_1 \in Id^+_{r}, l_3 \in Id_{q}^+\}\\
\end{array}$ 

\noindent
is an sp-bisimulation proving law (ii) because it contains the linking $(s, s) \oplus \overline{l}\rho \oplus l\rho$, where $l \in Id_{q}^+$ is such that
$ \pi_1(l\rho) = \dec(q)\rho = \pi_2(l\rho)$.
In a similar manner, law (iv) can be proved.
 \fine
 \end{proposition}

Note that in these laws some of the $a_i$'s can be the same action. For instance, the third law above (together with the identity laws for 
the choice operator and parallel one), allows us to prove that

\noindent
$\restr{\{a,b\}}(((\underline{a}.\underline{b}.a.p \para \overline{a}.p_1) \para \overline{b}.p_2 )\para \overline{a}.p_3)  \sim_{sp} 
\restr{\{a,b\}}(((\underline{a}.\underline{a}.b.p \para \overline{a}.p_1) \para \overline{b}.p_2 )\para \overline{a}.p_3).$

However, these laws of the preset do not ensure that 
all the possible equalities of this sort (i.e., related to the preset of a multi-party transition) can be derived.
For instance, considering Example \ref{ex-preset1}, if we assume that $b$ is not free in $p_1, p_2$ and $p_3$, then 
also $t =  \restr{a}((\underline{a}.\underline{a}.c.p_1 \para \overline{a}.p_2) \para \overline{a}.p_3)$ is sp-bisimilar to $p$, $q$
and $r$. But a specific law should be added to those in Proposition \ref{preset-law-fnm} to this aim.

In order to show which kind of laws we may single out about the postset of synchronized transitions, we discuss some simple examples.

\begin{example}\label{ex-postset1} 
Let us consider the term $q_1 = \restr{a}(a.p_1 \para \overline{a}.p_2)$.
The only initial transition
performable by $\dec(q_1)$ is a binary, $\tau$-labeled synchronization, reaching the marking
$((\dec(p_1) \oplus \dec(p_2))\sost{a'}{a}$.
The same behavior is also possible, among many others, by the following terms:

$q_2 = \restr{a}(a.(p_1\para p_2) \para \overline{a}.\nil)$,

$q_3 = \restr{a}(a.\nil \para \overline{a}.(p_1\para p_2))$,

$q_4 = \restr{a}(a.p_2 \para \overline{a}.p_1)$,

\noindent
where the continuations of the two synchronizing components of $q_1$ can be mixed up at will. In general, we can state that
for each $p_1', p_2'$ such that  $p_1' \para p_2' \sim_{sp} p_1 \para p_2$, also $q = \restr{a}(a.p_1' \para \overline{a}.p_2')$
is sp-bisimilar to $q_1$. This idea can be generalized by the following laws.
\fine
\end{example}

\begin{proposition}\label{postset1-law-fnm}{\bf (Laws of the postset-1)} 
For each $n \geq 1$, for all restriction-free processes $p, p_i \in \mathcal{P}_{FNM}$ for $i = 1, \ldots, n$, 
for all $a_0, \ldots a_n \in \mathcal{L}$, for each $A \subseteq \mathcal{L}$ such that $\{a_1, \ldots a_n\} \subseteq A$ ($a_0 \not\in A$),
for each $r \in \mathcal{P}^{grd}_{FNM}$ such that $fn(r) \cap \{a_1, \ldots, a_n\} = \emptyset$,
for each restriction-free process $q \in \mathcal{P}_{FNM}$ 
such that $fn(q) \cap \{a_1, \ldots, a_n\} = \emptyset$,
for all restriction-free $p', p'_i \in \mathcal{P}_{FNM}$ for $i = 1, \ldots, n$, such that
$p \para \Pi_{i=1}^n p_i \sim_{sp} p' \para \Pi_{i=1}^n p'_i$,
the following holds:\\
 
 $\begin{array}{lrcrllll}
(i) & \restr{A}(((\underline{a}_n.\ldots \underline{a}_1.a_0.p + r) \para \Pi_{i=1}^n \overline{a}_i.p_i) \para q)  &  \sim_{sp}  \\
&\restr{A}(((\underline{a}_{n}.\ldots \underline{a}_{1}.a_0.p' + r) \para \Pi_{i=1}^n \overline{a}_i.p'_i) \para q)\\
\end{array}$\\

For each $n \geq 0$, for all restriction-free processes $p, p_i \in \mathcal{P}_{FNM}$ for $i = 0, \ldots, n$, 
for all $a_0, \ldots a_n \in \mathcal{L}$, for each $A \subseteq \mathcal{L}$ such that $\{a_0, \ldots a_n\} \subseteq A$,
for each $r \in \mathcal{P}^{grd}_{FNM}$ such that $fn(r) \cap \{a_0, \ldots, a_n\} = \emptyset$,
for each restriction-free process $q \in \mathcal{P}_{FNM}$ 
such that $fn(q) \cap \{a_0, \ldots, a_n\} = \emptyset$,
for all restriction-free processes $p', p'_i \in \mathcal{P}_{FNM}$ for $i = 0, \ldots, n$, such that
$p \para \Pi_{i=0}^n p_i \sim_{sp} p' \para \Pi_{i=0}^n p'_i$,
the following holds:\\
 
  $\begin{array}{lrcrllll}
(ii) & \restr{A}(((\underline{a}_n.\ldots \underline{a}_1.a_0.p + r) \para \Pi_{i=0}^n \overline{a}_i.p_i) \para q)  &  \sim_{sp} \\
&\restr{A}(((\underline{a}_{n}.\ldots \underline{a}_{1}.a_0.p' + r) \para \Pi_{i=0}^n \overline{a}_i.p'_i) \para q)\\
\end{array}$\\

\proof Since $p \para \Pi_{i=1}^n p_i \sim_{sp} p' \para \Pi_{i=1}^n p'_i$ by hypothesis, by congruence w.r.t.
parallel composition, we also have that $(p \para \Pi_{i=1}^n p_i) \para q \sim_{sp} (p' \para \Pi_{i=1}^n p'_i) \para q$.
Let $R$ be an structure-preserving 
bisimulation proving this, and let $\rho$ be the substitution replacing each $a_i$ by $a_i'$ for each $a_i$ in $A$.
Let us consider $R' = \{l\rho \mid l \in R\}$. 
Let us also consider $Id_{q} = \{(s, s) \mid \exists m \in [\dec(q)\rangle, s \in m\}$ and $Id_{r} = \{(s', s') \mid \exists m \in [\dec(r)\rangle, s' \in m\}$.

Let $\overline{l_1}$ be an identity linking such that 
$\pi_1(\overline{l_1}) =\dec(\Pi_{i=1}^n \overline{a}_i.p_i)\rho = \pi_2(\overline{l_1})$.
Let $\overline{l_1}'$ be a linking obtained from $\overline{l_1}$ above
by replacing each link $(s, s) \in \overline{l_1}$ by the link $(s, s\delta)$, where $\delta$ is the substitution that 
replaces $p_i$ by $p_i'$ for $i = 1, \ldots, n$. Let us denote by $\underline{l}$ the link
$((\underline{a}_n.\ldots \underline{a}_1.a_0.p + r)\rho, (\underline{a}_n.\ldots \underline{a}_1.a_0.p' + r)\rho)$.

It is easy to observe that the relation 

$R'' = \{(\underline{l} \oplus \overline{l_1}' \oplus l_2\rho) \mid l_2 \in Id^+_{q}\} \cup 
 \{(l\rho \oplus \overline{l_1}' \oplus l_2\rho) \mid l \in Id_{r}^+, l_2 \in Id^+_{q}\} \cup R'$

\noindent
is the required sp-bisimulation. The second law can be proved similarly.
\fine
\end{proposition}

\begin{remark}\label{rem-pre-post}
Note that the laws (ii) and (iv) of Proposition \ref{preset-law-fnm} are a special case of the above laws (i) and (ii), respectively.
As a matter of fact, when $p = p'$ and the sequence $p_1', \ldots, p_n'$ is just a permutation of the sequence $p_1, \ldots, p_n$, 
we have that law (i) above is the same 
as law (ii) of Proposition \ref{preset-law-fnm}. Similarly, one can argue that law (iv) of Proposition \ref{preset-law-fnm} is
a special case of law (ii) above.
\fine
\end{remark}

\begin{example}\label{ex-postset2} 
The free mix up operation discussed in the previous example can be done only if the initial synchronization is unique, 
so that the system is deterministic w.r.t. the synchronizations. In general, this is not the case.
For instance, consider the term $q_1 = \restr{a}((a.p_1 + a.p_2) \para \overline{a}.p_3)$.
There are two initial $\tau$-labeled synchronizations
performable by $\dec(q_1)$, reaching either the marking
$((\dec(p_1) \oplus \dec(p_3))\sost{a'}{a}$ or the marking $((\dec(p_2) \oplus \dec(p_3))\sost{a'}{a}$.
The same behavior is also possible by the following term:

$q_2 = \restr{a}((a.(p_1 \para p_3) + a.(p_2 \para p_3)) \para \overline{a}.\nil)$

\noindent
where the continuations of the unique output subprocess is moved to the continuation
of the two, alternative input subprocesses. In general, we can state that
for each $r_1, r_2$ such that  $p_3 \sim_{sp} r_1 \para r_2$,  
also $q_2 = \restr{a}((a.(r_1 \para p_1) + a.(r_1 \para p_2)) \para \overline{a}.r_2)$
is structure-preserving bisimilar to $q_1$. This idea can be generalized by the following laws.
\fine
\end{example}

\begin{proposition}\label{postset2-law-fnm}{\bf (Laws of the postset-2)} 
For each $n \geq 1$,
for all $a_0, \ldots a_n \in \mathcal{L}$,  
for each $A \subseteq \mathcal{L}$ such that $\{a_1, \ldots a_n\} \subseteq A$ ($a_0 \not\in A$),
for each guarded process $v \in \mathcal{P}^{grd}_{FNM}$ 
such that $fn(v) \cap \{a_1, \ldots, a_n\} = \emptyset$,
for each restriction-free process $q \in \mathcal{P}_{FNM}$ 
such that $fn(q) \cap \{a_1, \ldots, a_n\} = \emptyset$,  for each $k \geq 1$,
for all restriction-free processes $v_i \in \mathcal{P}_{FNM}$ for $i = 1, \ldots, k$, 
for all restriction-free processes $p_i \in \mathcal{P}_{FNM}$ for $i = 1, \ldots, n$, 
for all restriction-free processes $r, r_i \in \mathcal{P}_{FNM}$  for $i = 1, \ldots, n$ such that
$ \Pi_{i=1}^n p_i \sim_{sp} r \para  \Pi_{i=1}^n r_i$,
the following holds:\\

 $\begin{array}{lrcrllll}
(i) & \restr{A}(((\sum_{i=1}^{k} \underline{a}_n.\ldots \underline{a}_1.a_0.v_i +v) \para \Pi_{i=1}^n \overline{a}_i.p_i) \para q)
&  \sim_{sp}  \\
& \restr{A}(((\sum_{i=1}^{k} \underline{a}_n.\ldots \underline{a}_1.a_0.(v_i \para r) +v) \para \Pi_{i=1}^n \overline{a}_i.r_i) \para q)\\
\end{array}$\\

For each $n \geq 0$, 
for all $a_0, \ldots a_n \in \mathcal{L}$,  for each $A \subseteq \mathcal{L}$ such that $\{a_0, \ldots a_n\} \subseteq A$,
for each guarded process $v \in \mathcal{P}^{grd}_{FNM}$ 
such that $fn(v) \cap \{a_0, \ldots, a_n\} = \emptyset$, 
for each restriction-free process $q \in \mathcal{P}_{FNM}$ 
such that $fn(q) \cap \{a_0, \ldots, a_n\} = \emptyset$, for each $k \geq 1$,
for all restriction-free processes $v_i \in \mathcal{P}_{FNM}$ for $i = 1, \ldots, k$, 
for all restriction-free processes $p_i \in \mathcal{P}_{FNM}$ for $i = 0, \ldots, n$, 
for each restriction-free processes $r, r_i \in \mathcal{P}_{FNM}$ for $i = 0, \ldots, n$, such that
$ \Pi_{i=0}^n p_i \sim_{sp} r \para  \Pi_{i=0}^n r_i$,
the following holds:\\
 
 $\begin{array}{lrcrllll}
(ii) & \restr{A}(((\sum_{i=1}^{k} \underline{a}_n.\ldots \underline{a}_1.a_0.v_i +v) \para \Pi_{i=0}^n \overline{a}_i.p_i) \para q)
&  \sim_{sp}  \\
& \restr{A}(((\sum_{i=1}^{k} \underline{a}_n.\ldots \underline{a}_1.a_0.(v_i \para r) +v) \para \Pi_{i=0}^n \overline{a}_i.r_i) \para q)\\
\end{array}$

\proof Let $\rho$ be the substitution replacing each $a_i$ by $a_i'$ for each $a_i$ in $A$.
Since, by hypothesis, we have that $\Pi_{i=1}^n p_i \sim_{sp} r \para  \Pi_{i=1}^n r_i$,, by congruence w.r.t. parallel composition, we also 
have that
$v_j \para \Pi_{i=1}^n p_i \sim_{sp} v_j \para (r \para  \Pi_{i=1}^n r_i)$ for $j = 1 \ldots, k$.
Let $R_j$ be an sp-bisimulation proving this, for $j = 1, \ldots, k$.
Let $Id_{q} = \{(s, s) \mid \exists m \in [\dec(q)\rangle, s \in m\}$ and let  $Id_{v} = \{(s', s') \mid \exists m \in [\dec(v)\rangle, s' \in m\}$.

Let $\overline{l_1}$ be the linking composed of the links $(\overline{a}_i.p_i, \overline{a}_i.r_i)\rho$ for $i = 1, \ldots, n$.
Let $(s_1, s_2)$ be such that $s_1 = (\sum_{i=1}^{k} \underline{a}_n.\ldots \underline{a}_1.a_0.v_i +v)\rho$
and $s_2 = (\sum_{i=1}^{k} \underline{a}_n.\ldots \underline{a}_1.a_0.(v_i \para r) +v) \rho$.
It is not difficult to prove that relation 

\noindent
$
R'  =  \{(s_1, s_2) \oplus \overline{l_1} \oplus l\rho \mid l \in Id_{q}^+\} \cup  \{l_1\rho \oplus \overline{l_1} \oplus 
l_2\rho \mid l_1 \in Id_{v}^+, l_2 \in Id_q^+\} \cup  \bigcup_{j=1}^{k} R'_j $

\noindent
where $R'_j = \{l_1\rho \oplus l_2\rho \mid l_1 \in R_j, l_2 \in Id_{q}^+\} $,
is a structure-preserving bisimulation proving law (i), because it contains the linking $\overline{l} = (s_1, s_2) \oplus \overline{l_1} \oplus l\rho$,
for $l$ such that $\pi_1(l) = \dec(q) = \pi_2(l)$, so that 

$\pi_1(\overline{l}) = 
\dec(\restr{A}(((\sum_{i=1}^{k} \underline{a}_n.\ldots \underline{a}_1.a_0.v_i +v) \para \Pi_{i=1}^n \overline{a}_i.p_i) \para q))$
and 

$\pi_2(\overline{l}) = \dec(\restr{A}(((\sum_{i=1}^{k} \underline{a}_n.\ldots \underline{a}_1.a_0.(v_i \para r) +v) \para \Pi_{i=0}^n \overline{a}_i.r_i) \para q))$.
\fine
\end{proposition}

\begin{example}\label{ex-postset3} 
The partial mix up operation discussed in the previous example can be easily done only if the output is unique. 
In general, no easy mix up can be performed.
For instance, consider the term $q_1 = \restr{a}((a.p_1 + a.p_2) \para (\overline{a}.p_3 + \overline{a}.p_4))$.
Note that there are four initial synchronizations
performable by $\dec(q_1)$. However, in this case, as both occurrences of the input action $a$ can be synchronized with both 
occurrences of the output action $\overline{a}$, there is no easy mix up of the continuations: it is necessary that 
$p_1$ and $p_2$ (or $p_3$ and $p_4$) share some behaviorally equivalent subcomponent. E.g., if 
$p_1 \sim_{sp} r \para r_1$ and 
$p_2 \sim_{sp} r \para r_2$, then 
$q_2 = \restr{a}((a.r_1 + a.r_2) \para (\overline{a}.(r \para p_3) + \overline{a}.(r \para p_4)))$ is sp-bisimilar to $q_1$. 
Similarly, if $p_3 \sim_{sp} r \para r_3$ and 
$p_4 \sim_{sp} r \para r_4$, then it is not difficult to see that also
$q_3 = \restr{a}((a.(p_1 \para r) + a.(p_2 \para r)) \para (\overline{a}.r_3 + \overline{a}.r_4))$ is sp-bisimilar to $q_1$. 
Specific laws should be added to 
deal with such situations.
\fine
\end{example}

The examples above give evidence that in order to capture more and more equalities, more and more (rather cumbersome) laws must be added.
In fact, we think that it is not possible to capture all the algebraic laws that hold for sp-bisimilarity.

%
\section{Sound Axiomatization}\label{comp-sp-ax-sec}
%

In this section we provide a sound, but incomplete,  axiomatization of structure-preserv-ing bisimulation equivalence over FNM.  
For simplicity's sake, the syntactic definition of open FNM ({\em cf.} Definition \ref{open-fnm-def}) is assumed here flattened,
with only one syntactic category, but we require that
each ground instantiation of an axiom must respect the syntactic definition of (closed) FNM given in 
Section \ref{fnm-sec}. 
This means that we can write the axiom $x + (y + z) = (x + y) + z$ (these terms cannot be written 
in open FNM according to Definition \ref{open-fnm-def}), but 
it is invalid to instantiate it to $C + (a.\nil + b.\nil \para \nil) = (C + a.\nil) + (b.\nil \para \nil)$ because these are not legal FNM 
processes (the constant $C$ and the parallel process $b.\nil \para \nil$ cannot be used as summands).

The set of axioms are outlined in Tables \ref{axiom-tab1} and \ref{axiom-tab2}. 
We call $E$ the set of axioms $\{${\bf A1, A2, A3, A4, S1, S2, S3, C1, C2, P1, P2, P3, R1, R2, R3, Pr1, Pr2, Ps1, Ps2, Ps3, Ps4}$\}$. 
By the notation $E \vdash p = q$ we mean that there exists an equational deduction proof 
of the equality $p = q$, by using the axioms in $E$. Besides the usual equational deduction rules of reflexivity, symmetry, transitivity,
substitutivity and instantiation (see, e.g., \cite{GV15} for an introductory book on the subject), in order to deal with constants we 
need also the following recursion
congruence rule:
\[
\frac{p = q \; \wedge \; A \eqdef p\{A/x\} \; \wedge \; B \eqdef q\{B/x\}}{A = B}
\]

Let us first comment on the axioms in Table \ref{axiom-tab1}, which are more standard and also already appeared in the literature 
in some form.

The axioms {\bf A1-A4} are the usual axioms for choice \cite{Salomaa,Mil84} where, however, {\bf A3-A4} have the side 
condition $x \neq \nil$; hence, it is not possible to prove $E \vdash \nil + \nil = \nil$, as expected, because 
these two terms have a completely different semantics; in fact, no other {\em sequential} process $p$ can be equated to $\nil$.

The axioms {\bf S1-S3} are the axioms for strong prefixing, similar to those originally proposed in \cite{GM90}.
Note that the annihilation axiom {\bf S3} requires that $x$ is $\nil$ or can be proved equal to a stuck place, such as $\nil + \nil$.
Note also that these axioms are actually a finite collection of axioms, one for each input prefix $a$: since the set $\mathcal{L}$ 
is finite, the instances of {\bf S1-S3} are finitely many. 

The {\em conditional} axioms (or inference rules) {\bf C1-C2} are about process constants, originally introduced in \cite{Gor17b}
and inspired to those for the fixpoint operator of finite-state CCS \cite{Mil84}. 
Note that axiom {\bf C1}
summarizes the algebraic laws {\em stuck} and {\em unfolding} of Proposition \ref{rec-laws-fnm} (cf. Remark \ref{rem-unf}).
Note also that these conditional axioms are actually a finite collection of axioms, one for each constant 
definition: since the set $\cons$ of process constants is finite, the instances of {\bf C1-C2} are finitely many. 

The axioms {\bf P1-P3} are the usual ones for parallel composition. They are typical for non-interleaving behavioral congruences
(see, e.g., \cite{Gor17b,Gor22}), while they are not used for standard bisimilarity, in favor of the so-called
{\em expansion theorem} (or {\em interleaving law}) \cite{Mil89,GV15}, according to which parallel composition can be seen as 
a derived operator expressible
by a suitable combination of the operators of action prefixing and choice.

Finally, we have axioms {\bf R1-R3} for the restriction operator. These axioms occur in many contexts as ingredients of suitable
{\em structural congruences} for the operational interleaving semantics (defined on labeled transitions systems) 
of some process algebras, 
notably the $\pi$-calculus (see, e.g., \cite{Parrow-hb,SW}) or Multi-CCS \cite{GV15}.

\begin{table}[t]
{\renewcommand{\arraystretch}{1.3}
\hrulefill\\[-.7cm]
\small{

\begin{center}
$\begin{array}{llrcll}
{\bf A1} &\; \;  \mbox{Associativity} &\; \;  x + (y + z) & = & (x + y) + z &\\
{\bf A2} &\; \;  \mbox{Commutativity} &\; \;  x + y & = & y + x& \\
{\bf A3} &\; \;  \mbox{Identity} &\; \;  x + \nil & = & x & \quad  \mbox{ if $x \neq \nil$}\\
{\bf A4} &\; \;  \mbox{Idempotence} &\; \;  x + x & = & x & \quad  \mbox{ if $x \neq \nil$}\\
\end{array}$

\hrulefill

$\begin{array}{llrcll}
{\bf S1} &\; \;  \mbox{Distribution} &\; \;  \underline{a}.(x + y) & = & \underline{a}.x + \underline{a}.y &\\
{\bf S2} &\; \;  \mbox{Absorption} &\; \;  \underline{a}.\tau.x & = & a.x& \\
{\bf S3} &\; \;  \mbox{Annihilation} &\; \;  \underline{a}.x & = & \nil + \nil & \quad  \mbox{ if $x = \nil$ or $x = \nil + \nil$}\\
\end{array}$

\hrulefill

$\begin{array}{llrcllll}
{\bf C1} &\; \;  \mbox{Unfolding} &\; \; \mbox{if $C \eqdef p$,} & \mbox{then} &
\mbox{$C = p + \nil$} &\\
{\bf C2} &\; \;  \mbox{Folding} &\; \;  \mbox{if $C \eqdef p\{C/x\} \; \wedge \; 
q = p\{q/x\}$,} & \mbox{then} & \mbox{$C = q$} & \\
\end{array}$

\hrulefill

$\begin{array}{llrcll}
{\bf P1} &\; \;  \mbox{Associativity} &\; \;  x \para (y \para z) & = & (x \para y) \para z &\\
{\bf P2} &\; \;  \mbox{Commutativity} &\; \;  x \para y & = & y \para x & \\
{\bf P3} &\; \;  \mbox{Identity} &\; \;  x \para \nil & = & x &\\
\end{array}$

\hrulefill

$\begin{array}{llrcll}
{\bf R1} &\; \;  \mbox{Remove} &\; \;  \restr{a}x & = & x  & \quad  \mbox{ if $a \not\in fn(x)$}\\
{\bf R2} &\; \;  \mbox{Reorder} &\; \;  \restr{a}(\restr{b}x) & = & \restr{b}(\restr{a}x) & \quad  \mbox{ if $a \neq b$}\\
{\bf R3} &\; \;  \mbox{$\alpha$-conversion} &\; \;  \restr{a}x  & = & \restr{b}(x\sost{b}{a})& \quad  \mbox{ if $b \not \in fn(x) \cup bn(x)$}\\
\end{array}$

\hrulefill

\end{center}}
}
\caption{Some axioms for structure-preserving bisimulation equivalence}\label{axiom-tab1}
\end{table}

Let us now comment on the axioms in Table \ref{axiom-tab2}, which are new, to the best of our knowledge.
These axioms are all conditional and, moreover, are actually {\em infinitary} schemata due to the potentially
unbounded number of participants
to the multi-party synchronization they model. In fact, they are parametric in $n$.

Axioms {\bf Pr1-Pr2} are related to the laws of the preset, discussed in Proposition \ref{preset-law-fnm}, while {\bf Ps1-Ps4} 
are the axiomatic counterpart of the laws in Proposition \ref{postset1-law-fnm} and \ref{postset2-law-fnm}.
Note that we have omitted
the axioms related to laws (ii) and (iv) in Proposition \ref{preset-law-fnm} because these axioms would have been redundant 
(cf. Remark \ref{rem-pre-post}).

\begin{table}[t]
{\renewcommand{\arraystretch}{1.3}
\hrulefill\\[-.7cm]
\small{

\begin{center}
$\begin{array}{llrcll}
{\bf Pr1} & \mbox{if $\{a_1, \ldots, a_n\} \subseteq A$, $a_0 \not\in A$, $fn(y) \cap \{a_1, \ldots, a_n\} = \emptyset$, $fn(z) \cap \{a_1, \ldots, a_n\} = \emptyset$ and} &  \\
& \mbox{$\delta$ a permutation on $\{1, \ldots, n\}$, then} \\
&\quad \restr{A}(((\underline{a}_n.\ldots \underline{a}_1.a_0.x + y) \para  \Pi_{i=1}^n \overline{a}_i.x_i) \para z)  = \\
& \qquad \qquad  \restr{A}(((\underline{a}_{\delta(n)}.\ldots \underline{a}_{\delta(1)}.a_0.x + y) \para  \Pi_{i=1}^n \overline{a}_i.x_i) \para z)\\
{\bf Pr2} & \mbox{if $\{a_0, \ldots, a_n\} \subseteq A$, $fn(y) \cap \{a_0, \ldots, a_n\} = \emptyset$, $fn(z) \cap \{a_0, \ldots, a_n\} = \emptyset$ and} &  \\
& \mbox{$\delta$ a permutation on $\{0, \ldots, n\}$, then} \\
&\quad \restr{A}(((\underline{a}_n.\ldots \underline{a}_1.a_0.x + y) \para  \Pi_{i=0}^n \overline{a}_i.x_i) \para z)  = \\
& \qquad \qquad  \restr{A}(((\underline{a}_{\delta(n)}.\ldots \underline{a}_{\delta(1)}.a_{\delta(0)}.x + y) \para  \Pi_{i=0}^n \overline{a}_i.x_i) \para z)\\
\end{array}$

\hrulefill

$\begin{array}{llrcllll}
{\bf Ps1} & \mbox{if $\{a_1, \ldots, a_n\} \subseteq A$, $a_0 \not\in A$, $fn(y) \cap \{a_1, \ldots, a_n\} = \emptyset$, $fn(z) \cap \{a_1, \ldots, a_n\} = \emptyset$ and} &  \\
& \mbox{$x \para \Pi_{i=1}^n x_i = x' \para \Pi_{i=1}^n x'_i$, then} \\
&\quad \restr{A}(((\underline{a}_n.\ldots \underline{a}_1.a_0.x + y) \para  \Pi_{i=1}^n \overline{a}_i.x_i) \para z)  = \\
& \qquad \qquad  \restr{A}(((\underline{a}_{n}.\ldots \underline{a}_{1}.a_0.x' + y) \para  \Pi_{i=1}^n \overline{a}_i.x'_i) \para z)\\
{\bf Ps2} & \mbox{if $\{a_0, \ldots, a_n\} \subseteq A$, $fn(y) \cap \{a_0, \ldots, a_n\} = \emptyset$, $fn(z) \cap \{a_0, \ldots, a_n\} = \emptyset$ and} &  \\
& \mbox{$x \para \Pi_{i=0}^n x_i = x' \para \Pi_{i=0}^n x'_i$, then} \\
&\quad \restr{A}(((\underline{a}_n.\ldots \underline{a}_1.a_0.x + y) \para  \Pi_{i=0}^n \overline{a}_i.x_i) \para z)  = \\
& \qquad \qquad  \restr{A}(((\underline{a}_{n}.\ldots \underline{a}_{1}.a_0.x' + y) \para  \Pi_{i=0}^n \overline{a}_i.x'_i) \para z)\\

{\bf Ps3} & \mbox{if $\{a_1, \ldots, a_n\} \subseteq A$, $a_0 \not\in A$, $fn(y) \cap \{a_1, \ldots, a_n\} = \emptyset$, $fn(z) \cap \{a_1, \ldots, a_n\} = \emptyset$ and} &  \\
& \mbox{$ \Pi_{i=1}^n x_i = w \para  \Pi_{i=1}^n w_i$, then} \\
&\quad \restr{A}(((\sum_{j=1}^{k} \underline{a}_n.\ldots \underline{a}_1.a_0.y_j +y) \para \Pi_{i=1}^n \overline{a}_i.x_i) \para z) = \\
& \qquad \qquad  \restr{A}(((\sum_{j=1}^{k} \underline{a}_n.\ldots \underline{a}_1.a_0.(y_j \para w) +y) \para \Pi_{i=1}^n \overline{a}_i.w_i) \para z)\\

{\bf Ps4} & \mbox{if $\{a_0, \ldots, a_n\} \subseteq A$, $fn(y) \cap \{a_0, \ldots, a_n\} = \emptyset$, $fn(z) \cap \{a_0, \ldots, a_n\} = \emptyset$ and} &\\
& \mbox{$ \Pi_{i=0}^n x_i = w \para  \Pi_{i=0}^n w_i$, then} \\
&\quad \restr{A}(((\sum_{j=1}^{k} \underline{a}_n.\ldots \underline{a}_1.a_0.y_j +y) \para \Pi_{i=0}^n \overline{a}_i.x_i) \para z) = \\
& \qquad \qquad  \restr{A}(((\sum_{j=1}^{k} \underline{a}_n.\ldots \underline{a}_1.a_0.(y_j \para w) +y) \para \Pi_{i=0}^n \overline{a}_i.w_i) \para z)\\
\end{array}$

\hrulefill 

\end{center}}
}
\caption{Some further conditional axiom schemata for structure-preserving bisimilarity}\label{axiom-tab2}
\end{table}

\begin{theorem}{\bf (Soundness)}\label{sound-th-fnm}
For every $p, q \in  \mathcal{P}_{FNM}$, if $E \vdash p = q$, then $p \sim_{sp} q$.
\proof
The proof is by induction on the proof of $E \vdash p = q$. The 
thesis follows by observing that all the axioms in $E$ are sound by the many propositions proved in Section \ref{comp-sp-alg-sec},
and that $ \sim_{sp}$ is a congruence.
\fine
\end{theorem}

Of course, this axiomatization is incomplete, as illustrated, e.g., in Example \ref{ex-postset3}.

\begin{example}
Let us consider the two unbounded producer/consumer systems in Figure \ref{upcs-fig}. It is easy to see that the net semantics of 
$\restr{a}(P_1 \para C_1)$, with $P_1 \eqdef prod.(P_1 \para D_1)$,
$D_1 \eqdef \overline{a}.\nil$, $C_1 \eqdef \underline{a}.del.C_1'$ and $C_1' \eqdef cons.C_1$, is a net isomorphic to that on the left,
while the net semantics of $\restr{b}(P_2 \para C_2)$, with $P_2 \eqdef prod.(P_2 \para D_2') + prod.(P_2 \para D_2'')$,
$D_2' \eqdef \overline{b}.\nil$, $D_2'' \eqdef \overline{b}.\nil$, $C_2 \eqdef \underline{b}.del.C_2'$ and $C_2' \eqdef cons.C_2$, 
is a net isomorphic to that on the right.

By using the axioms in $E$, it is possible to equate these two terms. First of all, by axiom {\bf R3}, we get that 
$\restr{a}(P_1 \para C_1)$ is equal to $\restr{b}((P_1 \para C_1)\sost{b}{a})$, so that, by applying the substitution, the resulting term
is $\restr{b}(P_{1_{\sost{b}{a}}} \para C_{1_{\sost{b}{a}}})$, with 

$\begin{array}{rcllll}
P_{1_{\sost{b}{a}}} & \eqdef & prod.(P_{1_{\sost{b}{a}}} \para D_{1_{\sost{b}{a}}}) & \quad
D_{1_{\sost{b}{a}}} & \eqdef & \overline{b}.\nil\\
C_{1_{\sost{b}{a}}} & \eqdef & \underline{b}.del.C_{1_{\sost{b}{a}}}'  & \quad
C_{1_{\sost{b}{a}}}' & \eqdef & cons.C_{1_{\sost{b}{a}}}.\\
\end{array}$

Now, by recursion congruence, we get that  $D_2' = D_2''$, so that, by substitutivity, we get $prod.(P_2 \para D_2') = prod.(P_2 \para D_2'')$,
and so, by axiom {\bf A4}, we get $prod.(P_2 \para D_2') + prod.(P_2 \para D_2') = prod.(P_2 \para D_2')$.
Let us define a new constant $P_2' \eqdef prod.(P_2' \para D_2')$; clearly, by recursion congruence, $P_2 = P_2'$.
Again, by recursion congruence, we also have that
$D_{1_{\sost{b}{a}}} = D_2'$, so that $prod.(x \para D_{1_{\sost{b}{a}}}) = prod.(x \para D_2')$ holds by substitutivity. Hence, by recursion congruence,
we get $P_{1_{\sost{b}{a}}} = P_2'$. 
By axiom {\bf C1}, we get $C_{1_{\sost{b}{a}}} = \underline{b}.del.cons.C_{1_{\sost{b}{a}}}$ as well as 
$C_{2} = \underline{b}.del.cons.C_{2}$, so that, by recursion congruence, we get  $C_{1_{\sost{b}{a}}} = C_2$.
Then, by substitutivity, $\restr{b}(P_{1_{\sost{b}{a}}} \para C_{1_{\sost{b}{a}}}) = \restr{b}(P_2 \para C_2)$,
so that, by transitivity, we get $\restr{a}(P_1 \para C_1) = \restr{b}(P_2 \para C_2)$.
\fine
\end{example} 

%
\section{Conclusion}\label{conc-comp-sec}
%

Structure-preserving bisimulation \cite{G15} is a very intuitive, rather manageable, truly concurrent
behavioral relation, whose process-oriented characterization is {\em causal-net bisimilarity} \cite{G15,Gor22},
a behavioral equivalence fully respecting causality and the branching
structure of systems. 
 
The decidability of sp-bisimilarity over finite (unbounded) P/T nets is an open problem, because the negative observation in \cite{Esp98}
does not apply to it. 
Nonetheless, structure-preserving bisimilarity is decidable on bounded nets in exponential time \cite{CG21a}. In fact, the 
set of reachable markings are finitely many for a bounded net, and the set of linkings definable on 
a pair of markings of equal size $k$ has size $k!$. Hence, there are finitely many 
relations (that are all finite) composed of linkings, and it is enough to exhaustively check whether each of them is a structure-preserving bisimulation
and, in the positive case, to check whether it contains a linking projected on the two initial markings of interest.

Note that the class of bounded finite P/T nets (i.e., the class onto which sp-bisimilarity is decidable) is much larger than
the class of nets that is used to give semantics to {\em regular CCS} \cite{Mil89,GV15} (also called RCS in \cite{Gor17}), 
because these process terms give origin to finite
P/T nets whose transitions have preset of size 1 or 2, and if the preset size is 1, then its postset size is 1 at most,
while if the preset size is 2, then its label is $\tau$ and the postset size is 2 at most.

Structure-preserving bisimilarity is decidable on BPP nets (i.e., nets whose transitions have preset of size 1), because on this class of 
nets it coincides with {\em team} bisimilarity \cite{Gor17b,Gor22}, which is decidable in polynomial time.

Van Glabbeek \cite{G15} argued that structure-preserving bisimilarity is the most appropriate behavioral 
equivalence for Petri nets, as it is the only one
respecting a list of desirable requirements he proposed. Among these, there is `compositionality', up to 
structure-preserving bisimilarity, of the operators (recursion not considered) of the process 
algebra CCSP, that Olderog proposed in his monograph  \cite{Old}
and equipped with a {\em safe net} (i.e., in each reachable marking each place may contain one token at most) semantics.
In this paper we have complemented his result by proving that structure-preserving bisimilarity
can be used to give a compositional semantics of the process algebra 
FNM \cite{Gor17}, which truly represents all (and only) the finite P/T nets, up to isomorphism. In this way, we have obtained for the first time
a compositional semantics, fully respecting causality and the branching
structure of systems, for the class of all the finite P/T Petri nets.

It is interesting to observe \cite{Gor17} that also the coarser step bisimilarity \cite{NT84} is a congruence for all the FNM operators, while
interleaving bisimilarity is not a congruence for the FNM operator of parallel composition, so that,
in order to give a satisfactory account of this process algebra, a non-interleaving behavioral semantics is strictly necessary.
Since structure-preserving bisimilarity may appear even too concrete, as it may fail to equate markings generating the same causal nets 
(cf. the definition of {\em i-causal net bisimilarity} in \cite{CG21a}), a challenging open problem is to see whether it is possible 
to define a compositional semantics for some other behavioral equivalence in between step bisimilarity (a bit too abstract) 
and structure-preserving bisimilarity (a bit too concrete).

To the best of our knowledge, algebraic properties of a truly concurrent behavioral equivalence for a calculus semantically 
richer than BPP\footnote{BPP is the acronym of {\em Basic Parallel Processes} \cite{Ch93}, a simple CCS \cite{Mil89,GV15} subcalculus 
(without the restriction operator) whose
processes cannot communicate. In \cite{Gor17} a variant of BPP, which requires guarded summation (as in 
BPP$_g$ \cite{Ch93}) and also that the body of each process constant is guarded (i.e., guarded recursion), is actually
shown to represent {\em all and only} the BPP nets, up to net isomorphism.
BPP is a subcalculus of FNM.}
have never been investigated before. Here we have shown that for sp-bisimilarity, besides the usual laws for choice, parallel composition 
and restriction, we have also less standard laws related to strong prefixing and process constants, as well as a number of 
additional original laws about the interplay 
between parallel composition and restriction. 

An interesting open problem is to find a {\em complete} axiomatization of sp-bisimilarity over FNM. However, note that such a complete set of axioms might be available only if sp-bisimilarity is decidable (that is not known yet), and in such a case, since we use infinitary axiom schemata, 
it seems that such a hypothetical set would be infinite. 

It is interesting to observe that in the special subcase of BPP \cite{Ch93,Gor17} the situation is much better. In fact, \cite{Gor17b} describes
a {\em finite}, sound and {\em complete}, axiomatization of {\em team} bisimilarity (which is an alternative formulation of sp-bisimilarity for BPP nets), 
while \cite{Gor22} describes
a {\em finite}, sound and {\em complete}, axiomatization of {\em h-team} bisimilarity (which is an alternative characterization of fully-concurrent bisimilarity \cite{BDKP91} for BPP nets).


\begin{thebibliography}{11}



\bibitem{ABS91}
C. Autant, Z. Belmesk, Ph. Schnoebelen,
\newblock  Strong bisimilarity on nets revisited, 
\newblock in Procs. PARLE'91, vol. II: Parallel Languages, LNCS 506, 295-312, Springer, 1991.





\bibitem{BD87}
E. Best, R. Devillers, 
\newblock  Sequential and concurrent behavior in Petri net theory,
\newblock {\em Theoretical Computer Science} 55(1):87-136, 1987.


\bibitem{BDKP91}
E. Best, R. Devillers, A. Kiehn, L. Pomello,
\newblock  Concurrent bisimulations in Petri nets,
\newblock {\em Acta Inf.}  28(3): 231-264, 1991.




\bibitem{CG21a}
A. Cesco, R. Gorrieri,
\newblock Decidability of two truly concurrent equivalences for finite bounded Petri nets,
 \newblock CoRR, abs/2104.14856, 2021, https://arxiv.org/abs/2104.14856
 
\bibitem{Ch93}
S. Christensen, 
\newblock {\em Decidability and Decomposition in Process Algebra},
\newblock Ph.D. Thesis, University of Edinburgh (1993)

\bibitem{DDM89}
P.~Degano, R.~De~Nicola, U.~Montanari,
\newblock Partial ordering descriptions and observations of nondeterministic concurrent systems,
\newblock in (J. W. de Bakker, W. P. de Roever, G. Rozenberg, Eds.)
\newblock {\em Linear Time, Branching Time and Partial Order in Logics and Models for Concurrency}, LNCS 354, 438-466, Springer, 1989.


\bibitem{DesRei98}
J. Desel, W. Reisig,
\newblock Place/Transition Petri nets,
\newblock  in {\em Lectures on Petri Nets I: Basic Models}, 
LNCS 1491, 122-173, Springer, 1998.


\bibitem{Esp98}
J. Esparza,
\newblock Decidability and complexity of Petri net problems: An introduction,
\newblock {\em Lectures on Petri Nets I: Basic Models},
\newblock   LNCS 1491, 374-428, Springer, 1998.

\bibitem{vGG89}
R.J. van Glabbeek, U. Goltz,
\newblock  Equivalence notions for concurrent systems and refinement of actions,
\newblock in Procs. MFCS'89, LNCS 379, 237-248, Springer, 1989.

\bibitem{GR83}
U. Goltz, W. Reisig,
\newblock The non-sequential behaviour of Petri nets,
\newblock {\em Information and Control} 57(2-3):125-147, 1983.



\bibitem{G15}
R.J. van Glabbeek,
\newblock  Structure preserving bisimilarity - Supporting an operational Petri net semantics of CCSP, 
\newblock in (R. Meyer, A. Platzer, H. Wehrheim, Eds.)
\newblock {\em Correct System Design} --- Symposium in Honor of Ernst-R\"udiger Olderog on the Occasion of His 60th Birthday,  
LNCS 9360, 99-130, Springer, 2015.

\bibitem{GM90}
R. Gorrieri, U. Montanari,
\newblock Towards hierarchical specification of systems: A proof system for strong prefixing,
\newblock {\em Int. J. of Foundations of Computer Science} 1(3): 277-293, 1990.

\bibitem{GV15}
R. Gorrieri, C. Versari,
{\em Introduction to Concurrency Theory: Transition Systems and CCS},
EATCS Texts in Theoretical Computer Science, Springer-Verlag, 2015.



\bibitem{Gor17}
R.~Gorrieri,
\newblock {\em Process Algebras for Petri Nets: The Alphabetization of Distributed Systems},  
\newblock EATCS Monographs in Computer Science, Springer, 2017.


\bibitem{Gor17b}
R.~Gorrieri,
\newblock Team bisimilarity, and its associated modal logic, for BPP nets, 
\newblock {\em Acta Informatica} 58(5):529-569, 2021. 



\bibitem{Gor21}
R. Gorrieri,
\newblock Place bisimilarity is decidable, indeed!,
\newblock CoRR, abs/2104.01392, 2021, {https://arxiv.org/abs/2104.01392}


\bibitem{Gor22}
R.~Gorrieri,
\newblock A study on team bisimulation and h-team bisimulation for BPP nets, 
{\em Theoretical Computer Science} 897:83-113, 2022.




\bibitem{Jan95}
P. Jan\u{c}ar,
\newblock Undecidability of bisimilarity for Petri nets and some related problems,
\newblock {\em Theoretical Computer Science} 148(2):281-301, 1995.

\bibitem{KM69}
R.M. Karp, R.E. Miller,
\newblock Parallel program schemata,
\newblock {\em Journal of Computer and System Sciences} 3(2):147-195, 1969.

%
\bibitem{Mil84} R. Milner.
\newblock A complete inference systems for a class of regular behaviors,
\newblock {\em J. Comput. System Sci.}  28: 439-466, 1984.

\bibitem{Mil89} 
R. Milner,
\newblock  {\it Communication and Concurrency},
\newblock Prentice-Hall, 1989.


\bibitem{MOP89}
A.W.~Mazurkiewicz, E.~Ochmanski, W.~Penczek.
\newblock Concurrent systems and inevitability.
\newblock {\em Theoretical Computer Science}, 64:281--304, 1989.

\bibitem{NT84}
M. Nielsen, P.S. Thiagarajan,
\newblock  Degrees of non-determinism and concurrency: A Petri net view,
\newblock in Procs. of the Fourth Conference on Foundations of Software Technology and Theoretical Computer Science (FSTTCS'84),
LNCS 181, 89-117, Springer-Verlag, 1984.

\bibitem{Old}
E.R.~Olderog,
\newblock {\em Nets, Terms and Formulas},
\newblock Cambridge Tracts in Theoretical Computer Science 23, Cambridge University Press, 1991.

\bibitem{Park81}
D.M.R. Park,
\newblock Concurrency and automata on infinite sequences,
\newblock In Proc. 5th GI-Conference on Theoretical Computer Science, LNCS 104, 167-183, 
Springer, 1981.

\bibitem{Parrow-hb}
J. Parrow,
\newblock An introduction to the $\pi$-calculus,
\newblock Chapter 8 of  {\em Handbook of Process Algebra} (J.A. Bergstra, A. Ponse, S.A. Smolka, eds.), 479-543, Elsevier, 2001.


\bibitem{Pet81}
J.L. Peterson,
\newblock {\em Petri Net Theory and the Modeling of Systems}, Prentice-Hall, 1981.

\bibitem{RT88}
A. Rabinovich,  B.A. Trakhtenbrot,
\newblock Behavior structures and nets,
\newblock {\em Fundamenta Informaticae} 11(4):357-404, 1988.

\bibitem{Salomaa}
A. Salomaa,
\newblock Two complete axiom systems for the algebra of regular events,
\newblock {\em Journal of the ACM} 13(1): 58-169, 1966. doi:10.1145/321312.321326

%
\bibitem{SW}
D. Sangiorgi, D. Walker,
\newblock {\em The $\pi$-calculus: A Theory of Mobile Processes},
\newblock Cambridge University Press, 2001.


\end{thebibliography}
\end{document}